\documentclass[3p,times,numbers,sort]{elsarticle} 
\usepackage{lineno,hyperref}
\modulolinenumbers[5]
\usepackage{amsmath}
\usepackage{commands}
\usepackage{amsfonts}
\usepackage{amsthm}
\usepackage{amssymb}
\usepackage{graphicx}
\usepackage{enumitem}
\usepackage{algorithm}
\usepackage{algpseudocode}
\usepackage{setspace}
\usepackage{mathtools}
\usepackage{subcaption}
\usepackage{ragged2e} 
\usepackage{natbib}
\usepackage{color}
\usepackage{fancyhdr}

\fancypagestyle{plain}{%
   \fancyhf{} 
   \cfoot{\vspace{0.5cm}  DISTRIBUTION STATEMENT A. Approved for public release: distribution unlimited.}\rfoot{\thepage}
   \renewcommand{\headrulewidth}{0pt}
}

\journal{Journal of Fluids and Structures}

\begin{document}

\begin{frontmatter}

\title{Stability of Helical Vortex Structures Shed from Flexible Rotors}

\author[add1]{Steven N. Rodriguez}
\author[add2]{Justin W. Jaworski}
\author[add1]{John G. Michopoulos}

\address[add1]{Computational Multiphysics Systems Laboratory, U.~S.~Naval Research Laboratory; Washington, DC, 20375, USA}
\address[add2]{Department of Mechanical Engineering and Mechanics, Lehigh University; Bethlehem, PA, 18015, USA}




\begin{abstract}
The presented investigation is motivated by the need to uncover connections between underlying rotor fluid-structure interactions and vortex dynamics to fatigue performance and characterization of flexible  rotor blades, their hub, and their supporting superstructure. Towards this effort, temporal stability characteristics of tip vortices shed from flexible rotor blades are investigated numerically. An aeroelastic free-vortex wake method is employed to simulate the helical tip vortices and the associated velocity field. A linear eigenvalue stability analysis is employed to quantify stability trends (growth-rate v.~perturbation wavenumber) and growth-rate temporal evolution of tip vortices. Simulations of a canonical rotor with rigid blades and its generation of tip vortices are first conducted to validate the stability analysis employed herein. Next, a stationary wind turbine is emulated using the National Renewable Energy Laboratory 5MW reference wind turbine base design to investigate the impact rotor aeroelasticity has on tip vortex stability evolution in time. Blade flexibility is shown to reduce the sensitivity of tip vortex destabilization to low wavenumber perturbations, also blade-pitch reduces growth-rate magnitude and alters the growth-rate peak dependence on perturbation wavenumber, all of which have in the past not been reported in the rotorcraft literature. The presented investigation aims to develop insight into the tip vortex kinematics and stability of the NREL 5MW reference wind turbine. However, the frameworks presented herein can be applied to generalized rotor designs to work towards identifying the impact tip-vortex kinematics and stability have on fatigue loading and adverse blade-vortex interaction effects, such as excessive noise emission and rotor vibrations.
\end{abstract}

\begin{keyword}
Tip vortices, helical vortex dynamics and stability, rotor near-wakes, rotor aeroelasticity
\end{keyword}

\end{frontmatter}

\fancypagestyle{pprintTitle}{%
\lhead{} \chead{}\rhead{}
\lfoot{\textit{Preprint submitted to the Journal of Fluids and Structures}}\cfoot{\vspace{0.5cm}  DISTRIBUTION STATEMENT A. Approved for public release: distribution unlimited.}}\rfoot{}
\renewcommand{\headrulewidth}{0.0pt}

\pagestyle{plain}{%
\lhead{}\chead{}\rhead{}
\lfoot{\footnotesize \textit{Preprint submitted to the Journal of Fluids and Structures}}\cfoot{ \vspace{0.5cm} DISTRIBUTION STATEMENT A. Approved for public release: distribution unlimited.}\rfoot{\thepage}
\renewcommand{\headrulewidth}{0.0pt}
}


\section{Introduction}

\begin{figure}[h!]
\center
\includegraphics[scale=0.8, trim=1cm 8cm 0 9cm]{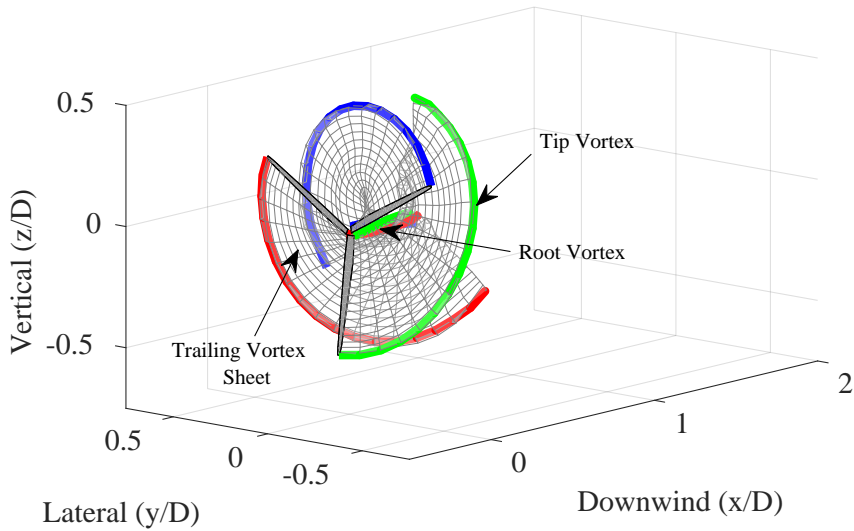}
\caption[Rotor near-wake aerodynamic vortex structures]{Rotor near-wake aerodynamic vortex structures}
\label{helical_vortex_structures}
\end{figure}

Unlike the vortices shed from fixed-wing aircraft, the wake of a rotor consists of a helical vortex structure that persists within proximity of the rotor plane. The helical vortex structure is a byproduct of the lift distribution across the span of the rotorblade and the mutual interaction of the vortex elements that rollup and generate strong tip and root vortices, as shown in Fig.~\ref{helical_vortex_structures}. However, bluff bodies at the center of rotors, such as nacelles on wind turbines, have historically been associated with the destabilization and dissipation of root vortices \cite{sherry2013characterisation}. Therefore, tip vortices are commonly the dominant aerodynamic structure in the wake of rotors and rotorcraft \cite{bhagwat2000correlation}. These tip vortices persist as strong coherent structures that can impact rotor blade loading and their associated performance significantly \cite{bhagwat2000correlation,leishman:14}. 

\begin{figure}[h!]
\center
\includegraphics[scale=0.4, trim=0 4cm 0 4cm]{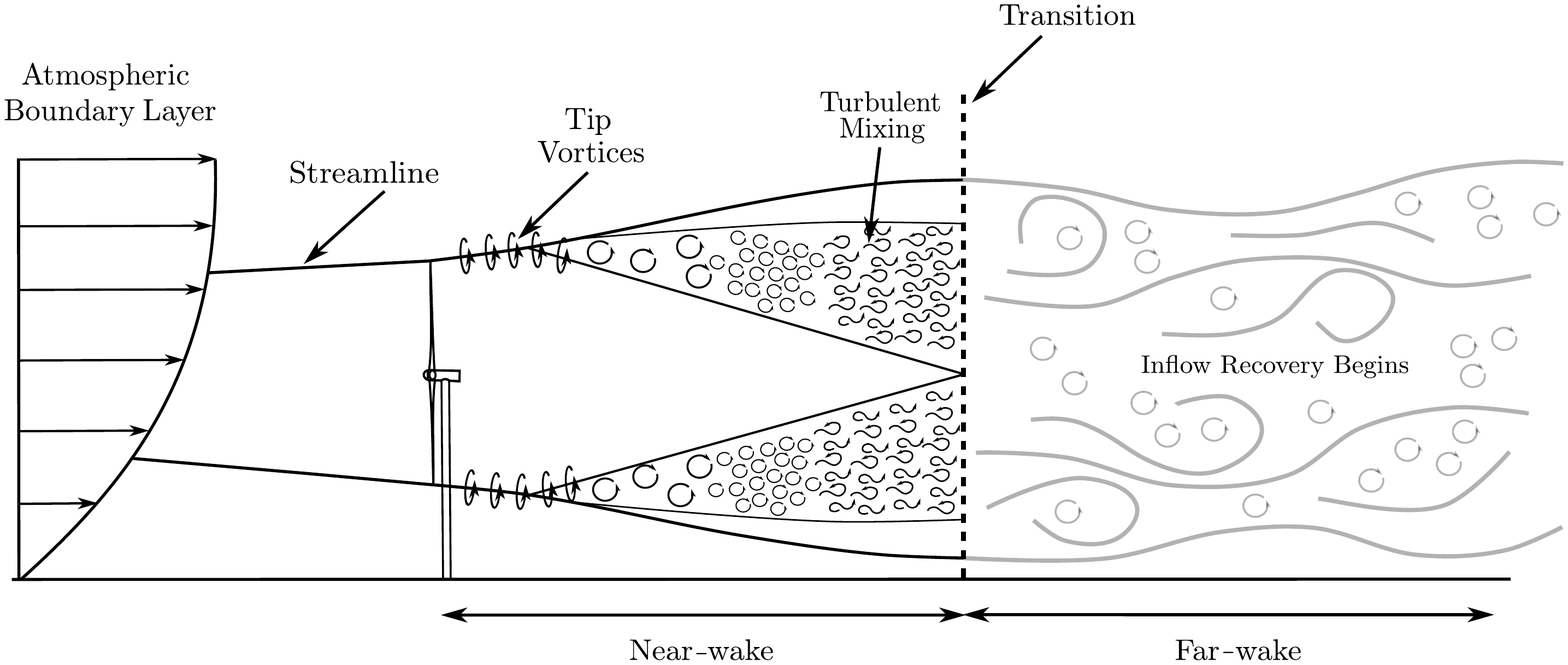}
\caption[Rotor-wake operational stages]{Example of wake stages generated by the operation of a wind turbine}
\label{wage_stages}
\end{figure}

As the wake ages and travels downstream of the rotor, the tip vortices destabilize and become part of a complex wake breakdown and transition into the far-wake region, in which inflow recovery begins, as shown in Fig.~\ref{wage_stages}. These wake stages have been extensively studied for decades for a variety of applications that range from coaxial rotorcraft to wind farms \citep{coleman1997survey, burton2011wind, hansen2006state, van2015rotor, okulov2015rotor, leishman:14, vermeer2003wind, porte2011large, lignarolo2015tip}.  The velocity field of the transition and far-wake regions have negligible impact on rotor blade loading and performance as compared to the near-wake region \cite{leishman:14, sorensen2011instability}. The aerodynamics of the transition and far-wake regions is usually of interest for applications in which a collection of rotors must operate in the wake of others, such as in wind farms. 

 Research concerned with the near-wake generally involves efforts aimed toward understanding the impact of near-by vortex structures on rotor performance and the vortex interactions with rotor blades \citep{leishman:14,bhagwat:4, ramasamy2010flowfield, ramasamy2015hover, yoon2014simulations,yoon2017computations,cesnik2004active}. These near-wake investigations are mainly motivated by the need to understand,  control and minimize adverse operational impact, such as excessive blade-vortex interaction noise, blade and supporting super-structure fatigue and material degradation, and poor rotor maneuverability. Furthermore, recent progress in additive manufacturing (AM) now enables the manufacturing of components in the rotor blade-supporting superstructure, which further emphasizes the need to understand fatigue and material degradation concerns due to AM-induced porosity and surface roughness \cite{jiang2018control}. Consequently, this underlines the need for an effort to understand how the aeroelastic loading is affected by the vortex structures shed from flexible rotors.  
 
 Exploration of the complex aerodynamics of rotor-wakes and associated temporal characteristics as the wake travels downstream of the rotor has generated a large field of active research problems that aim to improve rotor technology and their physics involving \emph{inter alia} turbulent rotor inflow modeling, near-wake and far-wake simulations, rotor-to-rotor near-wake interaction, rotor-wake instabilities, wake meandering, and wake stability \citep{sanderse2011review, ma2019analysis, leishman:14, hansen2006state, sorensen2011instability, vermeer2003wind}. 

The current work is focused on the research of kinematics and stability of shed helical vortices in the near-wake region. Investigations into the stability of coherent helical vortex structures generally study the influence of two types of disturbances: short-wave and long-wave perturbations \citep{quaranta2015long}. Short-wave perturbations are able to disturb vortex cores and are generally a byproduct of external strain fields or the curvature and torsion of the vortex itself \citep{quaranta2015long}. The present study is concerned with long-wave perturbations, which are generated by external disturbances with characteristic lengths much larger than the vortex core, such as gusts and atmospheric turbulence \citep{quaranta2015long}.

Fluid mechanical insight into the dynamics and stability of helical vortices has practical benefits in engineering applications, such as flow control, design and manipulation of early or prolonged wake breakdown, and the ability to anticipate unstable and catastrophic wake behaviors in rotor operations, such as vortex-ring state in helicopter flight \citep{leishman:2}. The benefits associated with understanding helical vortex dynamics and stability have sustained research activity on this topic for almost a century. The earliest recorded effort was conducted by \citet{levy1928steady}, which investigated the stability of a helical vortex filament subjected to long-wave sinusoidal perturbations. \citet{widnall1972stability} rectified their incorrect treatment of the Biot-Savart singularity along the vortex line by using the cutoff method and identified three modes of helical vortex instability:  short-wave, long-wave, and mutual-inductance modes. \citet{gupta1974theoretical} extended the work of Widnall to investigate the stability of multiple helical vortex filaments wound about a common axis. Their work concluded that the magnitude of their stability trends (growth-rate as a function of perturbation wavenumber) depends on the number of helical vortices in the domain. Specifically, an increase in the number of helical vortex structures will result in an increase of growth rate levels. It was also found that the divergence rates decreased as the perturbation wavenumbers increased. Decades later, following the advances in numerical and computational methods, Bhagwat and Leishman conducted a stability analysis of tip vortices generated numerically by a helicopter rotor \citep{bhagwat:4}. Bhagwat and Leishman determined that the stability trends depend on the number of intertwined helical vortices, as suggested previously by Gupta and Loewy. Bhagwat and Leishman also speculated that tip vortices are most unstable to perturbation wavenumbers equal to half-integer multiples of the number of helical vortex filaments, i.e. $\omega=N(k-1/2)$, where $\omega$ is the perturbation wavenumber, $N$ is the number of tip vortex filaments shed from rotor blades, and $k=1,2,\ldots$, denotes any natural number. Similar findings as reported by Bhagwat and Leishman were also found by \citet{ivanell2010stability} who used a large eddy simulation (LES) to study the stability of a wind turbine wake. More recent theoretical and numerical investigations 
into helical vortex dynamics and their stability have led to similar conclusions that the near-wake is unconditionally unstable and that the most unstable modes occur at wavenumber perturbations equal to half-integer multiples of the number of tip vortices in the domain \citep{okulov2007stability,rodriguez2016sweden, rodriguez2017tip, rodriguezJERT}. 

Experimental research of helical vortex dynamics and stability has historically lagged behind theoretical and numerical investigations due to the technological limitations of capturing the three-dimensional velocity field of these vortex structures. However, advancements in flow visualization and tracking in the past decade have enabled experimental investigations to reach a state where they can examine the validity of the conclusions postulated by theoretical and numerical studies. For example,  \citet{felli2011mechanisms} tracked the generation of the helical vortex structure of a two-bladed, three-bladed, and four-bladed propeller through velocity measurements and high-speed visualizations, and were the first to observe the onsets of short-wave, long-wave, and mutual-inductance instabilities of helical vortices predicted by \citet{widnall1972stability}. Another recent study by \citet{quaranta2015long} sought to conduct long-wave instability experiments of helical vortices for the purposes of comparing their results against classical theoretical works. Their work was generated by a single-bladed rotor, the velocity field and vorticity distributions were captured via particle image velocimetry (PIV) measurements, and the vortex was visualized by applying fluorescent dye to the rotor-blade tip. Their works showed consistent agreement with classical stability analyses, i.e.~the experimental helical vortex growth-rates caused by long-wave perturbations agreed well with the theoretical results of \citet{widnall1972stability} and \citet{gupta1974theoretical}. It was also observed that helical vortices are extremely receptive to small-amplitude spatial-temporal perturbations, which further reinforces the theoretical conclusions that helical vortices are unconditionally unstable. The reader is referred to \cite{felli2011mechanisms, quaranta2015long, sorensen2011instability, iungo2013linear, nemes2012generation, nemes2015mutual} for more recent works on experimental helical vortex stability.  

Recent developments of helical-vortex dynamics and stability suggest that fundamental knowledge regarding stability modes and mechanisms has begun to converge for simple, uniform, constant-pitch, and steady helical vortex structures. However, near-wakes encountered in real rotor engineering applications are rarely simplistic in nature, and generally involve multiphysics phenomena, such as aerodynamic-elasticity (aeroelasticity), which introduce additional layers of complexity to tip-vortex dynamics and stability. The present investigation examines the motion and stability of more realistic vortex structures generated by the multiphysics behavior of flexible rotors. \emph{Specifically, the work aims to investigate the temporal stability characteristics of tip vortices that are captured in the near-wake region that have been deformed by rotor-blade aeroelasticity}. Ultimately, the present work seeks to contribute fundamental knowledge of realistic rotor near-wake dynamics in the efforts to improve rotor-wake applications and technology, such as tip-vortex flow control, and that can work towards minimizing adverse blade-vortex interaction effects, such as excessive noise emission and large blade vibrations.

The remainder of this paper is structured as follows. Section \ref{FVM_sec} reviews the free-vortex wake method that is used to simulate rotor operation and to generate the helical wake. Section \ref{methods} introduces a linear-eigenvalue stability analysis, which is used to quantify temporal stability characteristics of tip vortices. This section also introduces a canonical three-bladed rotor configuration to validate the stability analysis presented in this paper against earlier investigations. Section \ref{flexible_sec} applies the stability analysis to tip vortices generated by a stationary zero-pitched rotor (i.e., rotor-plane perpendicular to inflow) and pitched rotor configuration of the National Renewable Energy Laboratory (NREL) 5MW reference wind turbine rotor with flexible blades under a range of static rotor (non-rigid body rotor kinematics) conditions, such as variable tip speed ratio, inflow speed, and blade pitch. Finally, the conclusions of this research paper are discussed in Section \ref{conclusion_sec}.

\section{Aeroelastic Free-Vortex Wake Method}\label{FVM_sec}

The numerical model used to simulate rotor operation and generate the helical vortex system is based upon the free-vortex wake method (FVM) presented by  \citet{rodriguezphd} and \citet{rodriguezJRE, rodriguezJRE2} developed to emulate the NREL 5MW reference wind turbine and its aeroelastic performance. Their model coupled a linear-kinematic beam theory for spinning structures to the free-vortex wake method code developed by Sebastian and Lackner, known as WInDS \citep{sebastian:10, snl:5}. This section provides a comprehensive context of the aeroelastic FVM framework. First, the aerodynamics and vortex modeling employing both the Vatistas vortex model \cite{leishman:14, snl:5} and vortex cut-off models \cite{snl:5} are presented. Next, an overview of the rotor structural dynamics and brief note on the fluid-structure interaction coupling scheme is presented.  Note that in the context of this paper all vectors are noted with a bold face,  (i.e., $ \mathbf{x}$) and matrices are noted as a bold faces inside square brackets (i.e., $\left[\mathbf{x}\right]$).

\subsection{Aerodynamics}
The flow physics modeled by the free-vortex wake method is governed by,
\begin{equation}
    \frac{d\mathbf{r}}{dt}=\mathbf{V}_{\infty}+\mathbf{V}_{\rm{induced}}+\mathbf{V}_{\rm{rbm}},
\end{equation}
where the left hand side tracks the velocity of the discrete filaments in the wake, and $\mathbf{V}_{\infty}$, $\mathbf{V}_{\rm{induced}}$, and $\mathbf{V}_{\rm{rbm}}$, are the freestream velocity, the induced velocity, and the velocity generated by rigid body motions (rbm) of the rotor (such as wave-induced motions of a floating offshore wind turbine), respectively. A uniform freestream velocity in the axial direction is assigned and prescribed in the present study. The rigid-body motion velocity may be superimposed onto the rotor kinematics, as has been done in \citet{rodriguezphd} and \citet{rodriguezJRE, rodriguezJRE2} for wind turbine blades in an offshore environment. However, no such rotor motion is imposed in this investigation, i.e.~$\mathbf{V}_{\rm{rbm}}=\mathbf{0}$. 
The induced velocity is computed using the Biot-Savart law where the before-mentioned vortex core models are employed. Two standard forms of the Biot-Savart law appear in the literature: the traditional form \citep{leishman:14} and another that is computationally more efficient \citep{phillips2000modern}. For the reader's convenience we present both and provide the relations used to arrive at either. 

\begin{figure}[h!]
    \centering
    \includegraphics[scale=0.75, trim=0 0 0 0]{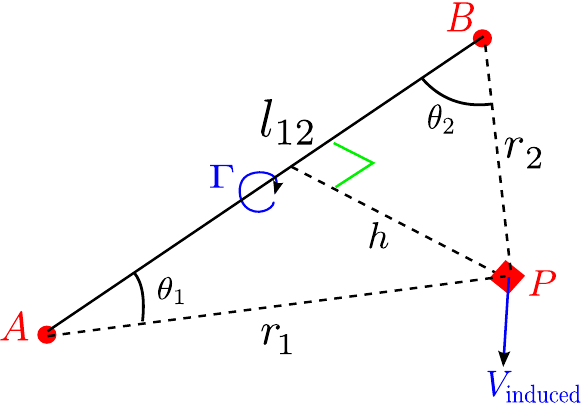}
    \caption{Induced velocity relationship of an $i$ \textsuperscript{th} semi-infinite vortex filament}
    \label{bs_law_fig}
\end{figure}

The traditional form of the Biot-Savart law for semi-infinite filaments is 
\begin{equation}
    \mathbf{V}_{\rm{induced}}=\frac{\Gamma}{4\pi}\int_l \frac{d\mathbf{l}\times \mathbf{r}}{|\mathbf{r}|^3}
    \label{traditional_BSL}
\end{equation}
where $\mathbf{r}$ is the position along a semi-infinite vortex filament. For a straight-segment filament the Biot-Savart can be written as 
\begin{equation}
\mathbf{V}_{\textup{induced},\:i}=\frac{\Gamma}{4\pi h}\left(\cos{\theta_1}-\cos{\theta_2}\right)\mathbf{e},
\end{equation}
where $i$ indicates the $i^{\textup{th}}$ filament, i.e.~the induced velocity contribution of the $i^{\textup{th}}$ vortex filament on a point in space. Geometric relationships of the $i$\textsuperscript{th} filament configurations illustrated in Fig.~\ref{bs_law_fig} lead to the following  expressions 
\begin{equation}
    h=\frac{|\mathbf{l}_{12}\times \mathbf{r}_1|}{l_{12}},\:\:\: \cos{\theta_1}=\frac{\mathbf{l}_{12} \cdot \mathbf{r_1}}{l_{12} r_1},\:\:\:\cos{\theta_2}=\frac{\mathbf{l}_{12} \cdot \mathbf{r}_2}{l_{12} r_2},\:\:\: \mathbf{e}=\frac{\mathbf{l}_{12}\times \mathbf{r}_1}{|\mathbf{l}_{12}\times \mathbf{r}_1|},
    \label{geometric_relations}
\end{equation}
where $\mathbf{e}$ is a column vector in $\mathbb{R}^3$.  Substitution of these expression into Eq.~\ref{traditional_BSL} and algebraic manipulation then yields Eq.~\ref{traditional_BSL2}, as presented in \cite{phillips2000modern},
\begin{equation}
    \mathbf{V}_{\textup{induced},\: i}=\frac{\Gamma}{4\pi}\frac{\mathbf{l}_{12}\times\mathbf{r}_1}{|\mathbf{l}_{12}\times \mathbf{r}_1|^2}\:\mathbf{l}_{12}\cdot \left(\frac{\mathbf{r}_1}{r_1}-\frac{\mathbf{r}_2}{r_2}\right)
    \label{traditional_BSL2}.
\end{equation}
Equation \ref{traditional_BSL2} can be rearranged to benefit numerical calculations by using the trigonometric relations
\begin{equation}
    \mathbf{l}_{12}=\mathbf{r}_1-\mathbf{r}_2,\:\:\:\mathbf{r}_1\cdot\mathbf{r}_2=r_1 r_2 \cos{\theta_1},\:\:\:|\mathbf{r}_1 \times \mathbf{r}_2|=r_1 r_2 \sin{\theta_1}.
\end{equation}
Substitution of these relations into Eq.~\ref{traditional_BSL2} yields
\begin{align}
\begin{split}
     \mathbf{V}_{\textup{induced},\:i}&=\frac{\Gamma}{4\pi}\frac{\mathbf{l}_{12}\times\mathbf{r}_1}{|\mathbf{l}_{12}\times \mathbf{r}_1|^2}\left(\left(\mathbf{r}_1-\mathbf{r}_2\right)\cdot \left(\frac{\mathbf{r}_1}{r_1}-\frac{\mathbf{r}_2}{r_2}\right)\right)\\&
 =\frac{\Gamma}{4\pi}\frac{\left(r_1+r_2\right) \left(\mathbf{l}_{12}\times\mathbf{r}_1\right)}{l_{12} r_1\left(l_{12}r_1+\mathbf{l}_{12}\cdot \mathbf{r}_1\right)}.
\end{split}
\end{align}
The form of the Biot-Savart law presented by \citet{phillips2000modern}, \citet{snl:5}, and \citet{sebastian:10} includes the vortex cutoff model \cite{snl:5} and the Vatistas vortex model \cite{leishman:14, snl:5}:

\begin{equation}
 \mathbf{V}_{\textup{induced},\:i} =
    \begin{cases}
             \frac{\Gamma}{4\pi}\frac{\left(r_1+r_2\right) \left(\mathbf{l}_{12}\times\mathbf{r}_1\right)}{l_{12} r_1\left(l_{12}r_1+\mathbf{l}_{12}\cdot \mathbf{r}_1\right)+\left(\delta_{c} l_{12}\right)^2}, &         \text{if } \textup{cutoff model},\\
            \frac{C_{\nu}\Gamma}{4\pi}\frac{\left(r_1+r_2\right) \left(\mathbf{l}_{12}\times\mathbf{r}_1\right)}{l_{12} r_1\left(l_{12}r_1+\mathbf{l}_{12}\cdot \mathbf{r}_1\right)},                                       &         \text{if } \textup{Vatistas model,}
    \end{cases}
    \label{modern_indvel}
\end{equation}
where $\delta_{c}$ is the vortex core cut-off radius, and where
\begin{equation}
    C_{\nu}=\left(\frac{\left(l_{12} r_1\right)^2-\left(\mathbf{l}_{12}\cdot \mathbf{r}_1\right)^2}{l_{12}^2}\right) \left(r_c^{2n}+\left(\frac{\left(l_{12} r_1\right)^2-\left(\mathbf{l}_{12}\cdot \mathbf{r}_1\right)^2}{l_{12}^2}\right)^{2n}\right)^{-1/n}.
\end{equation}
The geometric relations in Eq.~\ref{geometric_relations} enable  Eq.~\ref{modern_indvel} to be expressed in the traditional form as 

\begin{equation}
 \mathbf{V}_{\textup{induced},\:i} =
    \begin{cases}
             \frac{\Gamma}{4\pi}\frac{1}{\left(h+(\delta_{c} l_{12})^2\right)}\left(\cos{\theta_1}-\cos{\theta_2}\right)\mathbf{e}, &         \text{if } \textup{cutoff model},\\
            \frac{\Gamma}{4\pi}\frac{h}{\left(r_c^{2n}+h^{2n}\right)^{1/n}}\left(\cos{\theta_1}-\cos{\theta_2}\right)\mathbf{e},                                       &         \text{if } \textup{Vatistas model.}
            \label{traditional_indvel}
    \end{cases}
\end{equation}

The stability analysis presented in this investigation is derived from \citet{bhagwat:4}, who used the traditional form of the Biot-Savart law (Eq.~\ref{traditional_indvel} with the Vatistas model), which for consistency will be used for the remainder of this paper.

\subsection{Structural Dynamics}
The equations of motions to model individual rotor-blades were derived from classical beam theory for spinning structures \cite{leung1988spinning}. The equations of motion take into account axial, edgewise, flapwise, and torsional degrees-of-freedom, which are denoted by variables $u$, $v$, $w$, and $\phi$ respectively. The equations of motion for the rotor blade are as follows.\\ 

\noindent \textbf{Axial equation of motion}
\begin{equation}
m\left(2\Omega \dot{v}+\Omega^2\right) + EAu^{\prime \prime} = 0, 
\end{equation}

\noindent \textbf{Edgewise equation of motion}
\begin{equation}
m\left(\ddot{v} + 2\Omega \dot{u} - \Omega^2 v\right) + EI_yv^{\prime \prime \prime \prime} - P_{\Omega}v^{\prime \prime} = L_v, 
\end{equation}

\noindent \textbf{Flapwise equation of motion}
\begin{equation}
m\ddot{w}+EI_zw^{\prime \prime \prime \prime } - P_{\Omega}w^{\prime \prime}= L_w,
\end{equation}

\noindent \textbf{Torsional equation of motion}
\begin{equation}
m\left[J\ddot{\phi} + \Omega^2\left(I_y-I_z\right)\phi \right] - GJ\phi^{\prime \prime}= L_{\phi}.
\end{equation}
The primes denote derivatives take with respect to the spanwise variable, i.e.,~$d/dx$ \cite{rodriguezJRE}, and the overdots denote time derivatives, i.e.,~$d/dt$. The values $m$, $E$, $G$, $A$, $I_y$, $I_z$, and $J$ are the mass per unit length, elastic and shear moduli, cross-sectional area, edgewise and flapwise inertia, and the polar moment of inertia, respectively. Edgewise aerodynamic loading is taken into account by $L_v$, flapwise aerodynamic loading is taken into account by $L_w$, and moment aerodynamic loads are taken into account by $L_{\phi}$. These aerodynamic loads are retrieved by relating angle-of-attack FVM computations to lift, drag, and moment coefficient look-up tables for specific airfoil profiles reported in \cite{rodriguezJRE}. The aerodynamic loads on the rotor-blades are defined by the following relationships,
\begin{align}
L_v = \frac{1}{2}\rho_a c V_{\infty}^2c_d, \:\: L_w = \frac{1}{2}\rho_a c V_{\infty}^2c_l,  \:\:  L_{\phi} = \frac{1}{2}\rho_a c SV_{\infty}^2c_m,
\end{align}
where $\rho_a$ is the density of air, $c$ is the chord length of the airfoil, $S$ is the blade surface area, and $c_l$, $c_d$, $c_m$ are the spanwise section lift, drag, and moment coefficients, respectively.

Finally, it is important to note that our structural framework is based on idealizing the rotor-blades as independent cantilever beams without connections to a nacelle and tower, i.e., the current framework models a rotor in isolation. As a result, the current aeroelastic framework does not account for coupled modes between rotor-blades or rotor-blade boundary conditions that correspond to a nacelle connection and its connection to a flexible tower. As a result, it is expected that the natural frequencies of the current work differ from those reported by \citet{jonkman:6}, which report natural frequencies of the collective wind turbine and not independent rotor-blades. Specifically, the presented work arrives at a first natural frequency of $f_1=1.2$ Hz, which is the first flap-wise \emph{individual} rotor-blade mode. However, \citet{jonkman:6} have reported a first natural frequency of $f_1=0.69$ Hz, which corresponds to the first \emph{collective} flapwise mode. Despite difference in structural modeling, the presented work has been validated against work in \cite{jonkman:6} for a range of operational conditions, and has shown consistent results \cite{rodriguezJRE}. Future work will look into improving the current structural dynamics fidelity to account for a more comprehensive rotor model.

The structural dynamics equations are discretized in space by a linear Galerkin finite-element approach, as presented in \cite{cook2007concepts}. For details on the numerical framework employed here-in the reader is referred to \cite{rodriguezJRE}. The final semi-discretized linear system of equations is expressed by 
\begin{equation}
[\mathbf{M}]\ddot{\mathbf{D}}+[\mathbf{C}]\dot{\mathbf{D}}+[\mathbf{K}]\mathbf{D}=\mathbf{F},
\end{equation}
where $\mathbf{D}$, $\dot{\mathbf{D}}$, $\ddot{\mathbf{D}}$ represent the global vectors, which are the collection of local displacements, velocities, and accelerations. The matrices $\mathbf{M}$, $\mathbf{C}$, and $\mathbf{K}$, are the traditional global mass, damping, and stiffness matrices, as presented in \cite{cook2007concepts, bathe2}. The vector $\mathbf{F}$ represents the external forces applied to the structural system, which in this case are considered the aerodynamic loads. The semi-discrete equation of motion is integrated in time via the Newmark method presented in Ref.~\cite{bathe2}.

\subsection{Fluid-Structural Interaction Coupling}
The aeroelastic free-vortex wake framework employed in this investigation is a partitioned framework that solves the aerodynamic and structural equations of motion separately. The aerodynamics and structural dynamics are strongly-coupled via the Aitken $\Delta^2$ method \cite{weghs:10, erbts2014acc}. This coupling scheme enforces, at every time-step, kinematic and dynamic continuity conditions at the fluid-structure interface. A more detailed discussion on the coupling employed herein is discussed in Ref.~\cite{rodriguezJRE}.

\section{Methods for Stability and Dynamics Analysis} \label{methods}

The stability of helical vortex structures generated by rotor systems is typically considered in the context of either short-wave perturbations or long-wave perturbations \citep{quaranta2015long}. Short-wave perturbations disturb the vortex core structure, which may be generated by strain or torsion induced by a neighboring vortex. Long-wave perturbations consider the disturbance of the local helical geometry as a whole, without perturbing the vortex core. Long-wave perturbations may arise from atmospheric turbulence or any wave disturbance much larger than the vortex core radius and are the perturbation type considered in this work.

\subsection{Linear-Eigenvalue Stability Analysis} \label{stability_sec}

The current investigation considers long-wave perturbations on the rotor wake of a stationary wind turbine rotor (i.e., no rigid body motion due to offshore wave-induced forcing). The analysis used in this work was originally developed by \citet{bhagwat:4} to perturb the helicopter wake geometries harmonically in both space and time. Their investigation evaluated the tip-vortex stability of a hovering rotor, in which the tip-vortex geometry was generated by an FVM aerodynamic model. The current investigation also employs FVM. However, a major difference between the FVM employed by \citet{bhagwat:4} and the FVM used herein is that the current framework takes consideration of the impact that the trailing vortex sheet and its temporal changes have on the production of the tip vortex and roll-up effects. Hence, the wake geometry presented by this work is able produce additional physical insight into these effects on the stability analysis. 

It is also important to note that Bhagwat and Leishman investigated perturbations of the tip vortices only, neglected blade elasticity, and modeled a simple canonical rectangular rotor blade geometry. The work presented here will look into the stability of wakes generated by the flexible, non-uniform, and tapered rotor-blade geometry of the NREL rotor blade. However, the current stability analysis of the wake, like Bhagwat and Leishman, will only consider perturbing the tip vortices in isolation, even though the wake geometry was generated by accounting for the presence of the trailing vortex sheet. 

Due to the different modeling approach taken in this study to investigate tip-vortex stability, it can be anticipated that  study will recover similar trends reported by Bhagwhat and Leishman, but will also bring to light any deviation of these stability trends due to geometric distortion of tip-vortices as a result of rotor-blade deformation. The derivation of the tip vortex stability analysis is now presented.

\subsubsection{Perturbed Induced Velocity Formulation}

The free-vortex wake method depends on computing the local velocity of Lagrangian markers cast into the wake. Hence, the induced velocity field is perturbed by displacing the wake geometry by a small quantity, $\delta \mathbf{r}$, where $\delta$ is the perturbation operator and differs from the cutoff radius, $\delta_{c}$. To evaluate the stability of the system, the governing equation are perturbed as follows:
\begin{equation}\label{perturbed_goveq_2}
\frac{d(\mathbf{r}+\delta \mathbf{r})}{dt}=\mathbf{V}_{\rm{induced},\it{i}}(\mathbf{r}+\delta \mathbf{r})\:\:\: \rightarrow \:\:\: \frac{d\mathbf{r}}{dt}+\frac{d\left(\delta \mathbf{r}\right)}{dt}=\mathbf{V}_{\rm{induced},\it{i}}(\mathbf{r}+\delta \mathbf{r})
\end{equation}
\begin{equation}\label{perturbedeq}
\delta \dot{\mathbf{r}}=\mathbf{V}_{\rm{induced},\it{i}}(\mathbf{r}+\delta \mathbf{r})-\mathbf{V}_{\rm{induced},\it{i}}(\mathbf{r}).
\end{equation}
where the overdot in $\delta\dot{\mathbf{r}}$ represents the time derivative. The perturbed velocity can be expressed as an ordered series in $\delta \mathbf{r}$ as follows, where quadratic and higher order terms can be neglected:
\begin{equation}\label{indvel_series}
\mathbf{V}_{\rm{induced},\it{i}}(\mathbf{r}+\delta \mathbf{r})=\mathbf{V}_{\rm{induced},\it{i}}(\mathbf{r})+\delta \mathbf{V}_{\rm{induced},\it{i}} (\delta \mathbf{r}) + O((\delta\mathbf{ r})^2).
\end{equation}
Substituting Eq.~\ref{indvel_series} into Eq.~\ref{perturbedeq},
yields the perturbed governing equation:

\begin{equation}\label{perturbed_goveq_three}
\delta \dot{\mathbf{r}}=\delta \mathbf{V}_{\rm{induced},\it{i}} (\delta\mathbf{r}).
\end{equation}
The unperturbed velocity field is determined by the Biot-Savart law. Thus, substituting $\delta f= \frac{\partial f}{\partial x} \delta x$, where $f$ is some function of interest being perturbed that is dependent on a variable $x$, into the Biot-Savart law using the Vatistas or cutoff model for semi-infinite straight filaments from Eq.~\ref{traditional_indvel} will yield the perturbed induced velocity field:
\begin{align}
    \begin{split}
         \mathbf{V}^{\prime}&=\bar{\Gamma}\bigg[ \left(\cos{\theta_1}-\cos{\theta_2}\right)\mathbf{e} + \left(\delta\cos{\theta_1}-\delta\cos{\theta_2}\right)\mathbf{e} \\ & \:\:\:\:\:\:\:\:\:\:\:\:\:\:\:\:\:\:\:\:\:\:\:\:\:\:\:\:\:\:\:\:+ h_f\delta h \left(\cos{\theta_1}-\cos{\theta_2}\right)\mathbf{e} + \left(\cos{\theta_1}-\cos{\theta_2}\right)\delta\mathbf{e} \bigg ]
    \end{split}
\label{perturbed_goveq_four}
\end{align}
where for the Vatistas and cutoff models
\begin{equation}
 h_f =
    \begin{cases}
            -\frac{1}{\left(h+\left(\delta_{c} l_{12}\right)^2\right)}, &         \text{if } \textup{cutoff model},\\
            h^{-1}-\frac{2h^{2n-1}}{r_c^{2n}+h^{2n}}, &         \text{if } \textup{Vatistas model},
    \end{cases}
\end{equation}
and
\begin{equation}
 \bar{\Gamma} =
    \begin{cases}
    \frac{\Gamma}{4\pi}\frac{1}{\left(h+\left(\delta_{c} l_{12}\right)^2\right)}, &         \text{if } \textup{cutoff model},\\
           \frac{\Gamma}{4\pi}\frac{h}{\left(r_c^{2n}+h^{2n}\right)^{1/n}}, &         \text{if } \textup{Vatistas model}.
    \end{cases}
\end{equation}
Note that $\mathbf{V}^{\prime}=\mathbf{V}_{\rm{induced},\it{i}}+\delta\mathbf{V}_{\rm{induced},\it{i}}$, and the spatial parameters, $\cos{\theta_1}$, $\cos{\theta_2}$, $h$, and $\mathbf{e}$ must all be perturbed to arrive at the governing perturbed induced velocity. The final expression for the perturbed induced velocity field requires further simplification of the perturbed parameters $\delta h$, $\delta \left(\cos{\theta_1}\right)$, $\delta \left(\cos{\theta_2}\right)$, and $\delta \mathbf{e}$, in terms of the position vectors, $\mathbf{r}_A,\: \mathbf{r}_B,\: \rm{and}\:\mathbf{r}_P$. Applying this simplification yields the final form of the perturbed induced velocity on the point of interest, $P$,
\begin{equation} \label{rhs_perturb_vel}
\delta \mathbf{V}_{\textup{induced},\:i}=\bar{\Gamma}\left\lbrace [\mathbf{L}]\delta \mathbf{r}_A + [\mathbf{N}] \delta \mathbf{r}_B + [\mathbf{O}] \delta \mathbf{r}_P \right\rbrace,
\end{equation}
where
\begin{equation*}
\delta \mathbf{r}_{A}^{\rm{T}} =[\delta r_{Ax}, \delta r_{Ay}, \delta r_{Az}],\:\:\:\delta \mathbf{r}_{B}^{\rm{T}} =[\delta r_{Bx}, \delta r_{By}, \delta r_{Bz}],\:\:\:\delta \mathbf{r}_{P}^{\rm{T}} =[\delta r_{Px}, \delta r_{Py}, \delta r_{Pz}].
\end{equation*}
The coefficient matrices $[\mathbf{L}]$, $[\mathbf{N}]$, and $[\mathbf{O}]$ are now defined. First, consider the second term in Eq. \ref{perturbed_goveq_four} as 
\begin{equation}
\left(\delta\cos{\theta_1}-\delta \cos{\theta_2}\right)\mathbf{e}=
\bar{\mathbf{A}}_1 \mathbf{e}^{\rm{T}} \delta \mathbf{r}_A +\bar{\mathbf{B}}_1 \mathbf{e}^{\rm{T}}\delta+\bar{\mathbf{P}}_1 \mathbf{e}^{\rm{T}} \delta \mathbf{r}_P.
\end{equation}
where
\begin{align}
\bar{\mathbf{A}}_1^{\rm{T}} =\left(\frac{\partial(\cos{\theta_1})}{\partial r_{Ax}}-\frac{\partial(\cos{\theta_2})}{\partial r_{Ax}},\frac{\partial(\cos{\theta_1})}{\partial r_{Ay}}-\frac{\partial(\cos{\theta_2})}{\partial r_{Ay}},\frac{\partial(\cos{\theta_1})}{\partial r_{Az}}-\frac{\partial(\cos{\theta_2})}{\partial r_{Az}}\right),
\end{align}

\begin{align}
\bar{\mathbf{B}}_1^{\rm{T}} =\left(\frac{\partial(\cos{\theta_1})}{\partial r_{Bx}}-\frac{\partial(\cos{\theta_2})}{\partial r_{Bx}},\frac{\partial(\cos{\theta_1})}{\partial r_{By}}-\frac{\partial(\cos{\theta_2})}{\partial r_{By}},\frac{\partial(\cos{\theta_1})}{\partial r_{Bz}}-\frac{\partial(\cos{\theta_2})}{\partial r_{Bz}}\right),
\end{align}

\begin{align}
\bar{\mathbf{P}}_1^{\rm{T}} =\left(\frac{\partial(\cos{\theta_1})}{\partial r_{Px}}-\frac{\partial(\cos{\theta_2})}{\partial r_{Px}},\frac{\partial(\cos{\theta_1})}{\partial r_{Py}}-\frac{\partial(\cos{\theta_2})}{\partial r_{Py}},\frac{\partial(\cos{\theta_1})}{\partial r_{Pz}}-\frac{\partial(\cos{\theta_2})}{\partial r_{Pz}}\right).
\end{align}
The coefficient matrices are collected as
\begin{align}
\label{eqn:eqlabel}
\begin{split}
&[\mathbf{L}_1]=\bar{\mathbf{A}}_1\mathbf{e}^{\rm{T}},\\&[\mathbf{N}_1]=\bar{\mathbf{B}}_1\mathbf{e}^{\rm{T}},\\&[\mathbf{O}_1]=\bar{\mathbf{P}}_1\mathbf{e}^{\rm{T}}.
\end{split}
\end{align}
Next, the third term in Eq.~\ref{perturbed_goveq_four} is defined as
\begin{align}
\begin{split}
& h_f \delta h \left(\cos{\theta_1}-\cos{\theta_2}\right)\mathbf{e} 
=h_f\left(\cos{\theta_1}-\cos{\theta_2}\right) \bigg( \bar{\mathbf{A}}_2\mathbf{e}^{\rm{T}}\delta \mathbf{r}_A +\bar{\mathbf{B}}_2 \mathbf{e}^{\rm{T}}\delta \mathbf{r}_B +\bar{\mathbf{P}}_2\mathbf{e}^{\rm{T}}\delta \mathbf{r}_P \bigg),
\end{split}
\end{align}
where the coefficient matrices are collected and defined as
\begin{align}
\label{eqn:eqlabel}
\begin{split}
&[\mathbf{L}_2]=h_f\left(\cos{\theta_1}-\cos{\theta_2}\right)\bar{\mathbf{A}}_2 \mathbf{e}^{\rm{T}},\\&[\mathbf{N}_2]=h_f\left(\cos{\theta_1}-\cos{\theta_2}\right)\bar{\mathbf{B}}_2\mathbf{e}^{\rm{T}},\\&[\mathbf{O}_2]=h_f\left(\cos{\theta_1}-\cos{\theta_2}\right)\bar{\mathbf{P}}_2\mathbf{e}^{\rm{T}}, 
\end{split}
\end{align}
using
\begin{align}
\bar{\mathbf{A}}_2^{\rm{T}}=\left(\frac{\partial h}{\partial r_{Ax}}, \frac{\partial h}{\partial r_{Ay}},\frac{\partial h}{\partial r_{Az}}\right),
\end{align}
\begin{align}
\bar{\mathbf{B}}_2^{\rm{T}}=\left(\frac{\partial h}{\partial r_{Bx}}, \frac{\partial h}{\partial r_{By}},\frac{\partial h}{\partial r_{Bz}}\right),
\end{align}
\begin{align}
\bar{\mathbf{P}}_2^{\rm{T}}=\left(\frac{\partial h}{\partial r_{Px}}, \frac{\partial h}{\partial r_{Py}},\frac{\partial h}{\partial r_{Pz}}\right).
\end{align}
Finally, the fourth term in Eq. \ref{perturbed_goveq_four} is defined as 
\begin{align}
\left(\cos{\theta_1}-\cos{\theta_2}\right)\delta\mathbf{e}=
\left(\cos{\theta_1}-\cos{\theta_2}\right)[\bar{\mathbf{A}}_3]\delta \mathbf{r}_A+
\left(\cos{\theta_1}-\cos{\theta_2}\right)[\bar{\mathbf{B}}_3]\delta \mathbf{r}_B +\left(\cos{\theta_1}-\cos{\theta_2}\right)[\bar{\mathbf{P}}_3]\delta \mathbf{r}_P.
\end{align}
where 
\begin{align}
\begin{split}
&[\bar{\mathbf{A}}_3] = \Abarthree, \:\: [\bar{\mathbf{B}}_3] = \Bbarthree,  \:\: [\bar{\mathbf{P}}_3] = \Pbarthree.
\end{split}
\end{align}
the coefficient matrices are then defined as
\begin{align}
\label{eqn:eqlabel}
\begin{split}
& [\mathbf{L}_3]=\left(\cos{\theta_1}-\cos{\theta_2}\right)[\bar{\mathbf{A}}_3],\\&[\mathbf{N}_3]=\left(\cos{\theta_1}-\cos{\theta_2}\right)[\bar{\mathbf{B}}_3],\\&[\mathbf{O}_3]=\left(\cos{\theta_1}-\cos{\theta_2}\right)[\bar{\mathbf{P}}_3].
\end{split}
\end{align}
Finally, the global coefficient matrices of Eq.~\ref{rhs_perturb_vel} are obtained by summing the appropriate coefficient matrices:
\begin{equation}
[\mathbf{L}]=[\mathbf{L}_1]+[\mathbf{L}_2]+[\mathbf{L}_3],\:\:\:[\mathbf{N}]=[\mathbf{N}_1]+[\mathbf{N}_2]+[\mathbf{N}_3],\:\:\:[\mathbf{O}]=[\mathbf{O}_1]+[\mathbf{O}_2]+[\mathbf{O}_3]. \label{matrix_collections}
\end{equation}
Substituting Eq.~\ref{matrix_collections} into Eq.~\ref{rhs_perturb_vel}, defines the perturbed induced velocity field completely.

\subsubsection{Long-wave Harmonic Perturbation}

Now that the perturbed induced velocity is expressed by its independent parameters, the type of perturbation to be used on the Lagrangian marker velocity must be defined, i.e.,~the left hand side of Eq.~\ref{perturbed_goveq_three}. For this analysis the wake is perturbed in a harmonic fashion. In other words, the form of the perturbation is assumed to be a travelling wave such that a series of normal mode perturbations could describe any arbitrary disturbance \cite{bhagwat:4}. These perturbations are set in cylindrical coordinates for convenience, and their amplitudes are denoted by $\boldsymbol{\delta}_0$. The harmonic perturbation is defined by the following expression:
\begin{equation}\label{harm_eq}
\delta \mathbf{p}_k=\harmpercy e^{\alpha t + \textup{i} \omega \zeta_{k}}=\boldsymbol{\delta}_0e^{\alpha t + \textup{i} \omega \zeta_{k}}, 
\end{equation}
where $k=A\:,B\:,$ or $P$, which correspond to endpoints of the straight segment vortex filament and a point in space, as shown in Fig.~\ref{bs_law_fig}. It is important to note here that $\textup{i}=\sqrt{-1}$ and is not the filament index. Unidirectional perturbations are illustrated below in Figs.~\ref{radial_pert}-\ref{azimuthal_pert} for a helical vortex, at an arbitrary time, age, divergence rate, and a fixed frequency.
\begin{figure}[h!]
  \centering
  \includegraphics[scale=0.4, trim=0 5cm 0 5cm]{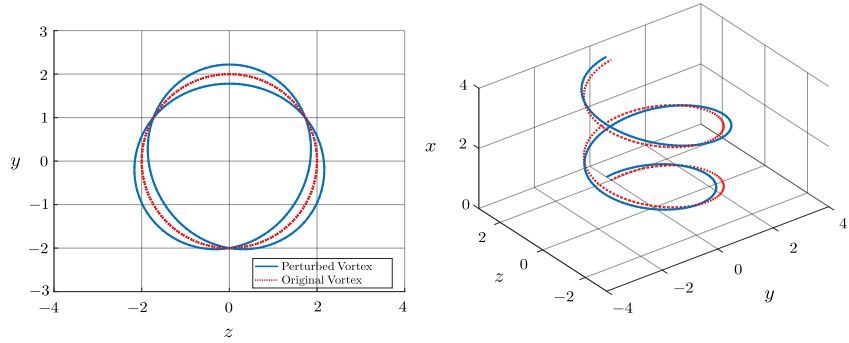}
  \caption[]{Illustration of a radial perturbation at $\omega=1.5\:\rm{rad}^{-1}$}
  \label{radial_pert}
\end{figure}

\begin{figure}[h!]
  \centering
  \includegraphics[scale=0.4, trim=0 5cm 0 5cm]{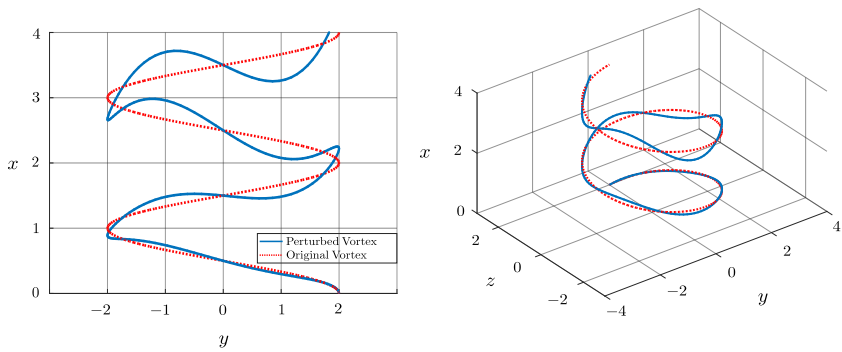}
  \caption[]{Illustration of an axial perturbation at $\omega=3\:\rm{rad}^{-1}$}
  \label{axial_pert}
\end{figure}

\begin{figure}[h!]
  \centering
  \includegraphics[scale=0.4, trim=0 5cm 0 4cm]{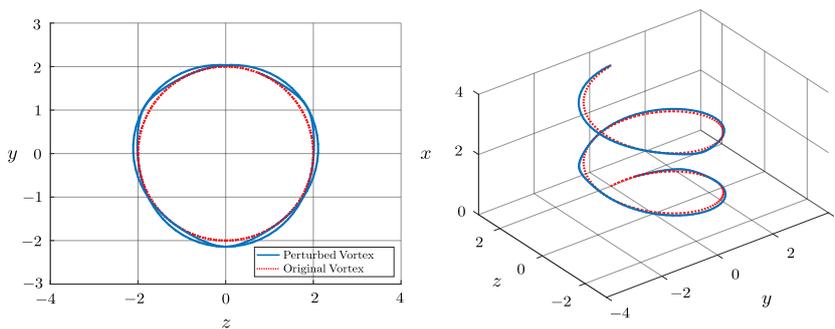}
  \caption[]{Illustration of an azimuthal perturbation at $\omega=1.5\:\rm{rad}^{-1}$}
  \label{azimuthal_pert}
\end{figure}

The cylindrical coordinate transformation matrix is defined by 
\begin{equation}
[\mathbf{T}]=\transmat,
\end{equation}
where the harmonic perturbation Eq.~\refeq{harm_eq} is transformed to Cartesian coordinates as follows
\begin{equation}\label{perturbed_vectors}
\delta \mathbf{r}_{A}= \mathbf{[T]} \delta \mathbf{p}_A,\:\:\:\delta \mathbf{r}_{B}= \mathbf{[T]} \delta \mathbf{p}_B,\:\:\:\delta \mathbf{r}_{P}=\mathbf{[T]}\delta \mathbf{p}_P.
\end{equation}
Also note the corresponding transformation matrix expressions:
\begin{equation}
\dot{\mathbf{r}}_k=\mathbf{[T]}\dot{\mathbf{p}}_k,
\end{equation}
\begin{equation}\label{lhs_perturb_vel}
\delta \dot{\mathbf{r}}_k=\mathbf{[T]}\delta\dot{\mathbf{p}}_k+[\mathbf{T}_2]\delta \mathbf{p}_k,
\end{equation}
\begin{equation}
[\mathbf{T}_2]=\transdotmat,
\end{equation}
\begin{equation}
\cos{\theta}=\frac{y}{\sqrt{y^2+z^2}},\:\:\:\sin{\theta}=\frac{z}{\sqrt{y^2+z^2}},\:\:\:\dot{\theta}=\frac{y\dot{z}-z\dot{y}}{y^2+z^2}.
\end{equation}

\subsubsection{Eigenvalue Analysis}
A complete set of equations has now been obtained  (Eqs.~\ref{rhs_perturb_vel}, \ref{lhs_perturb_vel}) that is necessary to formulate an eigenvalue stability analysis. First, substitute Eq.~\ref{lhs_perturb_vel} into the left hand side of the governing perturbed equation, Eq.~\ref{perturbed_goveq_three},
\begin{equation}\label{eig_val_gov_lhs}
\delta\dot{\mathbf{r}}_k=[\mathbf{T}]\delta \dot{\mathbf{p}}_k+[\mathbf{T}_2]\delta \mathbf{p}_k= \left(\alpha[\mathbf{T}]+[\mathbf{T}_2]\right)\boldsymbol{\delta}_0e^{\alpha t + \textup{i} \omega \zeta_k}.
\end{equation}
Next, substituting Eq.~\ref{perturbed_vectors} into Eq.~\ref{rhs_perturb_vel} will give
\begin{align}
\delta \mathbf{V}_{\rm{induced},\it{i}} & =\bar{\Gamma}\left([\mathbf{L}]\delta \mathbf{r}_A+ [\mathbf{N}] \delta \mathbf{r}_B + [\mathbf{O}] \delta \mathbf{r}_P \right) \nonumber \\ &=\bar{\Gamma}\left([\mathbf{L}][\mathbf{T}] \boldsymbol{\delta}_0 e^{\alpha t+\textup{i}\omega \zeta_A} + [\mathbf{N}][\mathbf{T}]\boldsymbol{\delta}_0 e^{\alpha t+\textup{i}\omega \zeta_B} + [\mathbf{O}][\mathbf{T}] \boldsymbol{\delta}_0 e^{\alpha t+\textup{i}\omega \zeta_P} \right) \nonumber \\
& =\bar{\Gamma}\left([\mathbf{L}]e^{\textup{i}\omega(\zeta_A-\zeta_P)} + [\mathbf{N}] e^{\textup{i}\omega(\zeta_B-\zeta_P)} + [\mathbf{O}] \right)[\mathbf{T}] \boldsymbol{\delta}_0e^{\alpha t+\textup{i}\omega \zeta_P} \nonumber \\ 
&=[\mathbf{V}_i][\mathbf{T}]\delta \mathbf{p}_{P},
\label{eig_val_gov_rhs}
\end{align}
where by only considering the real contribution of $e^{\textup{i}\omega\left(\zeta_A-\zeta_P\right)}$ and $e^{\textup{i}\omega\left(\zeta_B-\zeta_P\right)}$
\begin{equation}
[\mathbf{V}_i]=\bar{\Gamma}\left(\cos{\omega(\zeta_A-\zeta_P)}[\mathbf{L}]+\cos{\omega(\zeta_B-\zeta_P)}[\mathbf{N}]+[\mathbf{O}]\right).
\end{equation}
The subscript $i$ denotes the $i^{\rm{th}}$ filament (the segment between $\mathbf{r}_B$ and $\mathbf{r}_A$). The superposition of each individual filament's perturbed induced velocity field on a point of interest is expressed as $\delta\mathbf{V}=\sum_{i} \delta\mathbf{V}_{\rm{induced},\it{i}}$. Likewise, the expression $[\mathbf{V}]=\sum_{\it{i}} \mathbf{V}_i$ is the total perturbed induced velocity field in cylindrical coordinates on the point of interest. Substituting Eqs.~\ref{eig_val_gov_lhs} and \ref{eig_val_gov_rhs} into Eq.~\ref{perturbed_goveq_three} yields, after some rearrangement, 
\begin{equation}\label{eigval_one}
\alpha \delta \mathbf{r}_P= \left([\mathbf{V}]-[\mathbf{T}_2][\mathbf{T}]^{-1}\right)\delta\mathbf{r}_P.
\end{equation}
Setting $[\mathbf{W}]=\left([\mathbf{V}]-[\mathbf{T}_2][\mathbf{T}]^{-1}\right)$ and rearranging Eq.~\ref{eigval_one} further gives the final form of the eigenvalue problem for each \emph{point of interest} along the tip vortex: 
\begin{equation}
 \left([\mathbf{W}]-\alpha[\mathbf{I}]\right)\delta \mathbf{r}_P=\mathbf{0},
\label{final_eigvalpro}
\end{equation}
where solving for the maximum $\alpha$ along the age of the tip vortex, at a specified radial wavenumber $\omega$ and instant in time $t$, produces the stability trend, $\alpha\:\textup{v.}~\omega$, of the tip vortices.

\subsubsection{Validation} \label{validation_sec}

Validation of the stability analysis is performed on a three-bladed rotor with rectangular blades. The rotor specifications are listed in Table \ref{rotor_specs}.

{\renewcommand{\arraystretch}{1.5}
\begin{table}[h!]
\centering
\caption[]{Canonical rotor specifications modified from \cite{jonkman:6}}
\begin{tabular}{l  l} 
\hline\hline
Rotor orientation and configuration: & Upwind; 3 blades \\
Rotor and hub diameter: & 126 m, 3 m\\
Airfoil: & NACA64-A17\\
Chord length: & 4.5 m\\
Blade twist: & 0$^{\circ}$\\
Rotor plane tilt: & 0$^{\circ}$\\
\hline\hline
\end{tabular}
\label{rotor_specs}
\vspace{0.5cm}
\end{table}}

Simulations of the canonical rotor were generated using the aeroelastic free-vortex wake method presented by \citet{rodriguezJRE,rodriguezJRE2}, but rotor blades are constrained to be rigid for this validation case. Simulations for each rotor configuration were performed over 10 seconds with vortex shedding frequencies $f=10, \:20, \:30$ Hz at an inflow condition of $V_{\infty}=18$ m/s, tip-speed ratio (blade tip-speed /inflow speed) $\lambda=9$, and blade pitch angle $\theta_{\rm{bl}}=0^{\circ}$. Snapshots at $t=10$ s of these simulations at $f=20$ Hz for the Vatistas models ($n=1, 2, 3$), and $f=30$ Hz for the cutoff models ($\delta_{c}^2=0.1,\:10^{-2},\:10^{-6}$) are shown in Fig.~\ref{ThreeBlade_Wake}, where axis labels $\textup{X}=x/D$ and $\textup{Z}=z/D$ are the nondimensional distances, where $x$, $z$, and $D$ are distances in the downstream direction, distance in the vertical direction, and the rotor diameter, respectively. Finally, stability analyses were performed on the generated tip vortices.

It is important to note that previous work by \citet{rodriguezJERT} validated their stability analysis only against peak divergence trends of the flexible NREL 5MW reference wind turbine rotor. The current validation work supersedes previous validation attempts by \citet{rodriguezJERT} such that the current analyses confirms classical stability trends and vortex behaviors of zero-pitched rotor configurations with variable finite-core vortex modeling \cite{bhagwat:4}.

\subsubsection{Considerations of Numerical Stability}

Before proceeding to the validation of the stability analysis, a few notes on the numerical simulations must be presented. \citet{rodriguezJRE, rodriguezJRE2} showed that the FVM aeroelastic framework employed herein is capable of reproducing accurate, stable, and robust rotor-blade and rotor-wake performance metrics of the NREL 5MW reference wind turbine rotor reported by \citet{jonkman:6}, such as blade forces, blade deflection, and rotor thrust. The aerodynamic framework used in \citet{rodriguezJRE, rodriguezJRE2}, initially developed by \citet{sebastian:10}, has capabilities of faithfully generating the near-wake geometry of rotor wakes as validated in \citet{sebastian:10}. However, convection of Lagrangian markers can become numerically unstable as discussed in \citet{bagai1995rotor} and \citet{rodriguezphd}. This numerical issue is associated with numerical instabilities that have been studied by Leishman \emph{et al}.~\citep{bagai1995rotor, bagai1995rotor_exp, bhagwat:4}. Specifically, depending on the numerical method employed, the discretization of the induced velocity field can result in the addition of anti-dissipative terms, which can lead to exponential growth of non-physical disturbances caused by roundoff error \cite{leishman:2}. To circumvent non-physical disturbances caused by numerical artifacts, sophisticated FVM-specific numerical techniques, such as those presented in \cite{bagai1995rotor,bagai1995rotor_exp,bhagwat:4,bhagwat2001stability}, can be implemented to numerically stabilize the wake geometry.

The present investigation takes an alternate approach introduced by \citet{rodriguezphd} to avoid artificial disturbances that may corrupt the stability analysis of tip vortices. \citet{rodriguezphd} has shown that there are regions along the tip vortex where a stability analysis can be performed where the effects of artificial numerical instabilities can be minimized by truncating the wake downstream of the rotor. These regions are defined by truncating segments of the tip vortex which have already been dominated by artificial numerical instabilities.  \citet{rodriguezphd} also showed that using the stability analysis presented herein, one can quantitatively and qualitatively find the location of the onset of the numerical wake breakdown. The current validation analysis is conducted on one rotation (a wake age of $\zeta=2\pi$) of the tip vortices as they are shed off of the rotor blades, where ages longer than one full rotation are truncated. For additional information on variable window sizes employed in the stability analyses, the reader is referred to  \citet{rodriguezphd}.

\subsubsection{Stability of tip vortices shed from a canonical rotor}\label{stab_canon}

Three conditions must be met to validate the stability analysis as presented.  The stability trends (divergence rate versus perturbation wavenumber, $\alpha$ v.~$\omega$) must obey the following classical results: 1) peak divergence rates occur at perturbation wavenumbers $\omega=N_b\left(k-\frac{1}{2}\right)$, 2) divergence rates must converge to a constant value as the wavenumber goes to infinity, 3) because the canonical rotor configuration generates symmetric results, i.e. the rotor is not pitched and no out-of-phase periodicity is introduced, it is required that the stability trends of each individual tip vortex be identical.

\begin{figure}[h!]
	\centering
	\includegraphics[scale=0.85, trim=5cm 5cm 0 5cm]{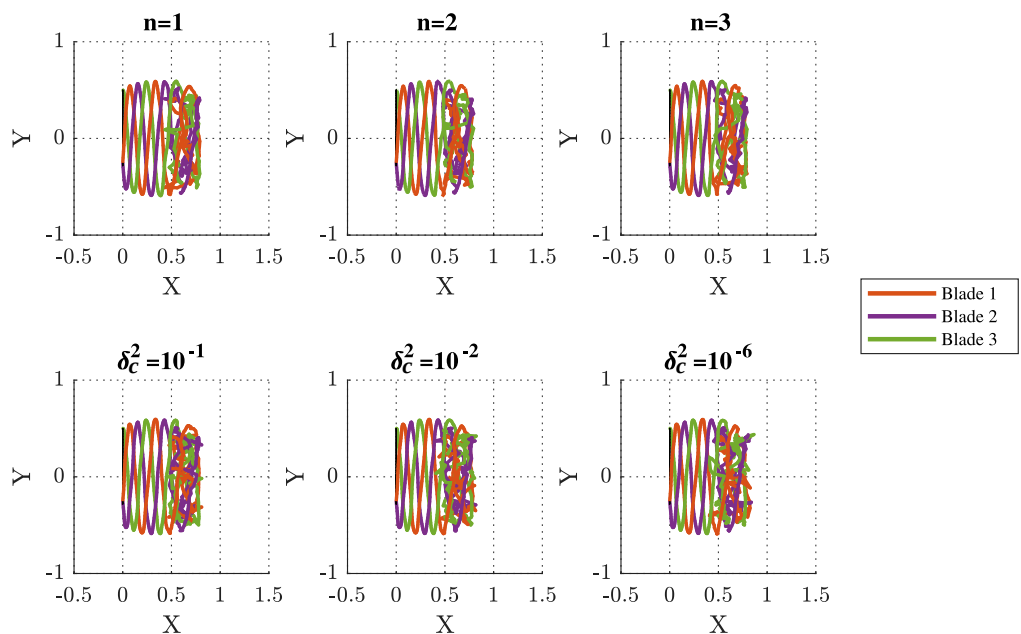}
	\caption{Three-bladed rotor wakes modeled by employing the Vatistas core model (first row) and the cutoff model (second row). Shedding frequency is $f=20$ Hz. }
	\label{ThreeBlade_Wake}
\end{figure}

\begin{figure*}[h!]
	\centering
	\begin{subfigure}[b]{0.475\textwidth}
		\centering
		\includegraphics[scale=0.5, trim=110 7cm 50 7cm]{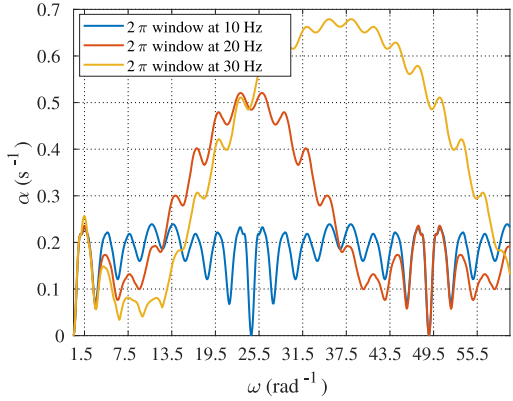}
		\caption[]%
		{Vatistas vortex core model, $n=1$}    
		\label{2bl_n1}
	\end{subfigure}
	\hfill
	\begin{subfigure}[b]{0.475\textwidth}  
		\centering 
		\includegraphics[scale=0.5, trim=80 7cm 25 7cm]{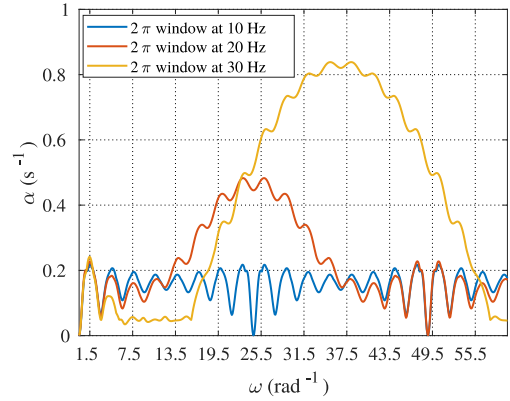}
		\caption[]%
		{Vatistas vortex core model, $n=2$}
		\label{2bl_n2}
	\end{subfigure}
	\vskip\baselineskip
	\begin{subfigure}[b]{0.475\textwidth}   
		\centering 
		\includegraphics[scale=0.5, trim=120 7cm 50 7cm]{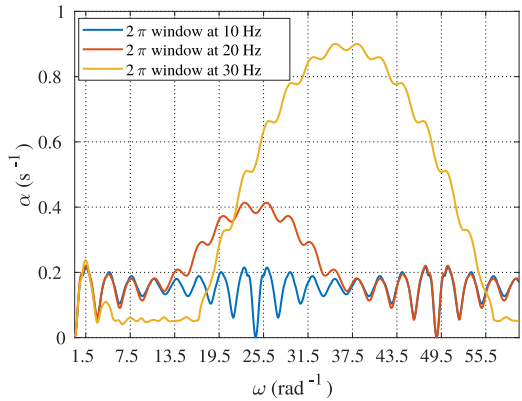}
		\caption[]%
		{Vatistas vortex core model, $n=3$}
		\label{2bl_n3}
	\end{subfigure}
	\begin{subfigure}[b]{0.475\textwidth}   
		\centering 
		\includegraphics[scale=0.5, trim=80 7cm 70 7cm]{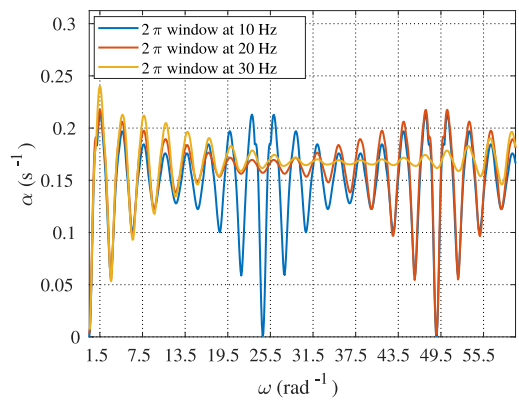}
		\caption[]%
		{Cutoff vortex core model, $\delta_{c}^2=10^{-1}$}
		\label{2bl_d01}
	\end{subfigure}
	\caption[]
	{Stability trends for one tip vortex shed from a three-bladed rotor at variable vortex shedding frequencies (10, 20, 30 Hz) with a one-rotation (2$\pi$) tip vortex window. Vatistas models for $n=1,\:2,\:\rm{and}\:3$ are employed, and the cutoff model is presented with a radius of $\delta_{c}^2=0.1$.}  
	\label{three_blade_stab_trends}
\end{figure*}

Stability results are now presented for the tip vortices produced by the  canonical rotor. Figure \ref{ThreeBlade_Wake} presents side views of the three-bladed rotor and its wake, where the Vatistas core modeling and cutoff core modeling have been used where indicated. First, the stability trends ($\alpha$ v.~$\omega$) are presented in Figure \ref{three_blade_stab_trends} for the Vatistas model and cutoff model at variable vortex shedding frequencies. Perhaps the most notable feature of Fig.~\ref{three_blade_stab_trends} is the periodic behavior in divergence rate oscillations as a function of wavenumber. Classical studies \cite{widnall1972stability, gupta1974theoretical, bhagwat:4} have traditionally looked at low wavenumbers that are consistent with the underlying long-wave perturbation qualitative assumption, i.e.~long-wave perturbations are defined as a perturbations much larger than the vortex core. However, because of the long-wave assumption, no investigation, to our knowledge, has presented stability trends for a high range of wavenumber perturbations to visualize the limits of long-wave perturbation analyses.

It is believed that this periodic behavior is caused by both the long-wave perturbation analysis and a numerical artifact from the time-marching scheme (second-order Runge-Kutta) of the free-vortex wake method currently used based on the following observations. Notice that for a relatively low vortex shedding frequency, such as $f=10$ Hz (blue trends in Fig.~\ref{three_blade_stab_trends}), the periodic behavior has lower amplitude and lower period than at shedding frequencies of $f=20$ and $30$ Hz. Higher shedding frequencies of $f=20$ and $30$ Hz show a high bell-shaped divergence rate trend. It is also observed that the width of this ``bell" increases with the shedding frequency. This result suggests that if the shedding frequency approaches infinity then the rotor-wakes simulated by the current free-vortex wake method would be most sensitive to high wavenumbers, i.e.~short-wave perturbations. Finally, the bell curve maxima occur at approximately the same location ($\omega\approx25$ rad$^{-1}$ at $f=20$ Hz; and $\omega\approx37$ rad$^{-1}$ at $f=30$ Hz) for rotor wake geometries employing the Vatistas model, which further reinforces the conjecture that the periodic bell-curve is an artifact of the long-wave perturbation analysis, the numerical integration, and their relationship to the Vatistas finite core modeling.

At wavenumbers between $0 \le \omega \le 7.5$ rad$^{-1}$, the stability trends generated by the Vatistas models ($n=1,\:2,\:\rm{and}\:3$) with shedding frequencies $f=10$ and $f=20$ Hz show very good agreement with reported classical stability trends, i.e., divergence-rate peaks occur at $\omega=N_b\left(k-1/2\right)$, where $k$ is any natural number. However, at $f=20\:\textup{and}\:30$ Hz, the stability trends do not exhibit the correct behavior beyond the second divergence rate peak. In fact, the stability trend becomes erratic as wavenumbers increase and approach the bell. This result may indicate that the shedding frequency has  neighboring vortices spaced too closely to remain numerically stable to record the vortex stability trends reported by classical studies with the Vatistas model. Limitations of the Vatistas model with regard to capturing the classical stability results at higher vortex shedding frequencies is further highlighted by the cutoff vortex model results in Fig.~\ref{2bl_d01}. The cutoff model shows very consistent oscillations of the divergence rates for all vortex shedding frequencies. It appears that as the vortex frequency increases the period of the bell behavior tends to infinity. This dependence on vortex shedding frequency suggests that an infinite vortex shedding frequency is required to achieve classical stability trend results for the cutoff model. The absence of the bell as seen for the Vatistas models also implies that the cutoff model is not susceptible to short-wave (high wavenumber) numerical instabilities.

\begin{figure}[h!]
	\centering
	\includegraphics[scale=0.5, trim=2cm 8cm 0 8cm]{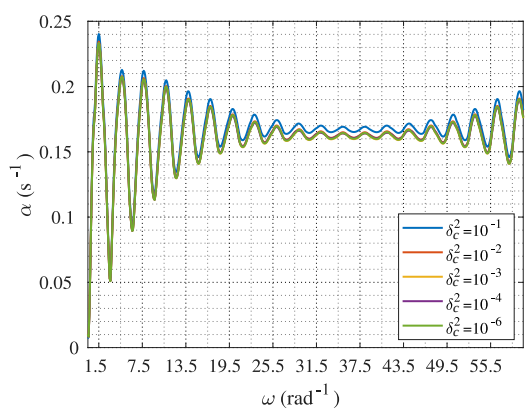}
	\caption{Stability trends for the three-bladed rotor wake with shedding frequency set at $f=30$ Hz and reducing the cutoff radii to $\delta_{c}^2\rightarrow 0$.}
	\label{3blade_delta_conv}
\end{figure}

The differences between the Vatistas and cutoff models motivated further investigation into reducing the cutoff radius, which becomes the Rankine vortex as $\delta_{c}^2 \rightarrow 0$. The cutoff radius reduction results are presented in Fig.~\ref{3blade_delta_conv} for a vortex shedding frequency of $f=30$ Hz. Results show marginal differences between cutoff radii as $\delta_{c}^2 \rightarrow 0$. Similarity between stability trends for variable cutoff radii further reinforces the notion that the Vatistas core model may be more sensitive to short-wave numerical instabilities.

\begin{figure*}[h!]
        \centering
        \begin{subfigure}[b]{0.475\textwidth}
            \centering
            \includegraphics[scale=0.4, trim=110 7cm 50 7cm]{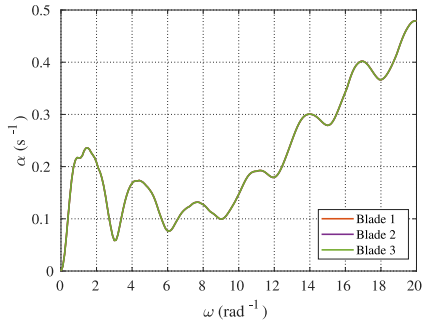}
            \caption[]%
            {Vatistas vortex core model, $n=1$}    
            \label{3bl_n1}
        \end{subfigure}
        \hfill
        \begin{subfigure}[b]{0.475\textwidth}  
            \centering 
            \includegraphics[scale=0.4, trim=80 7cm 25 7cm]{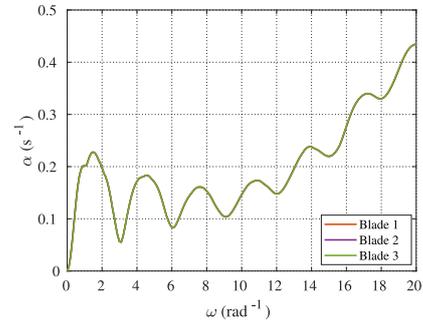}
            \caption[]%
            {Vatistas vortex core model, $n=2$}
            \label{3bl_n2}
        \end{subfigure}
        \vskip\baselineskip
        \begin{subfigure}[b]{0.475\textwidth}   
            \centering 
            \includegraphics[scale=0.4, trim=120 7cm 50 7cm]{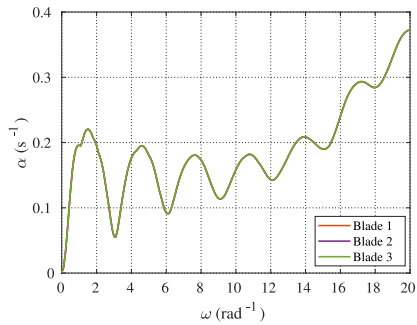}
            \caption[]%
            {Vatistas vortex core model, $n=3$}
            \label{3bl_n3}
        \end{subfigure}
        \begin{subfigure}[b]{0.475\textwidth}   
            \centering 
            \includegraphics[scale=0.4, trim=80 7cm 70 7cm]{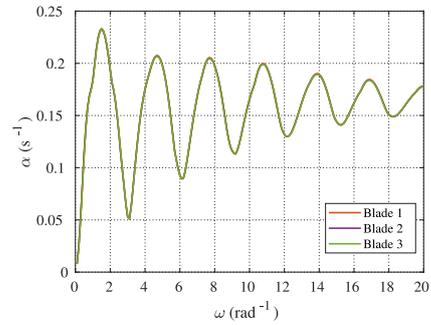}
            \caption[]%
            {Cutoff vortex core model, $\delta_{c}^2= 10^{-6}$}
            \label{3bl_d01}
        \end{subfigure}
     \caption{Stability trends for all the individual tip vortices shed from a three-bladed rotor from Vatistas and cutoff finite core models} 
     \label{three_blade_stab_trends_zoom}
    \end{figure*}

The above stability trends are for single tip vortex shed from one blade. Recall that the stability trends for each tip vortex on a multi-bladed rotor must be identical for the analysis to be valid. The stability trends for each individual tip vortex for a three-bladed rotor are presented in Fig.~\ref{three_blade_stab_trends_zoom}. Here it is demonstrated that the stability trends for the tip vortex from each blade are identical, as expected due to the zero-pitched rotor configuration. For both the Vatistas and the cutoff models, the divergence rates peak at perturbation wavenumbers  $\omega=N_b\left(k-1/2\right)$, i.e., the classical stability trend criteria, for wavenumber perturbations up to $\omega=13.5$ rad$^{-1}$. After $\omega=13.5$ (rad$^{-1}$) peak divergence rates correspond to wavenumbers that are shifted forward by some value $\epsilon$ (i.e., peaks occur at some $\omega=N_b\left(k-1/2\right) + \epsilon$), which deviates from the classical stability trend.

\subsubsection{Validation Study Conclusions}
The resulting stability trends for the three-bladed canonical rotor satisfy generally the criteria established at the beginning of Section \ref{stab_canon} to validate the stability analysis presented. However, it is important to remember a few numerical caveats: First, vortex shedding frequencies of the free-vortex wake method impact stability trends. For the Vatistas model, too high of a vortex shedding frequency will deconstruct the classical stability trend. However, for the vortex cutoff model, no such deconstruction of the classical stability trend occurs. Second, the present stability analysis employing a wake generated by the Vatistas model or the cutoff model and the current time integration scheme only satisfies the classical stability trend for low wavenumbers (approximately between $0 \le \omega \le 7.5$ rad$^{-1}$ ). Third, stability trends obtained from wakes employing the Vatistas model experience large periodic divergence (the bell behavior) as the perturbation wavenumber increases, whereas the cut-off model does not show large periodic divergence. However, the cut-off model does show periodic trend behaviors as the wavenumber increases. Even though the cut-off model has shown to be less susceptible to numerical artifacts,  the presented work will employ the Vatistas finite core model with $n=2$ to remain consistent with previous validated aeroelastic works of \citet{rodriguezJRE, rodriguezJRE2, rodriguezphd}, and prior free-vortex tip-vortex stability analyses by \citet{bhagwat2001stability, bhagwat:4}.

\section{Stability and Dynamics of Tip Vortices Shed from Flexible Rotors}\label{flexible_sec}
Tip vortices shed from the NREL 5MW Reference Wind Turbine rotor, with a rigid rotor, and a flexible rotor, are now investigated. Three cases spanning low-level and high-level operating conditions of the NREL 5MW reference wind turbine are considered to evaluate a range of rotor aeroelasticity and its impact on tip vortex dynamics and stability. The operational conditions considered in the current work were first presented in \cite{sebastian2012analysis}. The tip-speed ratio $\lambda$ ( $\textup{blade-tip speed}/V_{\infty}$), blade pitch $\theta_{\textup{bl}}$, inflow velocity $V_{\infty}$, and rotor diameter-based Reynolds number $Re_{D}=\rho D V_{\infty}/\mu$, where for all cases $\rho=1.23$ kg/$\textup{m}^3$, $\mu=1.20\times10^{-5}$ kg/(m$\:\cdot\:$s), $D=126$ m \cite{jonkman:6}, are used to describe case conditions. Table \ref{case_definitions} lists the parameters for all cases. In addition, tip vortices shed from a zero-pitched rotor (ZPR) configuration (rotor-plane is perpendicular to the horizontal uniform inflow) and the NREL-designed pitched rotor (PR) configuration (rotor with a 5$^{\circ}$ pitch) will be investigated. All simulations were run for 60 s with a vortex shedding frequency of $f=12$ Hz, using the Vatistas finite core model with $n=2$, where these parameters were chosen according to the convergence analysis presented in \cite{rodriguezJRE}. Simulation snapshots are presented with  $\textup{X}=x/D$, $\textup{Y}=y/D$,  and $\textup{Z}=z/D$ axis labels, where $x$, $y$, $z$, and $D$ are distances in the downstream direction, distance in the lateral direction, distance in the vertical direction, and the rotor diameter, respectively. In addition, aeroelastic results (blade deformations) presented are non-dimensionalized by the blade-length, $l_b=61.5$ m, where edgewise ($v$), flapwise ($w$), and torsional ($\phi$) degrees-of-freedom are presented but axial deformations ($u$) were omitted as they were of negligible magnitude. Finally, the stability analyses employed the window cutoff specified by \citet{rodriguezphd}, such that case 1 employs a $2\pi$ window due to small tip vortex pitch rate, and case 2 and 3 employ a $4\pi$ window due to a larger tip vortex pitch rate. Thus, it is important to remember that by performing the stability analysis on a windowed portion of the wake, conclusions to be drawn apply only to early ages of tip vortices and not as a whole vortex structure that includes far-wake regions. To satisfy the window specifications and avoid initial aeroelastic numerical transients, the stability analyses are conducted on tip vortices between simulation instants $t=16.6$ s and $t=60$ s, where the stability analysis is performed within a total time window of 43.4 seconds. 

{\renewcommand{\arraystretch}{1.5}
	\begin{table}[h!]
		\centering
		\caption[]{Parametric conditions considered}
		\begin{tabular}{l  l  l  l  l  } 
			\hline\hline
			& $\lambda$ & $V_{\infty}$ m/s & $\theta_{\textup{bl}}$ $^{\circ}$ & $Re_{D}$\\
			\hline
			Case 1: & 9.63 & 6 & 0 &7.75$\times 10^{7}$\\
			Case 2: & 7 &  11.4 & 0 &14.7$\times 10^{7}$\\
			Case 3: & 4.43 & 18 & 15  & 23.2 $\times 10^{7}$\\
			\hline\hline
		\end{tabular}
		\label{case_definitions}
		\vspace{0.5cm}
\end{table}}

\subsection{Case 1}

\begin{figure*}[h!]
	\centering
	\begin{subfigure}[b]{0.495\textwidth}
		\centering
		\includegraphics[scale=0.575, trim=7cm 10cm 6cm 9cm]{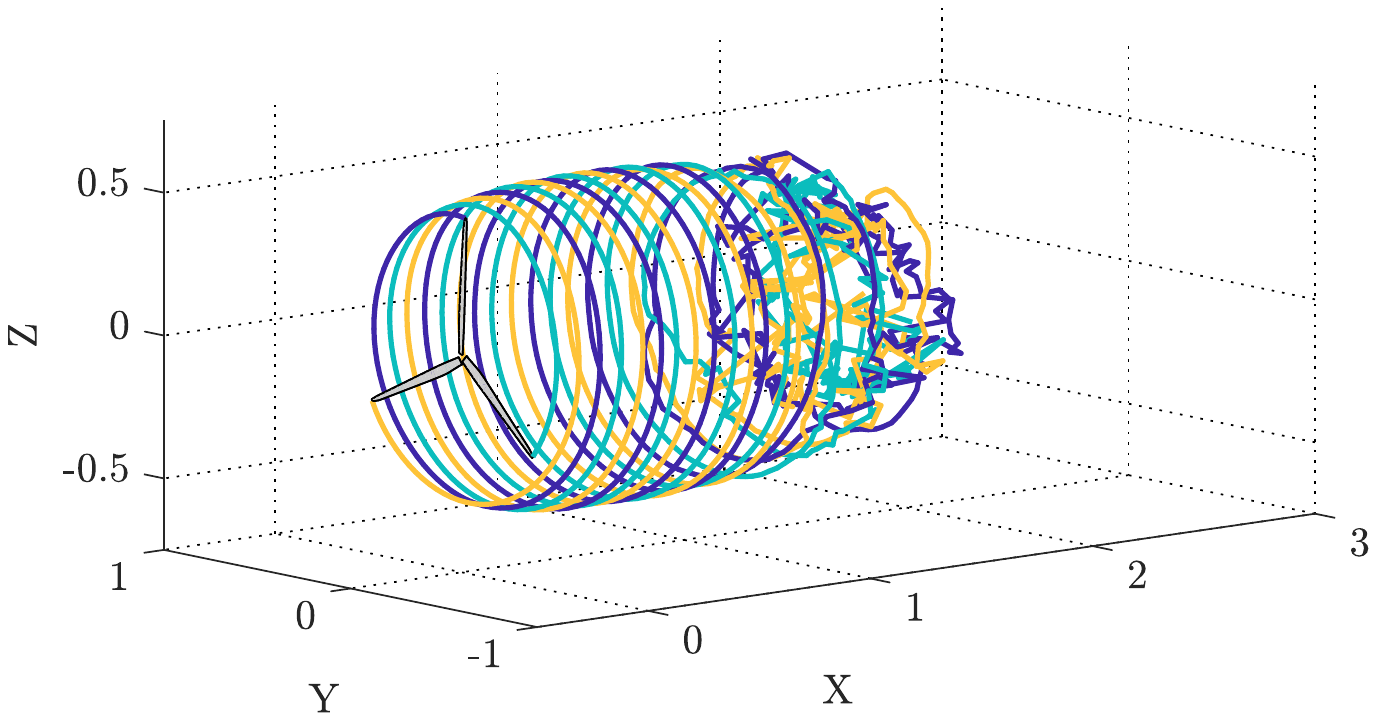}
		\caption[]%
		{ Zero-pitched rigid rotor configuration }    
	\end{subfigure}
	\hfill
	\begin{subfigure}[b]{0.495\textwidth}  
		\centering 
		\includegraphics[scale=0.575, trim=4cm 10cm 25 9cm]{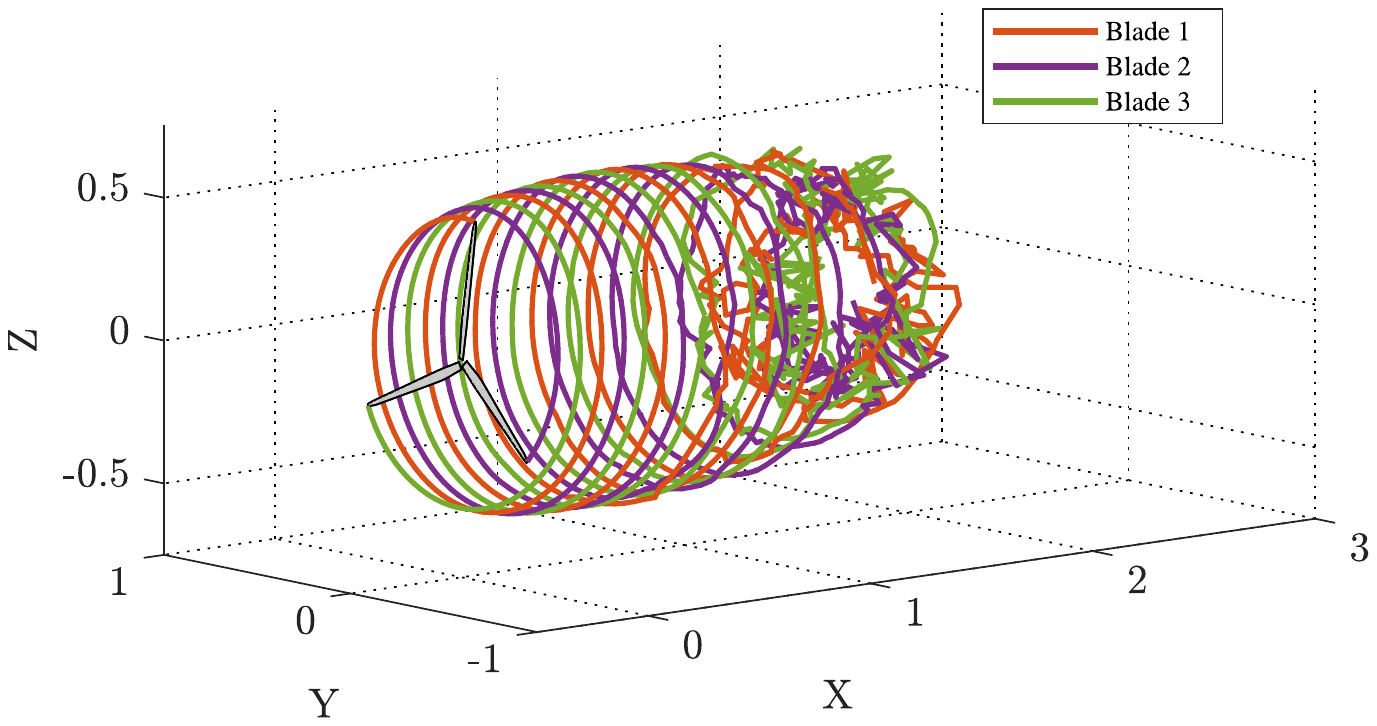}
		\caption[]%
		{Pitched rigid rotor configuration}
	\end{subfigure}
	\hfill
	\begin{subfigure}[b]{0.495\textwidth}  
		\centering 
		\includegraphics[scale=0.575, trim=4cm 10cm 25 8cm]{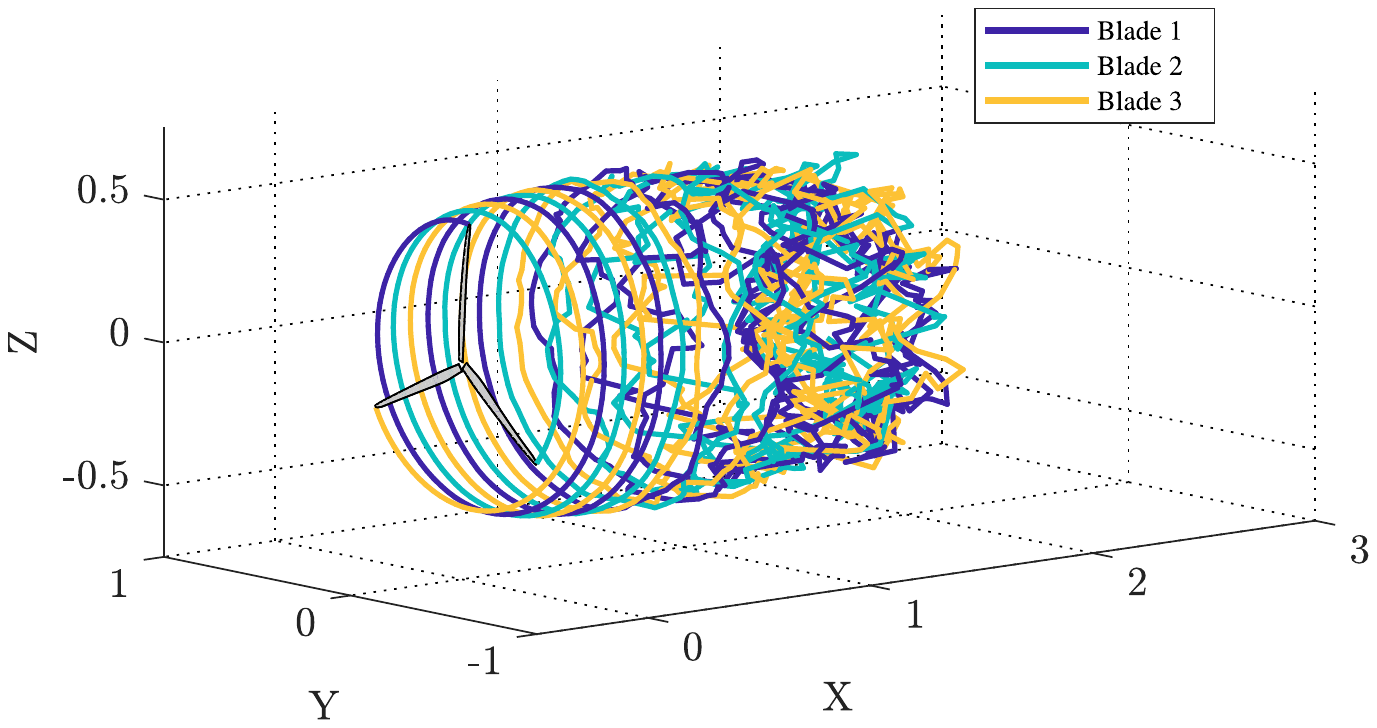}
		\caption[]%
		{Zero-pitched flexible rotor configuration}
	\end{subfigure}
	\hfill
	\begin{subfigure}[b]{0.495\textwidth}  
		\centering 
		\includegraphics[scale=0.575, trim=4cm 10cm 25 8cm]{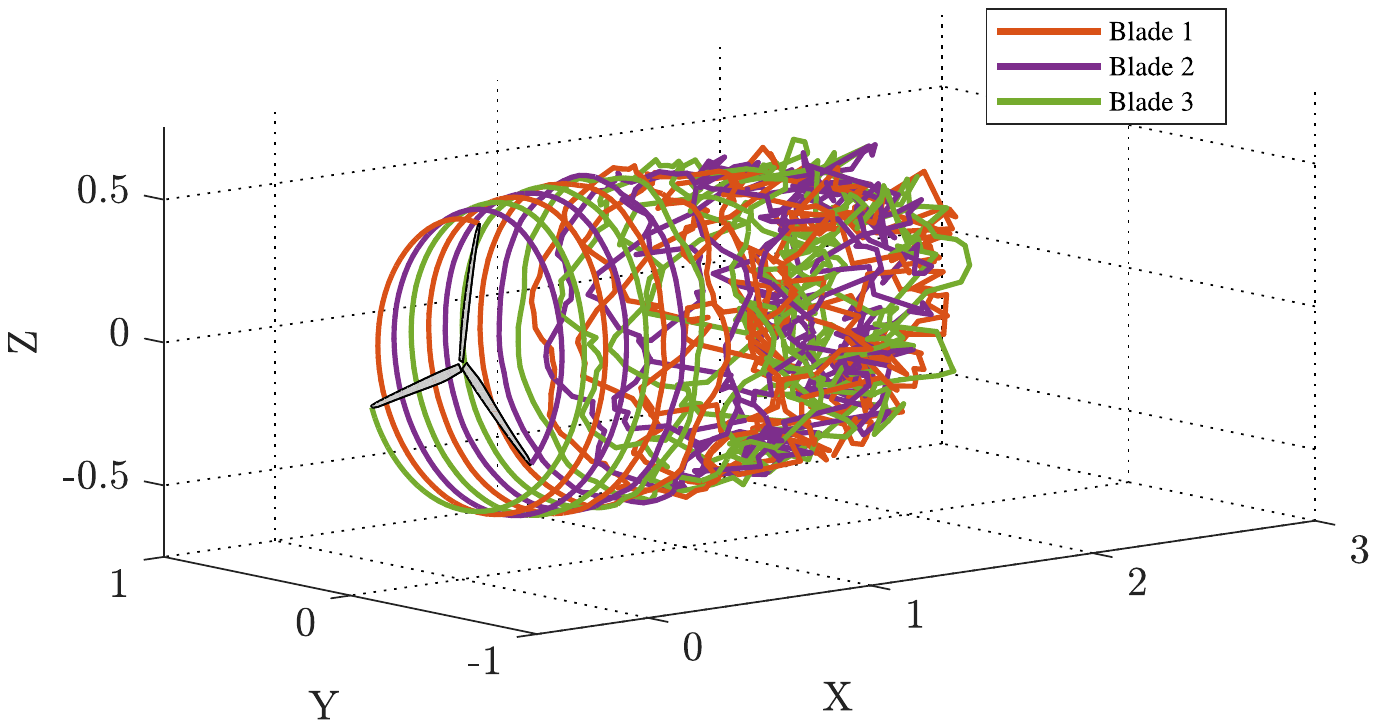}
		\caption[]%
		{Pitched flexible rotor configuration}
	\end{subfigure}
	\caption[]
	{Case 1 tip-vortex snapshot at $t=$ 60 s for rigid and flexible rotor operation} 
	\label{snapshot_brsymtilt}
\end{figure*}

First, case 1 of the NREL 5MW wind turbine is investigated. Snapshots of the wakes generated by ZPR and PR configurations at $t=60$ s are shown in Fig.~\ref{snapshot_brsymtilt}.  The difference in rigid ZPR and PR configurations is highlighted in the location at which the numerical onset of qualitative wake breakdown occurs. The ZPR configuration appears coherent approximately between $0\le$ X $\le$ 1, whereas the PR configuration maintains coherence approximately between $0\le$ X $\le$ 0.7. Each of the flexible rotor configurations both exhibit earlier breakdown than the rigid configurations. The earlier wake breakdown in flexible rotors is likely due to the initial transient impact of the blade deformation and its subsequent influence on the formation of the tip vortices, as was observed in \cite{rodriguezphd}. In other words, the dynamics of the initial transient introduce spatial perturbations that have deformed the tip vortices in the wake causing an earlier wake breakdown compared to rigid cases. However, for both flexible rotor configurations (ZPR and PR) the location at which the numerical wake break-down begins is approximately the same, which indicates that rotor-configuration is not a major contribution to wake-breakdown at relatively low-inflow conditions.  This result is likely due to the wake breakdown being largely dominated by the computational start-up transients of the blade deformation rather than being dominated by periodic aerodynamic loading, as discussed in \cite{rodriguezJRE}.

\begin{figure*}[h!]
	\centering
	\begin{subfigure}[b]{0.495\textwidth}
		\centering
		\includegraphics[scale=0.55, trim=6cm 8cm 6cm 7cm]{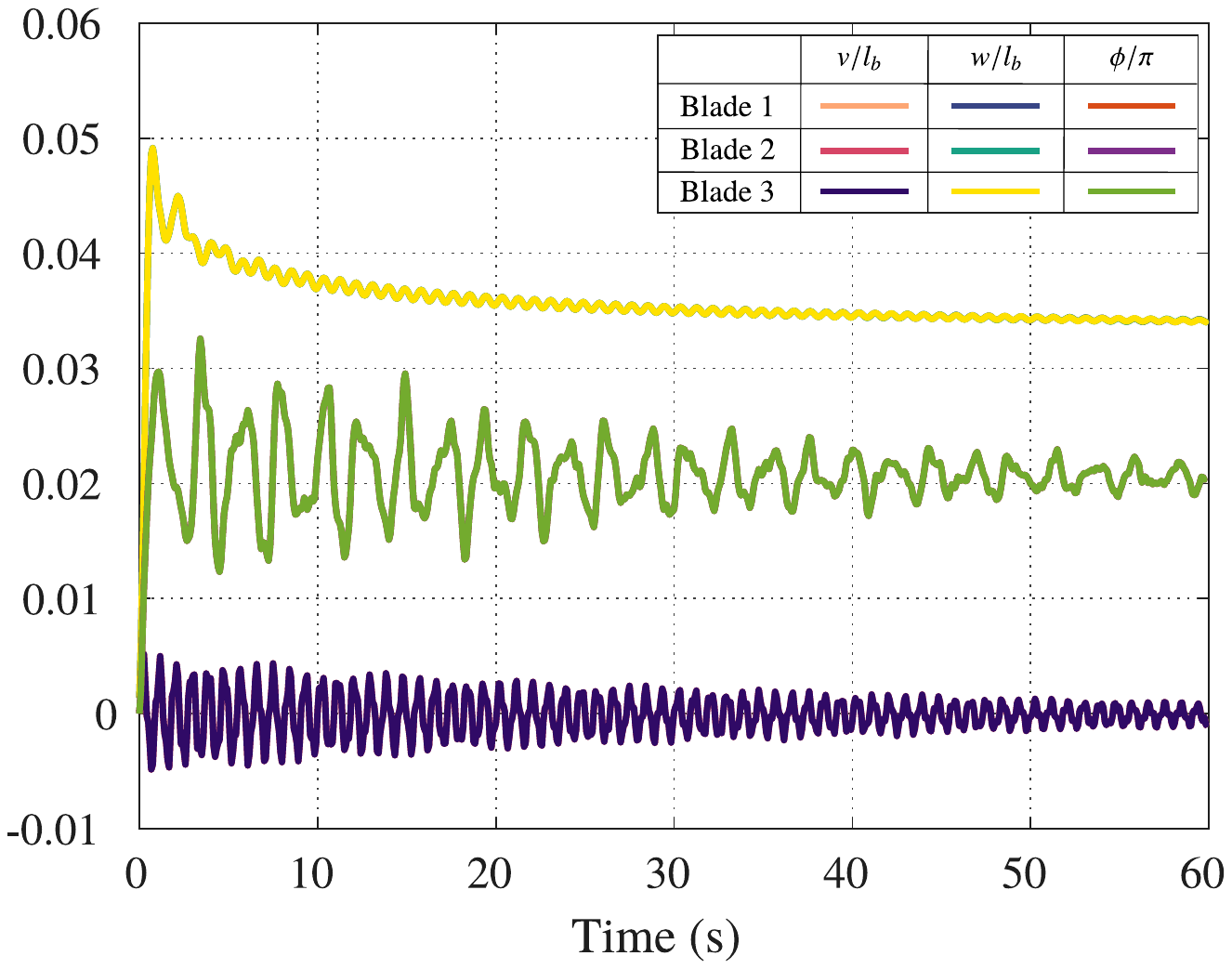}
		\caption[]%
		{ Zero-pitched rotor configuration }    
		\label{br_symmetric_aeroelaticresults}
	\end{subfigure}
	\hfill
	\begin{subfigure}[b]{0.495\textwidth}  
		\centering 
		\includegraphics[scale=0.55, trim=3cm 8cm 25 7cm]{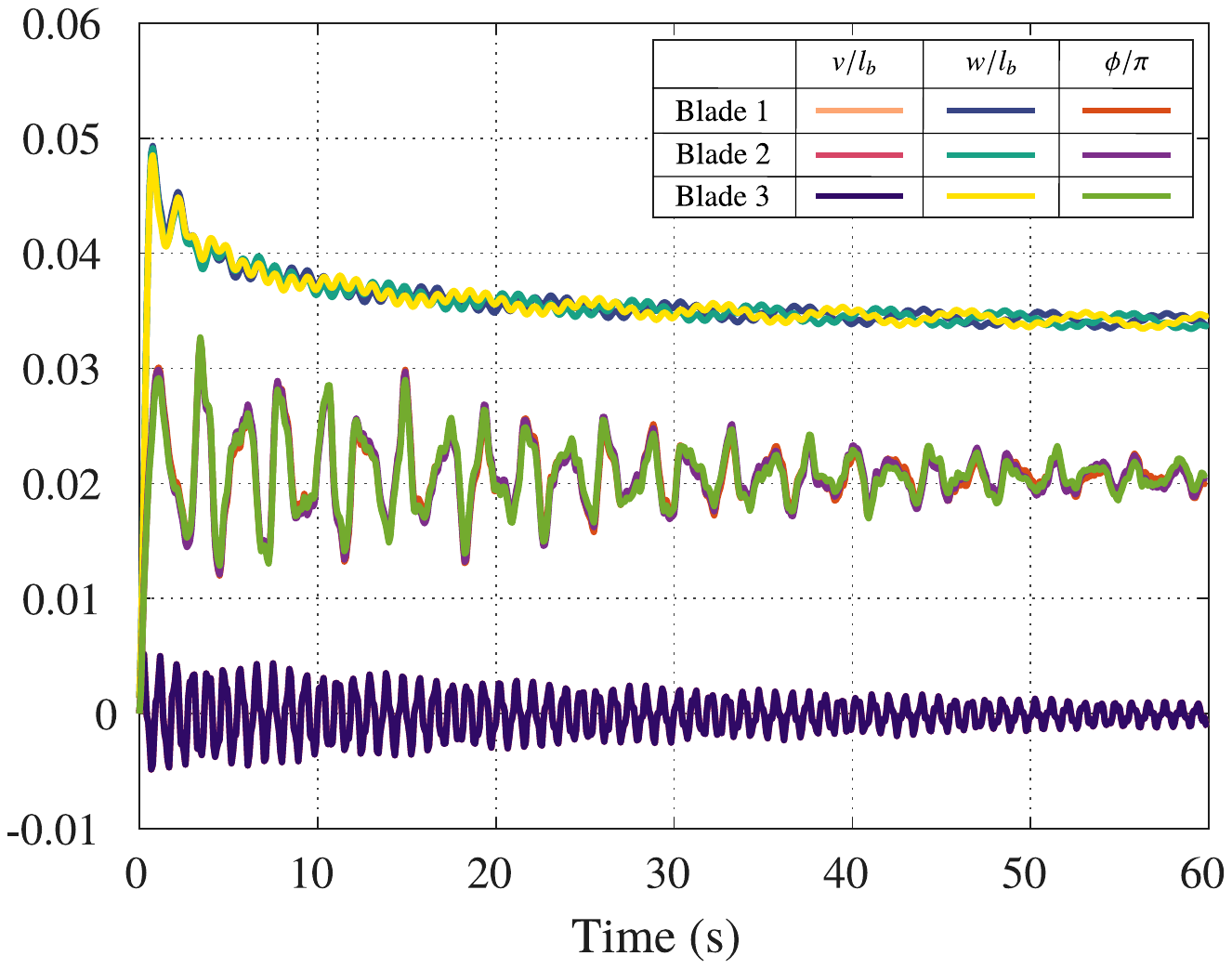}
		\caption[]%
		{Pitched rotor configuration}
		\label{br_operational_aeroelaticresults}
	\end{subfigure}
	\caption[]
	{Case 1 tip-vortex snapshot at $t=$ 60 s for rigid and flexible rotor operation } 
	\label{BR_aeroelastic_results}
\end{figure*}

Figure \ref{BR_aeroelastic_results} presents the aeroelastic responses at the tips of the rotor-blades for both ZPR and PR configurations. The aeroelastic responses for both configurations show that initial transient deformations reach about 5\% of the blade length and begin converging to a steady-state deformation of 3.3\% of the blade length at about $t=15$ s into the simulation. The ZPR configuration shows identical behavior for all rotor-blades across all degrees-of-freedom, while the PR configuration shows blades exhibiting an out-of-phase behavior in the flapwise degrees-of-freedom. The PR out-of-phase behavior is attributed to the periodic changes in the angle-of-attack and corresponding aerodynamic loading as the blade moves into and away from the inflow condition due to the rotor pitch. The time-histories of the edgewise deformation exhibit negligible responses, and torsional responses reflect about 2\% of $\pi$. However, torsional responses are effectively negligible with regard to its aerodynamic impact on the wake dynamics as the torsional frequency response amplitude is much lower relative to that of the flapwise degree-of-freedom. In other words, flapwise deformations are the leading order kinematics in the fluid-structure interaction for both ZPR and PR configurations for case 1. 

\begin{figure}
	\centering
	\begin{subfigure}[b]{0.475\textwidth}
		\centering
		\includegraphics[scale=0.5, trim=6.5cm 8cm 6cm 10cm]{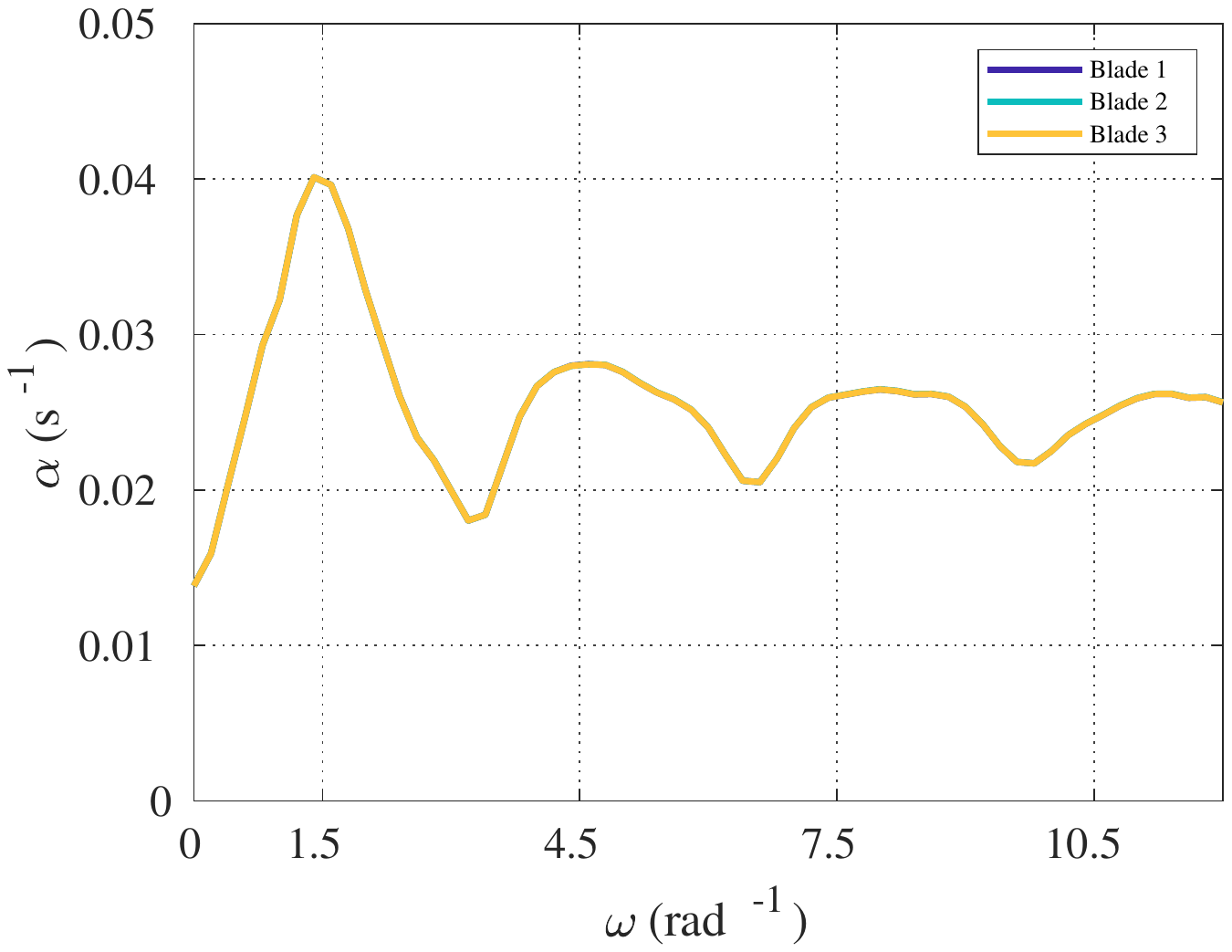}
		\caption[]%
		{Rigid zero-pitched rotor configuration}    
		\label{avo_br_sym_rigid}
	\end{subfigure}
	\hfill
	\begin{subfigure}[b]{0.475\textwidth}  
		\centering 
		\includegraphics[scale=0.5, trim=7.5cm 8cm 6cm 10cm]{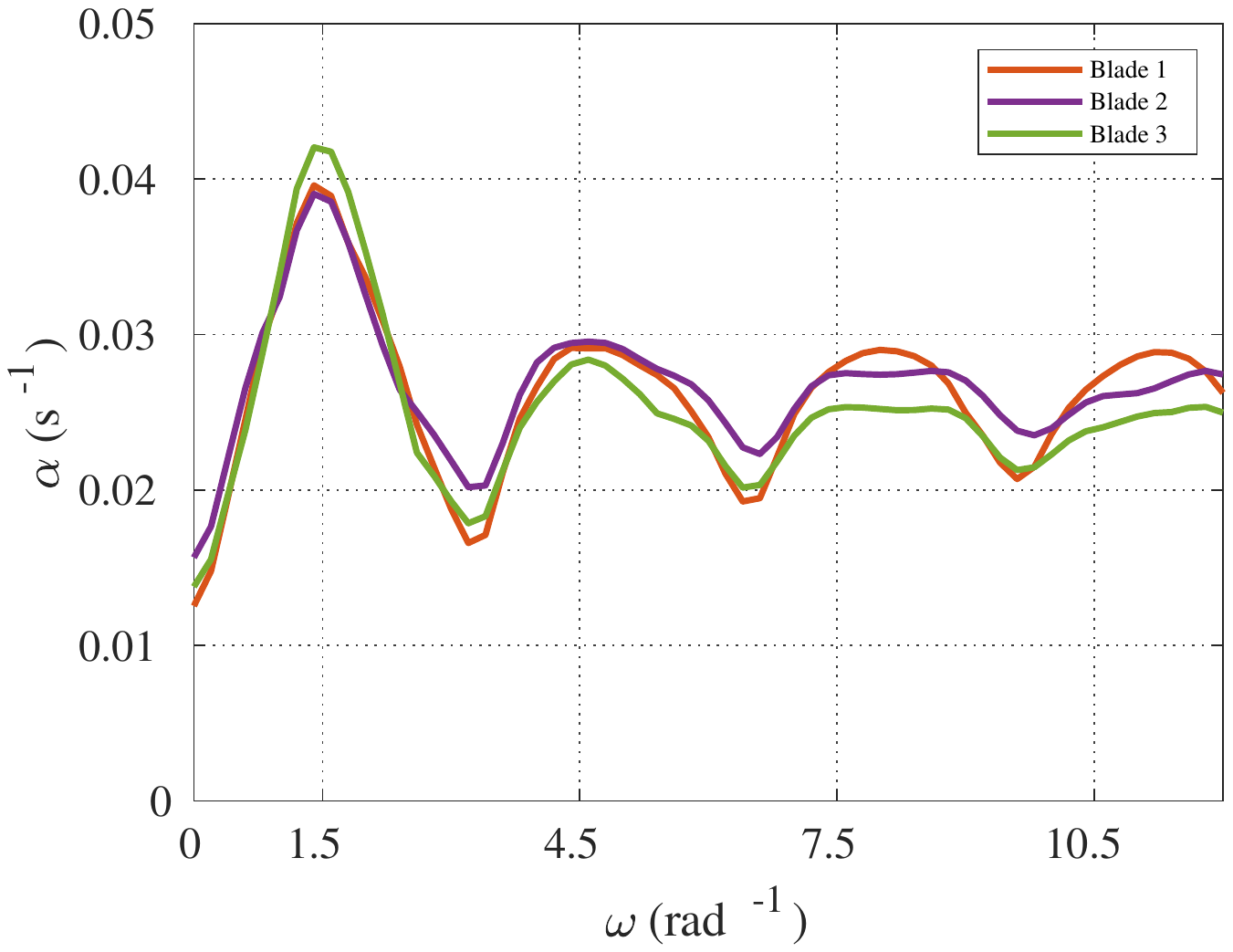}
		\caption[]%
		{Rigid pitched rotor configuration}
		\label{avo_br_tilt_rigid}
	\end{subfigure}
	\begin{subfigure}[b]{0.475\textwidth}  
		\centering 
		\includegraphics[scale=0.5, trim=8cm 8cm 6cm 8cm]{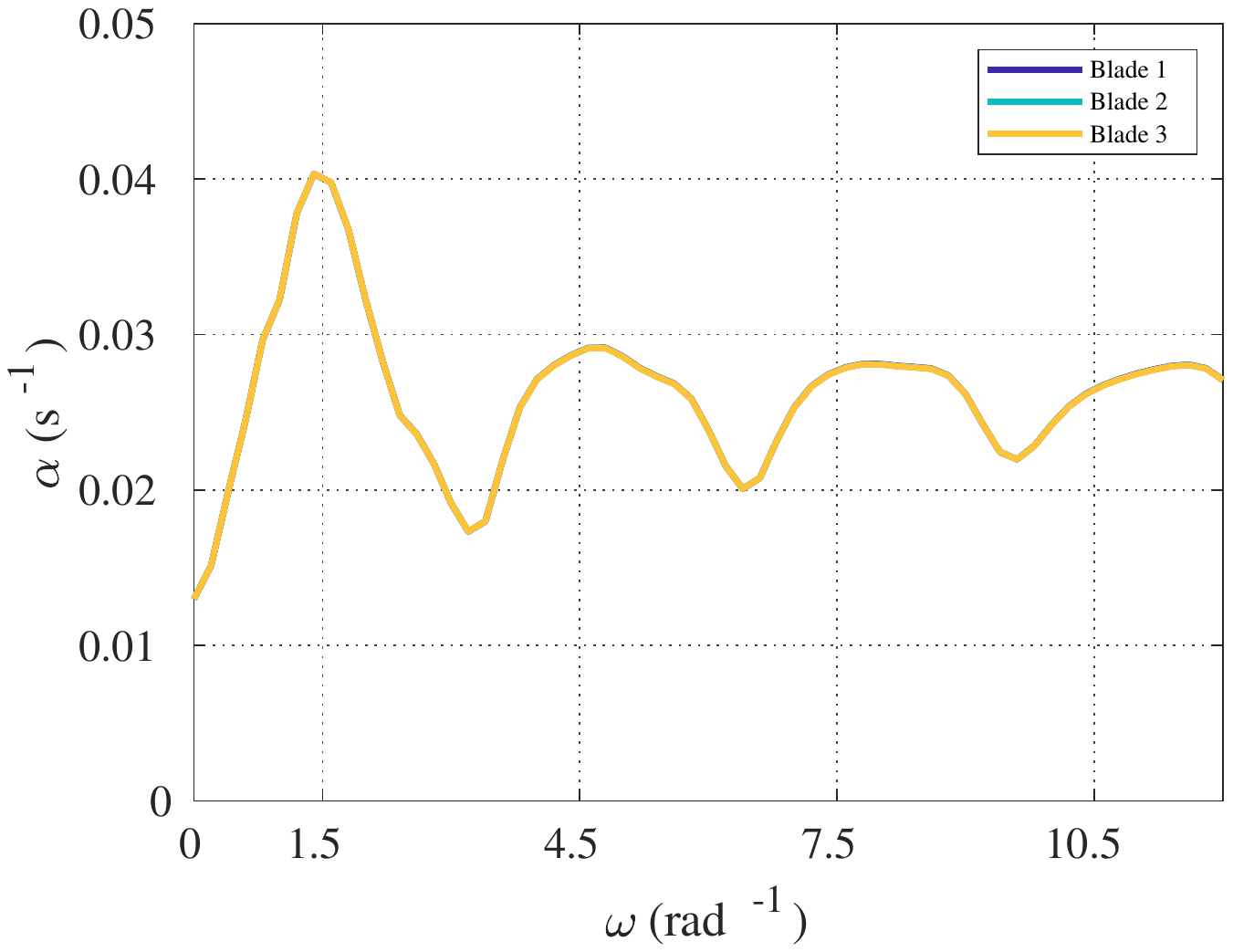}
		\caption[]%
		{Flexible zero-pitched rotor configuration}
		\label{avo_br_sym_flex}
	\end{subfigure}
	\begin{subfigure}[b]{0.475\textwidth}  
		\centering 
		\includegraphics[scale=0.5, trim=6cm 8cm 6cm 8cm]{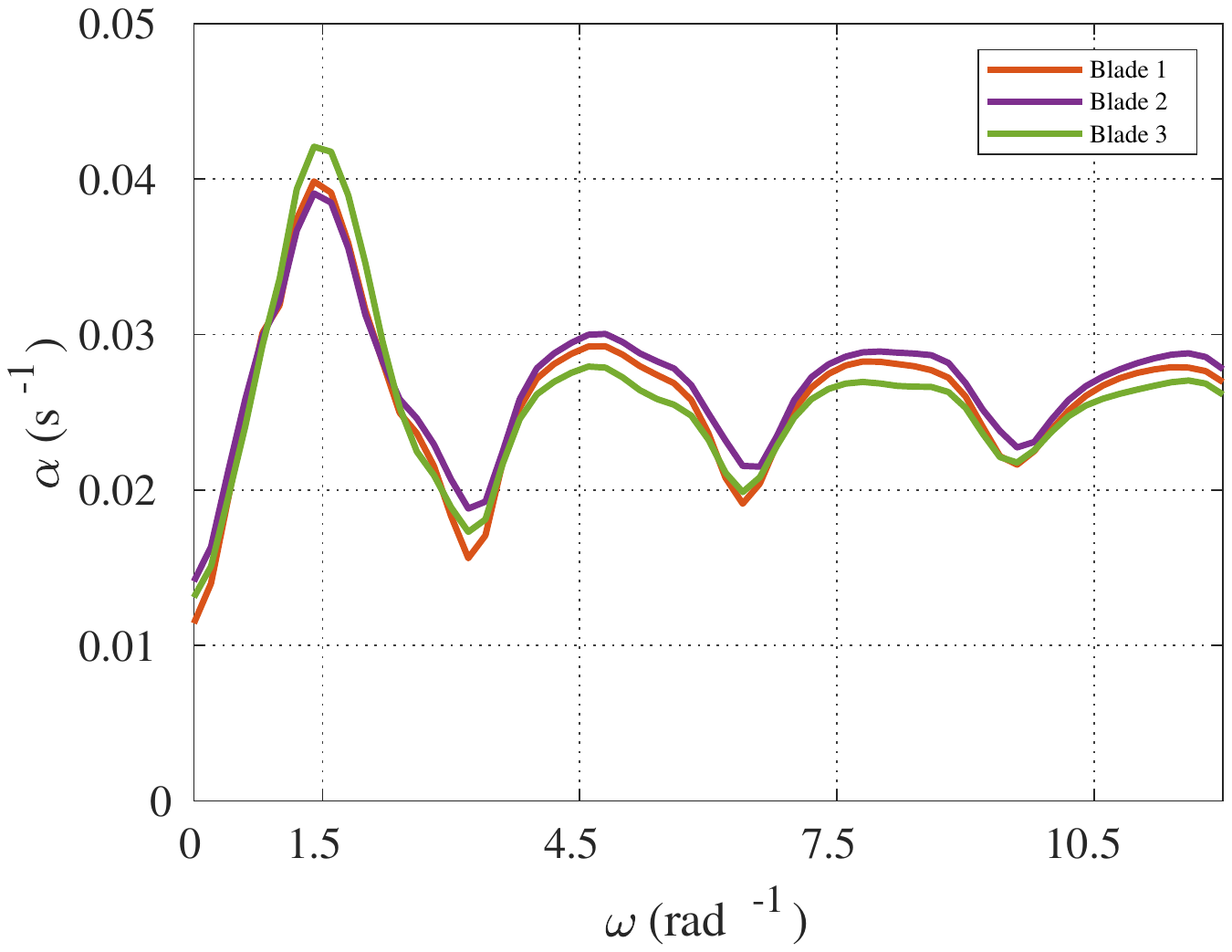}
		\caption[]%
		{Flexible pitched rotor configuration}
		\label{avo_br_tilt_flex}
	\end{subfigure}
	\caption[]%
	{Stability trend snapshots at $t=40$ s of tip vortices shed from rigid and flexible ZPR (left column) and PR configurations (right column).}
	\label{br_symtilt_avo}
\end{figure}

The qualitative evaluation of Fig.~\ref{snapshot_brsymtilt} posed above is reinforced quantitatively by the stability analysis performed on the snapshot $t=40$ s in Fig.~\ref{br_symtilt_avo}. As expected, via the classical stability trend criteria, the maximum eigenvalue for all configurations as shown in Fig.~\ref{br_symtilt_avo} corresponds to a perturbation wavenumber of $\omega=1.5$ rad$^{-1}$. The PR configurations exhibit slightly higher peaks than the ZPR configuration counterparts, which bolsters the notion that PR configurations generate more unstable tip vortices for rigid rotors. However, it is interesting to note that wake breakdown appears to occur at similar locations downstream for both flexible rotor configurations, despite the flexible PR configuration having larger maximum eigenvalues than the flexible ZPR configuration. It is important to note, though, that similar locations of wake breakdown may be a result from the initial transient perturbations from blade deformation that dominate the wake formation downstream. However, as time progresses the ``windowed" stability analysis captures the inherent tip vortex stability of each corresponding rotor configuration, which results in higher maximum eigenvalues for PR configurations in the $2\pi$ window.

\begin{figure*}[h!]
	\centering
	\begin{subfigure}[b]{0.475\textwidth}
		\centering
		\includegraphics[scale=0.6, trim=6cm 8cm 6cm 9cm]{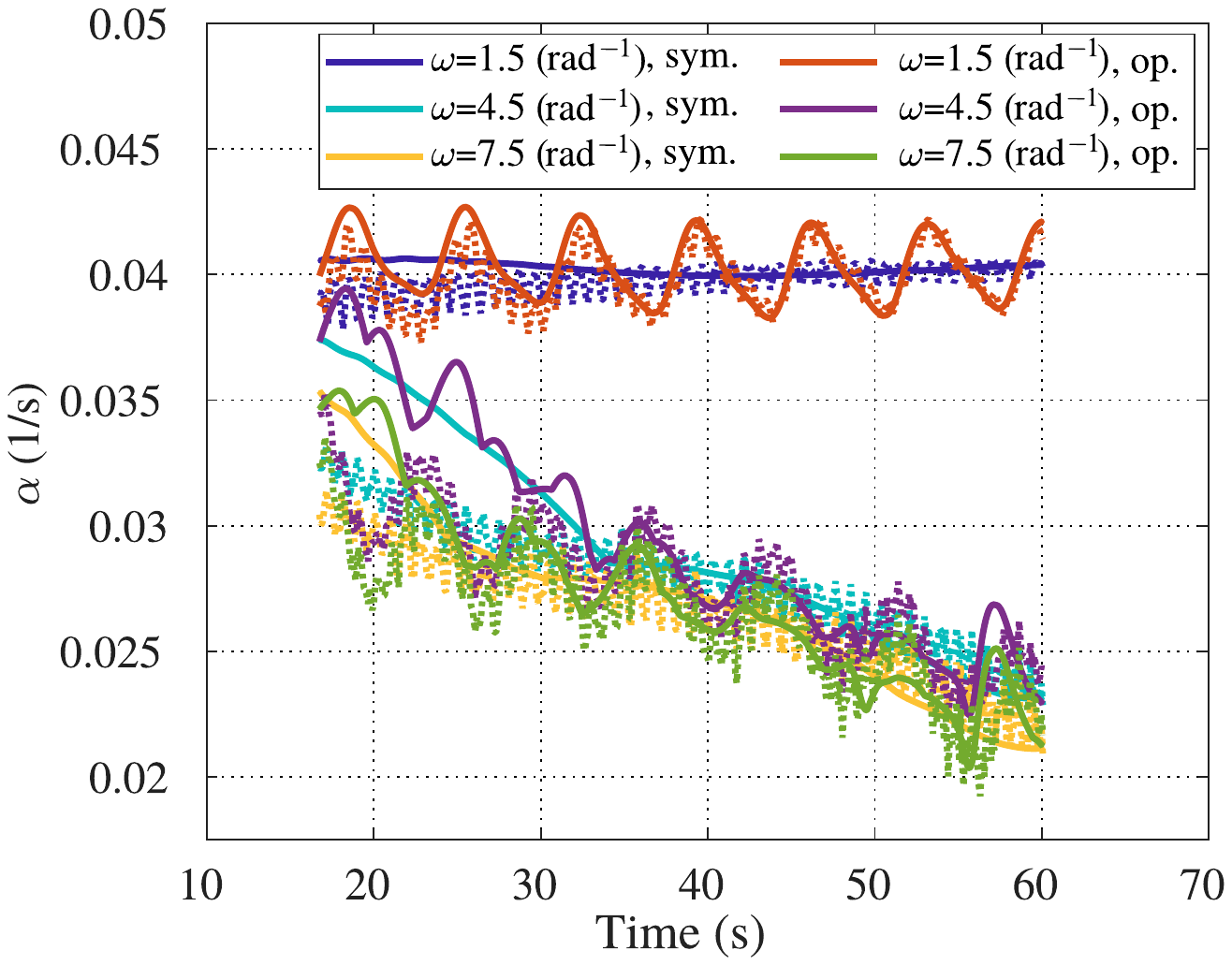}
		\caption[]%
		{Eigenvalue time-histories for blade 1}    
		\label{avt_br_symtilt_oms}
	\end{subfigure}
	\hfill
	\begin{subfigure}[b]{0.475\textwidth}  
		\centering 
		\includegraphics[scale=0.6, trim=6cm 8cm 6cm 9cm]{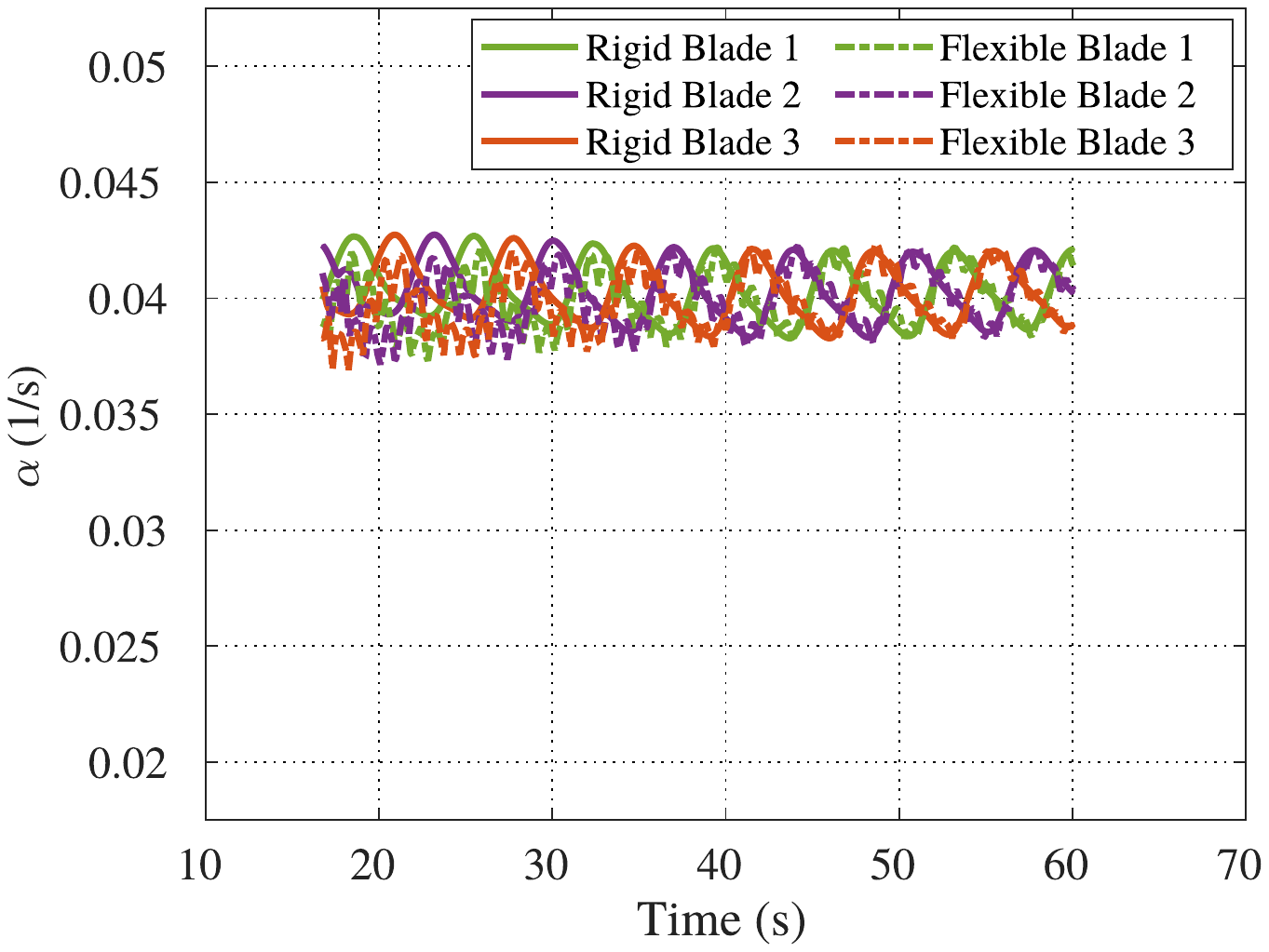}
		\caption[]%
		{Eigenvalue time-history for blade 1, 2, and 3}
		\label{avt_br_symtilt_blds}
	\end{subfigure}
	\caption[]
	{Case 1 time history growth rates: a) eigenvalues corresponding to perturbation wavenumbers $\omega$=1.5,  4.5, 7.5 rad$^{-1}$ for ZPR and PR configurations(legend provides color scheme of rotor configuration and dashed lines represent time-history of growth-rates from flexible blades)); and b) operational configuration eigenvalues corresponding to $\omega$=1.5 rad$^{-1}$ for all three blades (solid lines = rigid rotors, dashed lines = flexible rotors).} 
	\label{avt_br}
\end{figure*}

To further highlight the impact blade flexibility has on tip vortex stability, eigenvalues corresponding to perturbation wavenumbers $\omega=1.5,\:4.5$ and $7.5$ rad$^{-1}$, i.e. $N_b=3$ and $k=1,2,3$ for $\omega=N_b\left(k-1/2\right)$, are tracked in time for ZPR and PR configurations and results are presented in Fig.\ref{avt_br}. Despite the rotor configuration, it is shown in Fig.~\ref{avt_br_symtilt_oms} that within the $2\pi$ window of tip vortices under consideration, blade flexibility initially generates less unstable (lower positive eigenvalues) tip vortices than rigid rotors. As time progresses, the eigenvalues generated from the stability of rigid rotor tip vortices tend to approach the temporal eigenvalue characteristics of tip vortices generated by flexible rotors. The impact of blade flexibility on the stability of tip vortices is likely an indication that aeroelastic interactions allow the wake to more quickly approach its equilibrium state.

The conjecture presented earlier that PR configurations break down earlier than ZPR configurations is supported  by Fig.~\ref{avt_br_symtilt_oms}, which shows that PR configurations periodically reach higher maximum eigenvalues ($\alpha$ corresponding to $\omega=1.5$ rad$^{-1}$) than ZPR configurations. The periodic behavior seen in Fig.~\ref{avt_br_symtilt_oms}, not present in the ZPR configuration, corresponds to a change in the angle-of-attack of individual blades as they pass through a full rotation in the tilted rotor plane, which was reflected in the aeroleastic flapwise response. Figure \ref{avt_br_symtilt_blds} further illustrates the out-of-phase behavior of the maximum eigenvalues (corresponding to perturbation wavenumber $\omega=1.5$ rad$^{-1}$) of each tip vortex shed off of individual blades. It is seen that the each individual tip vortex contains the same stability trend but with a phase shift that corresponds to the periodic change in angle of attack of each rotor blade as it moves around the tilted rotor plane.

\begin{figure*}[h!]
	\centering
	\begin{subfigure}[b]{0.475\textwidth}
		\centering
		\includegraphics[scale=0.55, trim=6cm 8cm 6cm 7cm]{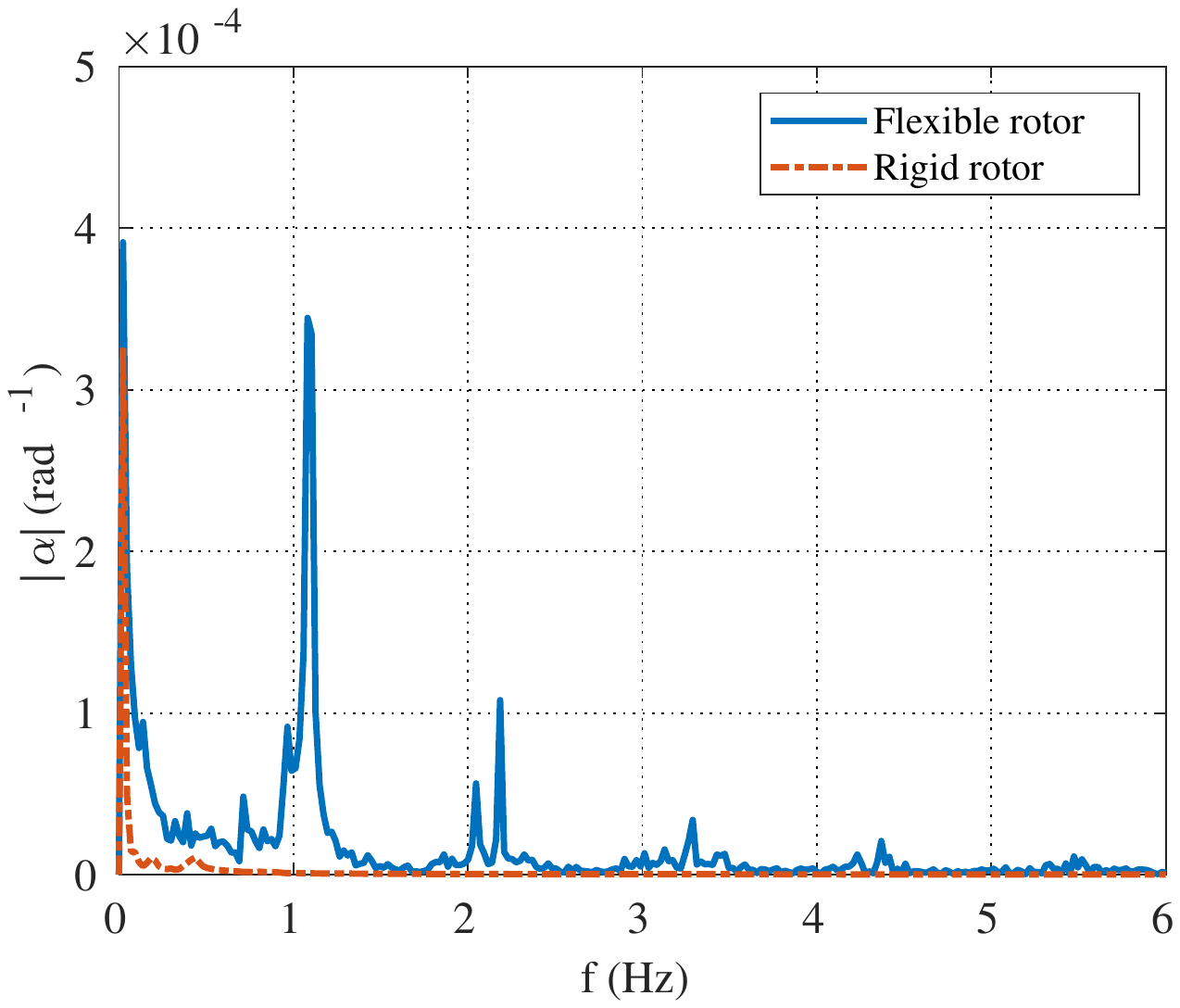}
		\caption[]%
		{Zero-pitched rotor configuration}    
		\label{br_sym_FFT}
	\end{subfigure}
	\hfill
	\begin{subfigure}[b]{0.475\textwidth}  
		\centering 
		\includegraphics[scale=0.55, trim=6cm 8cm 6cm 7cm]{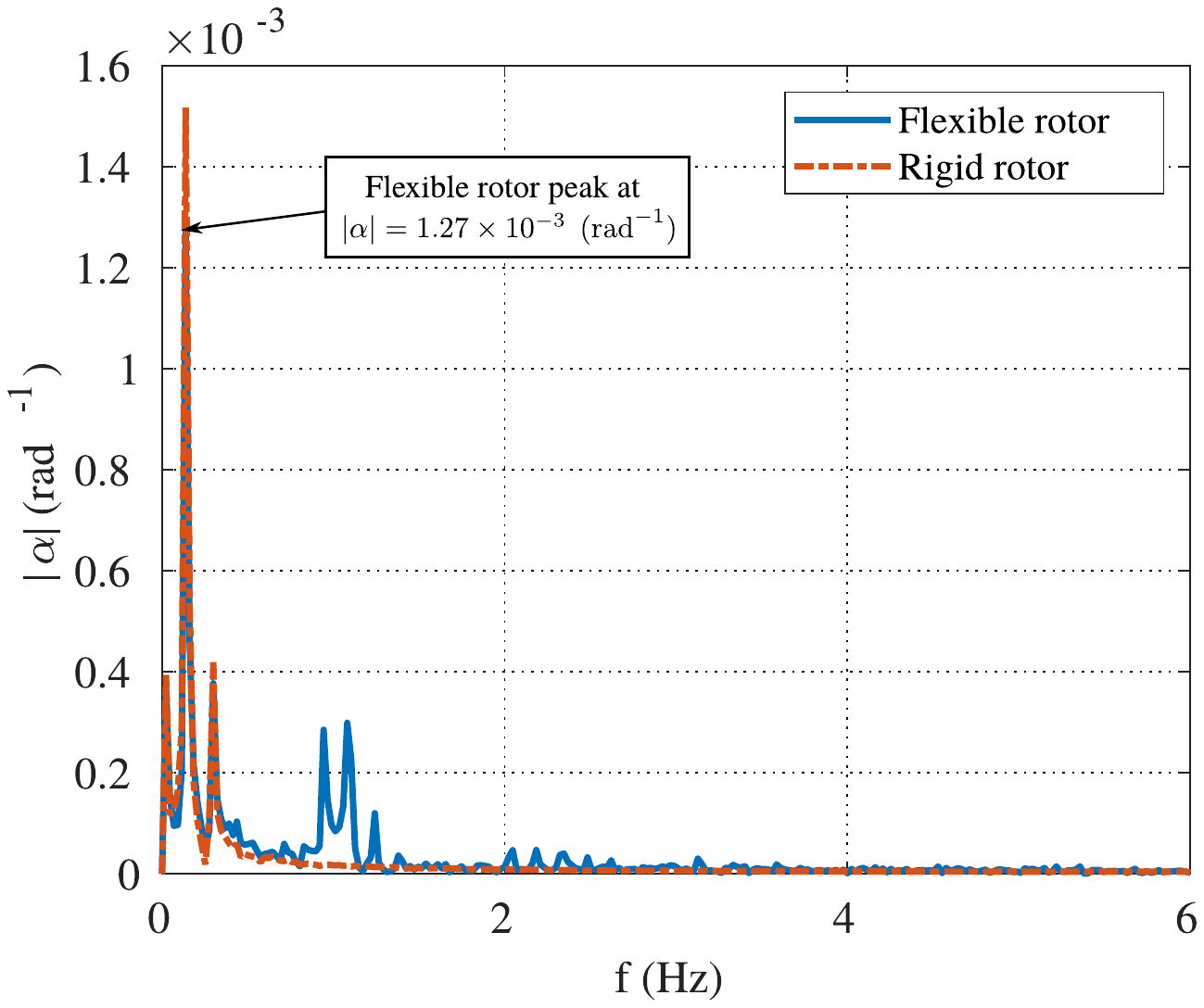}
		\caption[]%
		{Pitched rotor configuration}
		\label{br_tilt_FFT}
	\end{subfigure}
	\caption[]
	{FFT-derived frequency spectra of eigenvalue time history signal for perturbation wavenumber $\omega=1.5$ rad\textsuperscript{-1} : a) ZPR configuration; b) PR configuration} 
	\label{br_FFT}
\end{figure*}

The impact that the PR configuration has on stability trends is highlighted by the fast-Fourier transform (FFT) of eigenvalues corresponding to $\omega=1.5$ rad$^{-1}$ in Fig.~\ref{br_FFT}. The FFTs of both ZPR and PR configurations are compared to evaluate rotor-plane tilt impact on tip vortex stability, in Fig.~\ref{br_sym_FFT} and Fig.~\ref{br_tilt_FFT}, respectively. The FFT for the rigid ZPR configuration reflects a low-frequency contribution of $f\approx 0.02$ Hz that is not trivial to interpret. This contribution is seen in Fig.~\ref{avt_br_symtilt_oms} where the time-history of the eigenvalue shows a slight decay and growth, which may likely be caused by the meandering of the initial transient effects of the simulation, i.e., the dynamics of the agglomeration of filaments downstream in Fig.~\ref{snapshot_brsymtilt}. The flexible ZPR configuration also exhibits the low-frequency contribution of $f\approx 0.02$ Hz but also exhibits a significant contribution of $f\approx 1.1$ Hz, which is near the first natural frequency of the NREL rotor blade, $f_1=1.2$ Hz. The FFT-derived frequency spectra of the PR configuration highlights a dominant contribution of $f=\Omega/2\pi\approx0.14$ Hz. This frequency was first presented by \citet{rodriguezJERT} and then by \citet{rodriguezphd}, where it was determined that stability trends fluctuate at a period $\Lambda=2\pi/\Omega$. However, \citet{rodriguezJERT} and \citet{rodriguezphd} concluded the stability trend fluctuation was a byproduct of the blade passing frequency, and not the rotor plane tilt. Figure \ref{br_FFT} shows that the fluctuating frequency, $f=\Omega/2\pi$, corresponds to the impact of the pitched rotor on the rotor blade angle of attack and stability trends, and does not correspond to the rotation rate alone.

\subsection{Case 2}

\begin{figure*}[h!]
	\centering
	\begin{subfigure}[b]{0.495\textwidth}
		\centering
		\includegraphics[scale=0.575, trim=7cm 10cm 6cm 9cm]{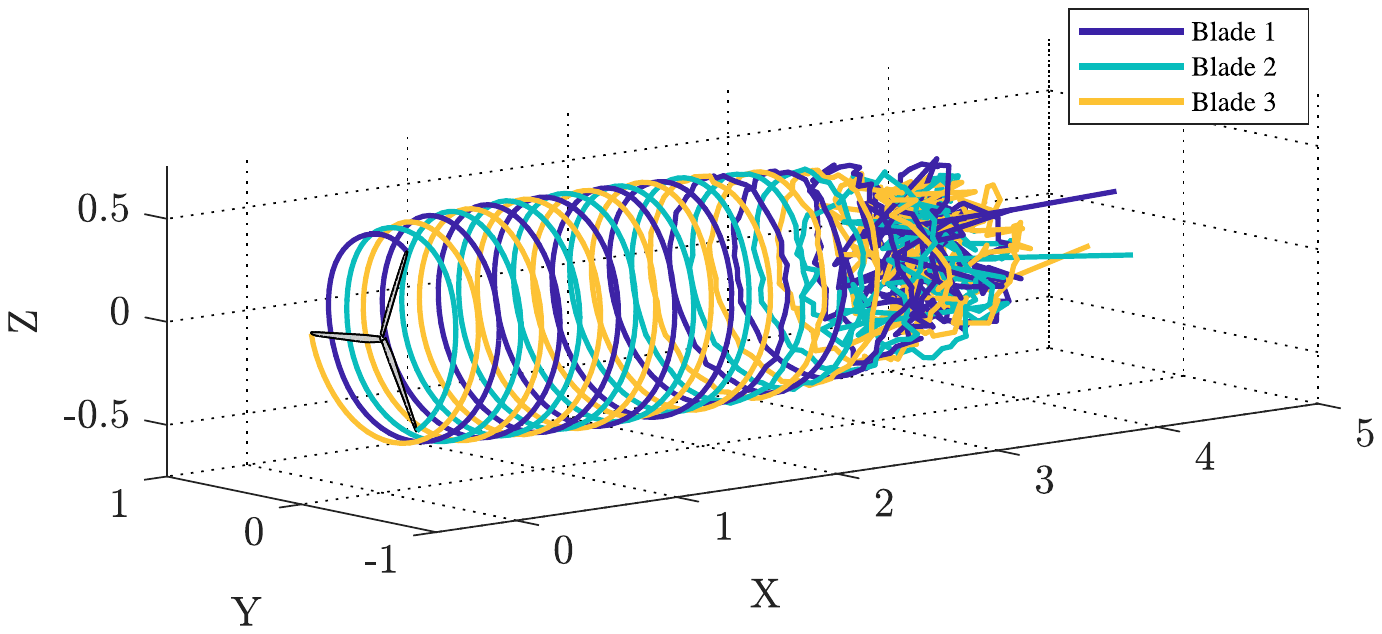}
		\caption[]%
		{ Zero-pitched rigid rotor configuration }    
	\end{subfigure}
	\hfill
	\begin{subfigure}[b]{0.495\textwidth}  
		\centering 
		\includegraphics[scale=0.575, trim=4cm 10cm 25 9cm]{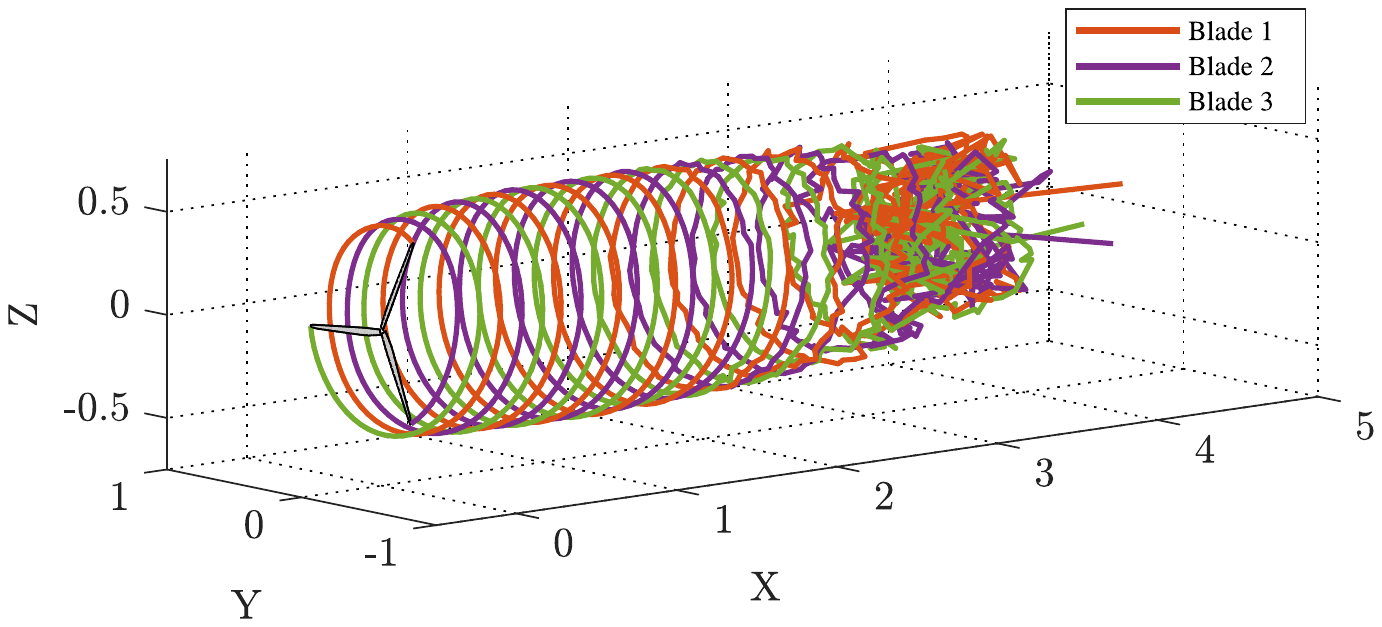}
		\caption[]%
		{Pitched rigid rotor configuration}
	\end{subfigure}
	\hfill
\begin{subfigure}[b]{0.495\textwidth}  
	\centering 
	\includegraphics[scale=0.575, trim=4cm 10cm 25 8cm]{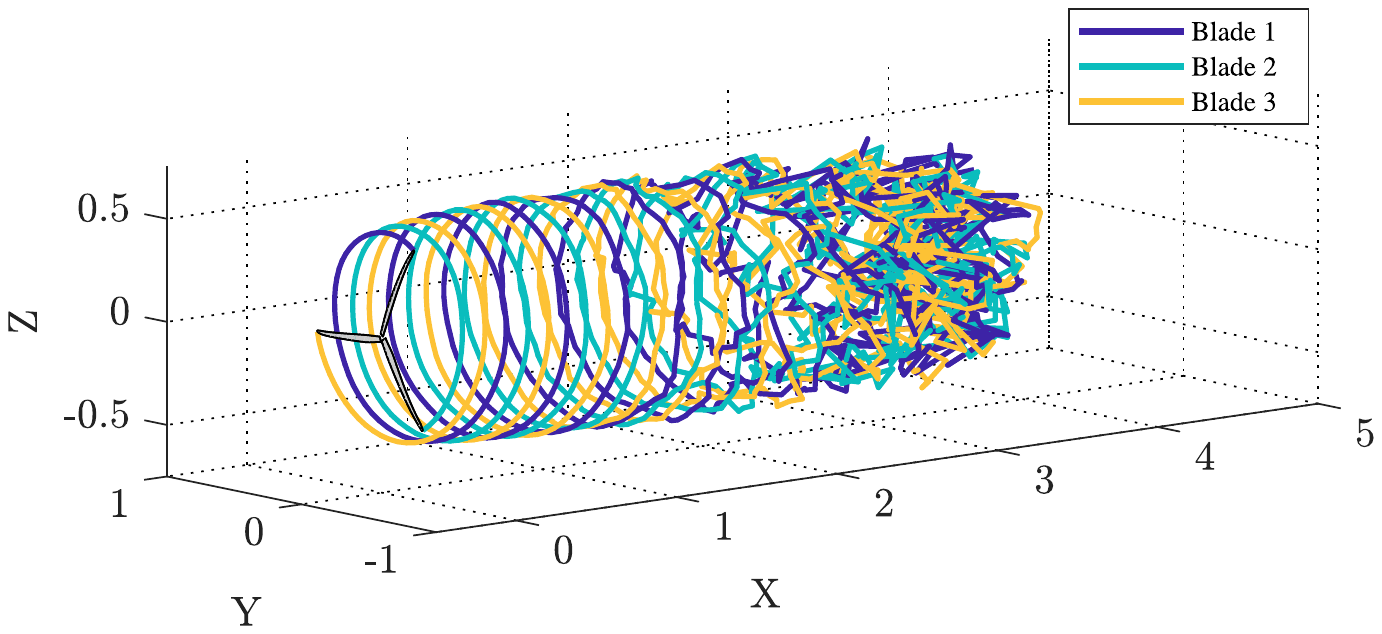}
	\caption[]%
	{Zero-pitched flexible rotor configuration}
\end{subfigure}
	\hfill
\begin{subfigure}[b]{0.495\textwidth}  
	\centering 
	\includegraphics[scale=0.575, trim=4cm 10cm 25 8cm]{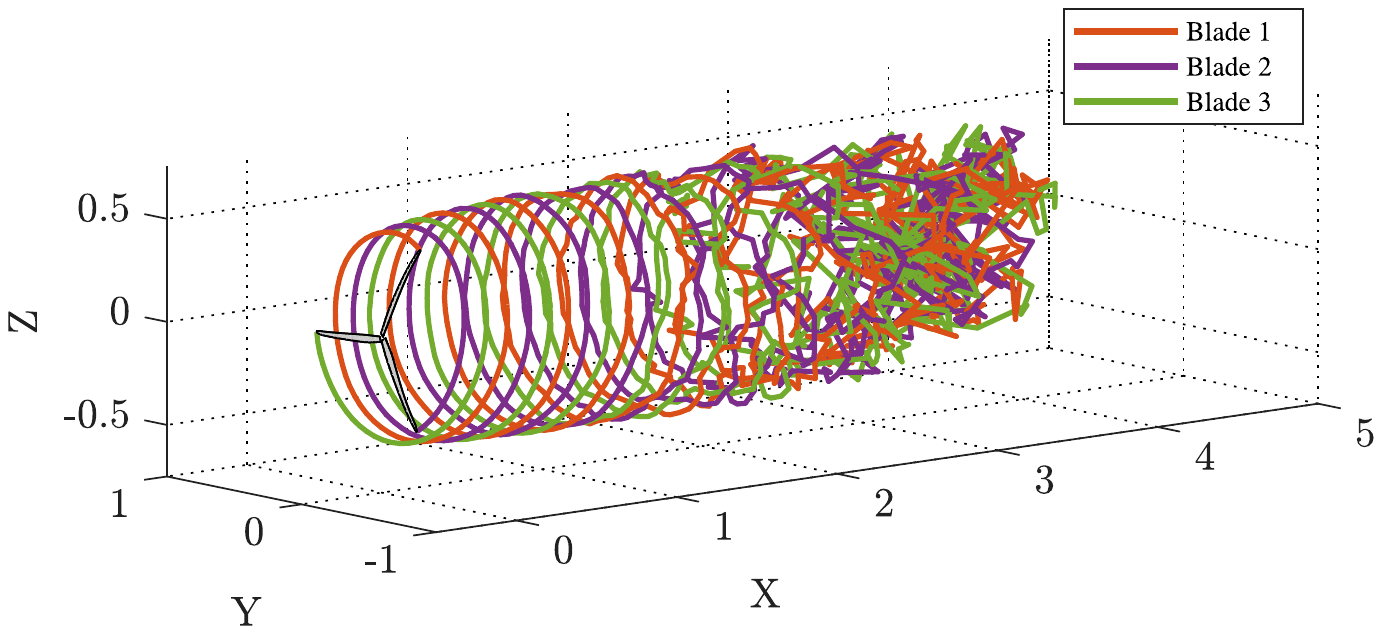}
	\caption[]%
	{Pitched flexible rotor configuration}
\end{subfigure}
	\caption[]
	{Case 2 tip-vortex snapshot at $t=$ 60 s for rigid and flexible rotor operation} 
	\label{snapshot_rsymtilt}
\end{figure*}

Case 2 conditions are now investigated. Snapshots of the ZPR and PR configurations at $t=60$ s are shown in Fig.~\ref{snapshot_rsymtilt}. Differences in wake breakdown location between rigid ZPR and PR are less visible qualitatively than case 1 rotor-wakes. Differences in wake breakdown location are difficult to identify likely because of the decrease in tip-speed ratio, which increases tip-vortex to tip-vortex spacing (pitch), thereby generating less unstable vortices and mitigating the effect of perturbation propagation across tip vortices compared to case 1 operation. However, one major difference between ZPR and PR wakes is the transition into the agglomeration of tip vortices in the far wake. The ZPR configuration exhibits a smooth transition into the far wake as opposed to the operational configuration which exhibits distortion in tip vortices as early as $X\approx 1.8$ downstream. Similar to case 1, the flexible rotor configurations exhibit earlier breakdown than both the rigid configurations. These earlier occurrences of wake breakdown in flexible rotors is likely due to the initial transient impact that blade deformation has on the formation and geometry of the tip vortices. Rotor configuration (ZPR vs.~PR) for flexible rotors seems to not have a substantial impact with regard to the onset of wake breakdown location. The minimal influence of rotor configuration on wake breakdown in this case is likely due to the wake breakdown being largely dominated by the initial transients of the blade deflection rather than being dominated by periodic aerodynamic loading as discussed in \cite{rodriguezJRE}.

\begin{figure*}[h!]
	\centering
	\begin{subfigure}[b]{0.495\textwidth}
		\centering
		\includegraphics[scale=0.55, trim=7cm 8cm 6cm 9cm]{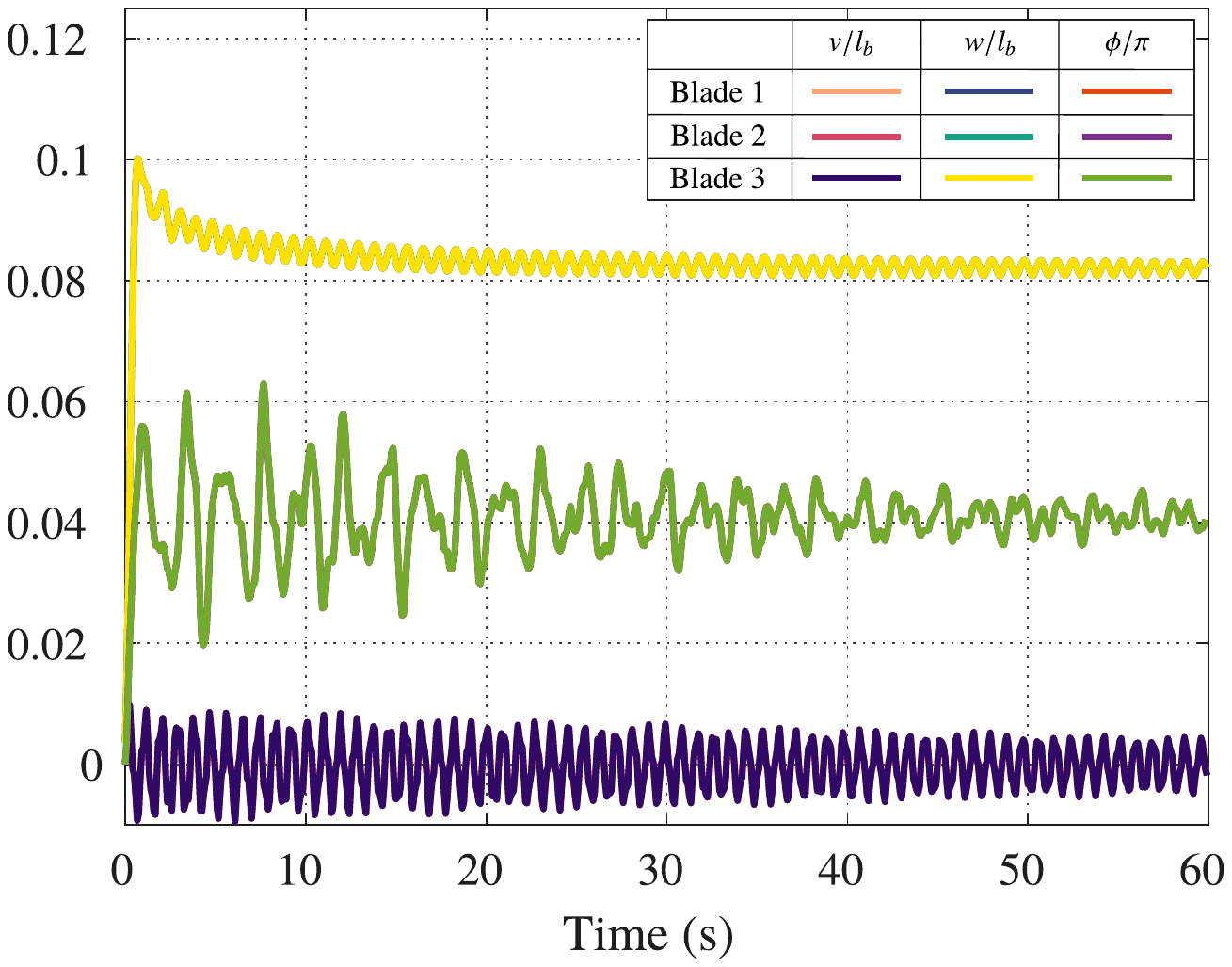}
		\caption[]%
		{ Zero-pitched rotor configuration}    
		\label{}
	\end{subfigure}
	\hfill
	\begin{subfigure}[b]{0.495\textwidth}  
		\centering 
		\includegraphics[scale=0.55, trim=4cm 8cm 25 9cm]{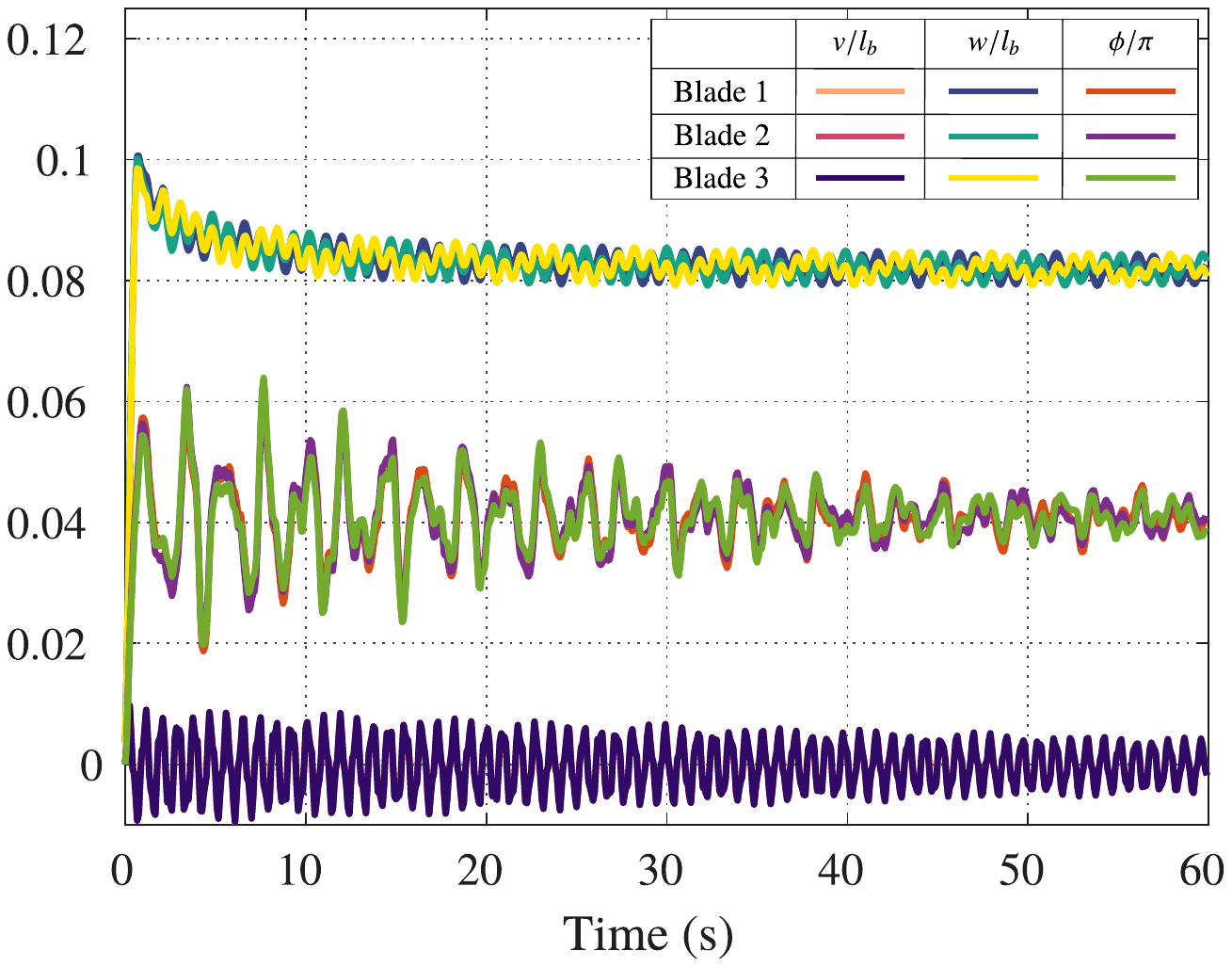}
		\caption[]%
		{Pitched rotor configuration}
		\label{}
	\end{subfigure}
	\caption[]
	{Case 2 tip-vortex snapshot at $t=$ 60 s for rigid and flexible rotor operation } 
	\label{R_aeroelastic_results}
\end{figure*}

Figure \ref{R_aeroelastic_results} presents the aeroelastic responses at the tips of the rotor-blades for both ZPR and PR configurations in case 2. It is important to note the case 2 corresponds to the highest aerodynamic loading condition of the NREL 5MW reference wind turbine as defined by \cite{jonkman:6}. \citet{rodriguezJRE} (Fig.~16 in \cite{rodriguezJRE}) also show that case 2 exhibits the highest angle-of-attack distribution along the length of the rotor-blades across all wind speeds for which the NREL 5MW wind turbine was designed. In this case, as in case 1, the dominant aeroelastic response for both configurations is the flapwise deformation. Here, the flapwise degree-of-freedom exhibits an initial transient deformation of about 10\% the blade length and begins converging to a steady-state deformation of 8\% the blade length at about $t=15$ s into the simulation. The ZPR configuration shows identical behavior for all rotor-blades across and degrees-of-freedom, while the PR configuration shows blades exhibiting an out-of-phase periodic behavior in the flapwise degrees-of-freedom as discussed in case 1. The time-histories of the edgewise deformation exhibit negligible responses, and torsional responses reflect about 4\% of $\pi$. Torsional responses are again effectively negligible with regard to its aerodynamic impact on the wake dynamics as the torsional frequency response amplitude is much lower relative to that of the flapwise degree-of-freedom, which again is the dominant kinematic response.

\begin{figure}
	\centering
	\begin{subfigure}[b]{0.475\textwidth}
		\centering
		\includegraphics[scale=0.5, trim=6.5cm 8cm 6cm 10cm]{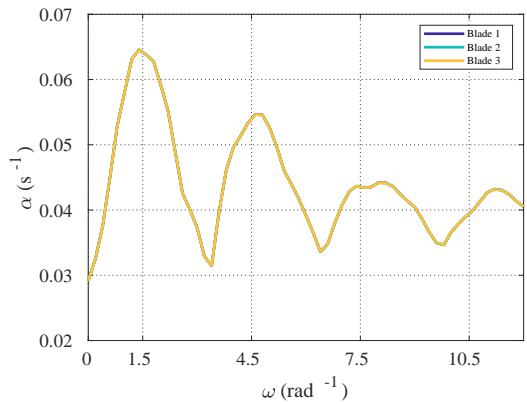}
		\caption[]%
		{Rigid zero-pitched rotor configuration }    
		\label{avt_r_symtilt_oms}
	\end{subfigure}
	\hfill
	\begin{subfigure}[b]{0.475\textwidth}  
		\centering 
		\includegraphics[scale=0.5, trim=7.5cm 8cm 6cm 10cm]{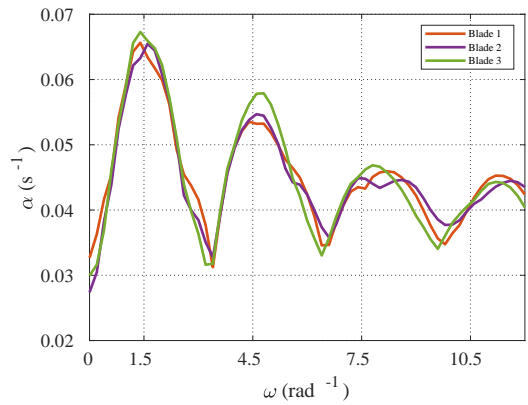}
		\caption[]%
		{Rigid pitched rotor configuration}
		\label{avt_r_symtilt_blds}
	\end{subfigure}
	\begin{subfigure}[b]{0.475\textwidth}  
		\centering 
		\includegraphics[scale=0.5, trim=8cm 8cm 6cm 8cm]{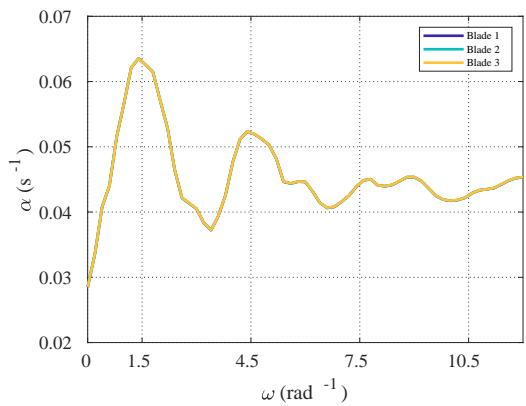}
		\caption[]%
		{Flexible zero-pitched rotor configuration}
		\label{avt_r_symtilt_blds}
	\end{subfigure}
	\begin{subfigure}[b]{0.475\textwidth}  
		\centering 
		\includegraphics[scale=0.5, trim=6cm 8cm 6cm 8cm]{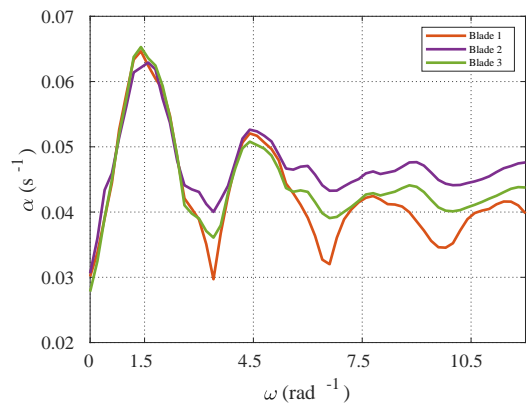}
		\caption[]%
		{Flexible pitched rotor configuration}
		\label{avt_r_symtilt_blds}
	\end{subfigure}
	\caption[]%
	{Stability trend snapshots at $t=40$ s of tip vortices shed from rigid and flexible ZPR (left column) and PR configurations (right column).}
	\label{r_symtilt_avo}
\end{figure}

Snapshots at $t=40$ s of the stability analysis performed on tip vortices for two rotations ($4\pi$ window) are presented in Fig.~\ref{r_symtilt_avo}. Maximum eigenvalues occur at the expected perturbation wavenumber of $\omega=1.5$ rad$^{-1}$. Across both rigid and flexible rotors, the PR configurations exhibit higher eigenvalue peaks than their ZPR counterpart, which again supports the notion that PR configurations tend to further destabilize tip vortices. However, it is interesting to note that wake breakdown appears to occur at similar locations downstream for both ZPR and PR configurations as seen in Fig.~\ref{snapshot_rsymtilt}, despite PR configurations having larger maximum eigenvalues than the ZPR configuration counterparts as shown in Fig.~\ref{r_symtilt_avo}. It is important to note, though, that similar locations of wake breakdown may be a result from the initial transient perturbations that dominate the wake formation downstream, and not the rotor configuration. However, as time progresses the ``windowed" stability analysis captures the inherent tip vortex stability of each corresponding rotor configuration, which results in higher maximum eigenvalues for PR configurations in the $4\pi$ window. Furthermore, the stability analysis for flexible rotors result in distorted (non-smooth) stability trends as perturbation wavenumbers increase, which was not as prevalent in case 1 and is likely due an increase in blade deformation. The differences between stability trends in rigid and flexible cases serve as an indicator that blade flexibility may reduce, shift, or distort tip vortex instability at specified perturbation frequencies.

\begin{figure*}[h!]
	\centering
	\begin{subfigure}[b]{0.475\textwidth}
		\centering
		\includegraphics[scale=0.6, trim=6cm 8cm 6cm 8cm]{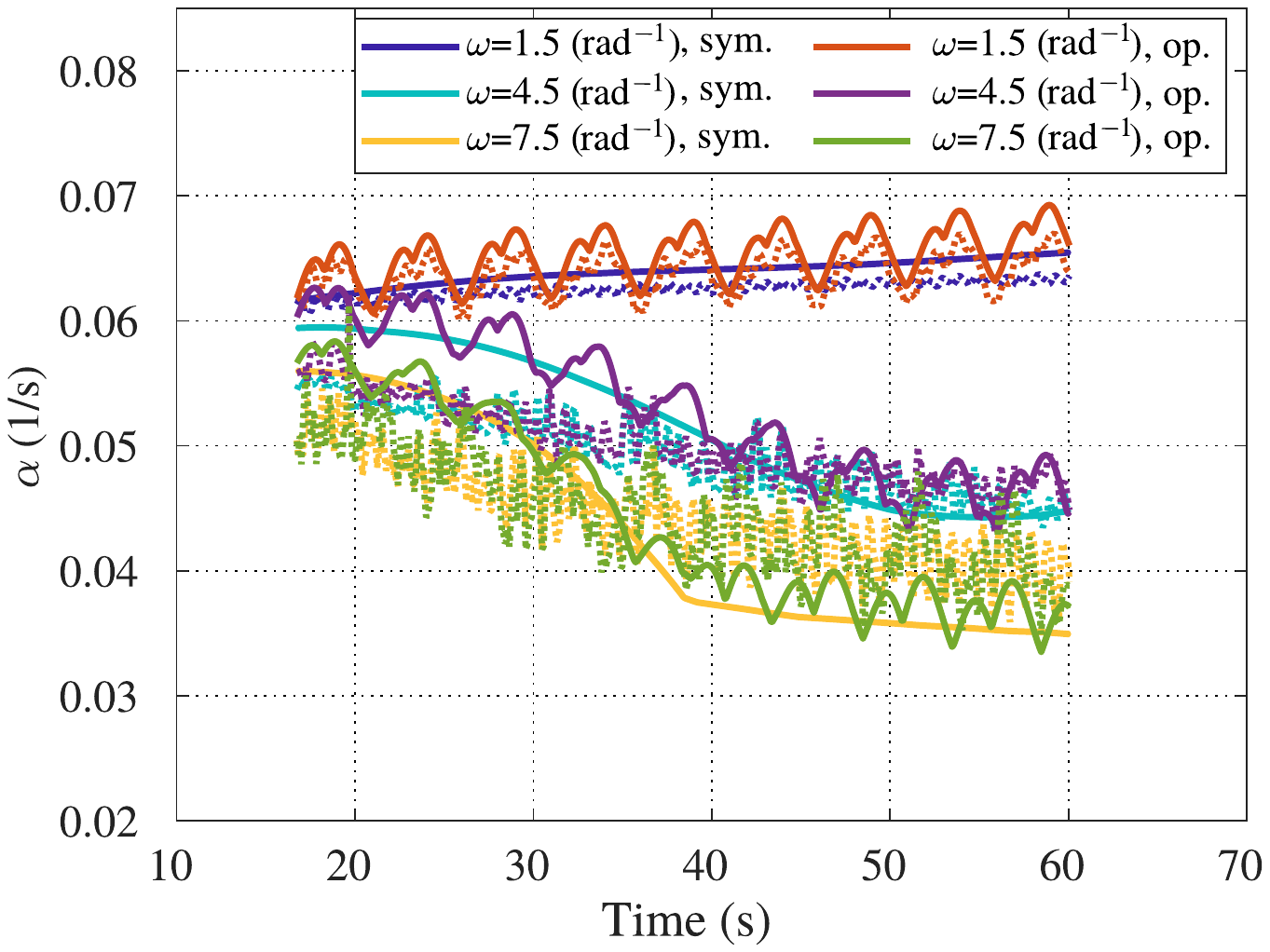}
		\caption[]%
		{Eigenvalue time-histories for blade 1}    
		\label{avt_r_symtilt_oms}
	\end{subfigure}
	\hfill
	\begin{subfigure}[b]{0.475\textwidth}  
		\centering 
		\includegraphics[scale=0.6, trim=6cm 8cm 6cm 8cm]{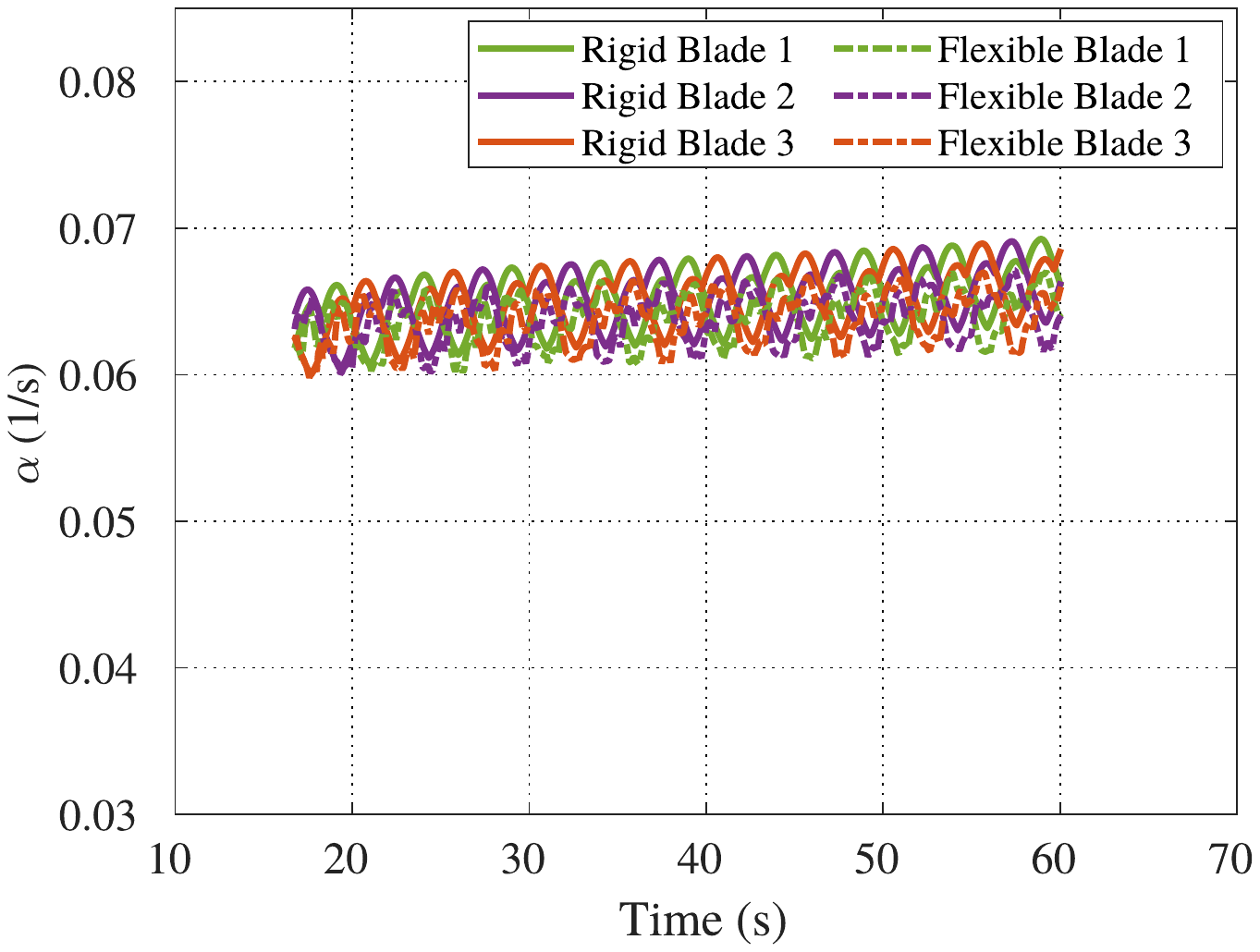}
		\caption[]%
		{Eigenvalue time-history for blades 1, 2, and 3}
		\label{avt_r_symtilt_blds}
	\end{subfigure}
	\caption[]
	{Case 2 time history operation growth rates: a) eigenvalues corresponding to perturbation wave numbers $\omega$=1.5, 4.5,7.5 rad\textsuperscript{-1} for ZPR and PR configurations (legend provides color scheme of rotor configuration and dashed lines represent time-history of growth-rates from flexible blades); b) Operational configuration eigenvalues corresponding to $\omega$=1.5 rad\textsuperscript{-1} for all three blades} 
	\label{avt_r}
\end{figure*}

The time histories of eigenvalues corresponding to perturbation wavenumbers $\omega=1.5,\:4.5$ and $7.5$ rad$^{-1}$, i.e., $N_b=3$ and $k=1,2,3$ for $\omega=N_b\left(k-1/2\right)$ are now tracked in time for ZPR and PR configurations in Fig.~\ref{avt_r}. Independent of the configuration, Fig.~\ref{avt_r_symtilt_oms} shows that within the $4\pi$ window of tip vortices under consideration, blade flexibility enables less unstable (lower positive eigenvalues) tip vortices at $\omega=1.5$ rad$^{-1}$ than rigid rotors. As time progresses, the presence of blade flexibility may generate tip vortices that are more sensitive to higher wavenumber perturbations as shown in the time histories for stability trends for $\omega=4.5$ and $7.5$ (rad$^{-1}$). The same higher wavenumber sensitivity was seen in tip vortices for conditions in case 1, but for case 2  this sensitivity is more pronounced in Fig.~\ref{avt_r}.

Figure \ref{avt_r_symtilt_oms} quantitatively supports the qualitative observation stated earlier that PR configurations breakdown earlier than ZPR configurations. As in case 1, it is shown in Fig.~\ref{avt_r_symtilt_oms} that operational configurations periodically reach higher maximum eigenvalues ($\alpha$ corresponding to $\omega=1.5$ rad$^{-1}$) than symmetric operational configurations. This periodic behavior seen in Fig.~\ref{avt_r_symtilt_oms}, not present in the symmetric configuration, corresponds to a change in the angle-of-attack of individual blades as they pass through a full rotation in the tilted rotor plane, as was also observed in case 1 operation. Figure \ref{avt_r_symtilt_blds} further illustrates the out-of-phase behavior of the maximum eigenvalues (corresponding to perturbation wavenumber $\omega=1.5$ rad$^{-1}$) of each tip vortex shed off of individual blades.

\begin{figure*}[h!]
	\centering
	\begin{subfigure}[b]{0.475\textwidth}
		\centering
		\includegraphics[scale=0.55, trim=6cm 8cm 6cm 7cm]{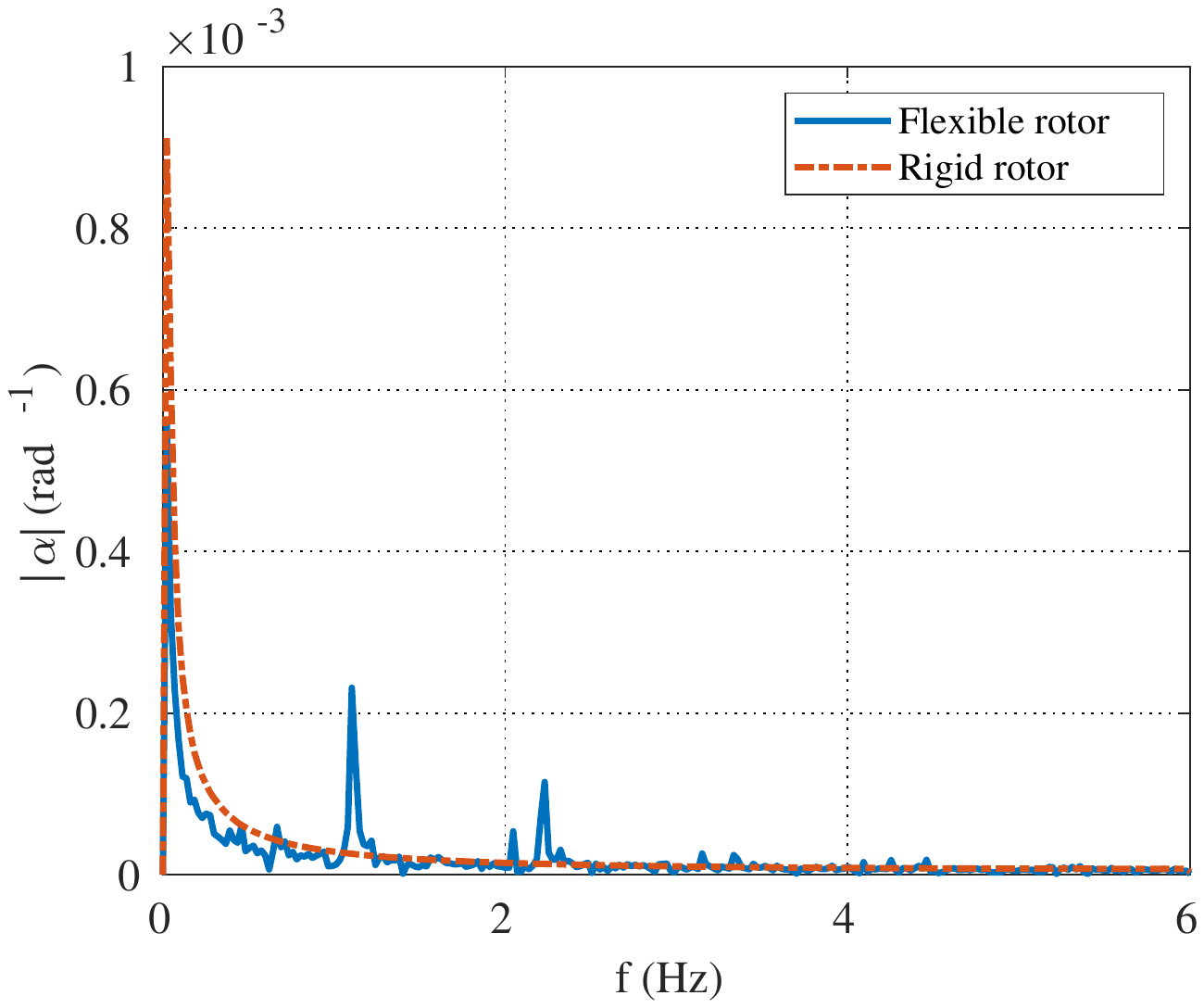}
		\caption[]%
		{ Zero-pitched rotor configuration}    
		\label{r_sym_FFT}
	\end{subfigure}
	\hfill
	\begin{subfigure}[b]{0.475\textwidth}  
		\centering 
		\includegraphics[scale=0.55, trim=6cm 8cm 6cm 7cm]{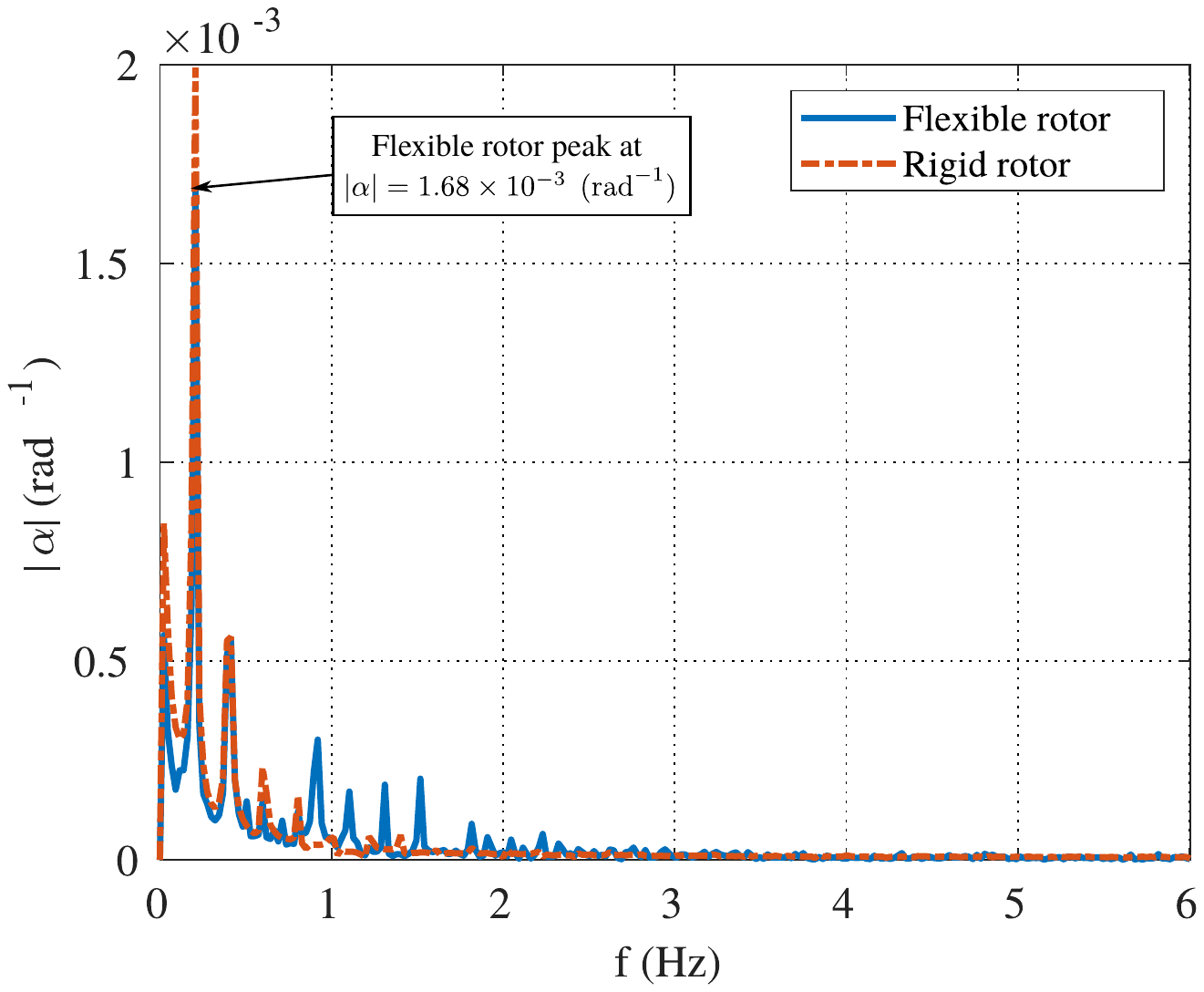}
		\caption[]%
		{Pitched rotor configuration}
		\label{r_tilt_FFT}
	\end{subfigure}
	\caption[]
	{FFT-derived frequency spectra of eigenvalue time history signal for perturbation wave number $\omega=1.5$ rad\textsuperscript{-1}: a) ZPR configuration; b) PR configuration} 
	\label{r_FFT}
\end{figure*}

The frequency spectra of the eigenvalue time-histories for case 2 ZPR and PR configurations is shown in Fig.~\ref{r_FFT}. For the ZPR configuration, the FFT shows a dominant low-frequency contribution at $f\approx 0.02$ Hz that was also seen in case 1. As mentioned previously, identifying this low-frequency contribution is not trivial but seems to be related to the agglomeration of tip vortices downstream. This low frequency contribution is seen in Fig.~\ref{avt_br_symtilt_oms} through a small and almost negligible eigenvalue fluctuation in time. For the ZPR configuration, the FFT is showing a secondary frequency of $f\approx 1.1$ Hz contribution near the first flapwise natural frequency of the NREL rotor blade ($f_1\approx 1.2$ Hz), which was also present in case 1. The PR configuration FFT is dominated by the frequency at which the angle-of-attack changes in the tilted rotor plane, which for case 2 is $f=\Omega/2\pi\approx 0.2$ Hz.  

\subsection{Case 3}
\begin{figure*}[h!]
	\centering
	\begin{subfigure}[b]{0.495\textwidth}
		\centering
		\includegraphics[scale=0.575, trim=7cm 10cm 6cm 9cm]{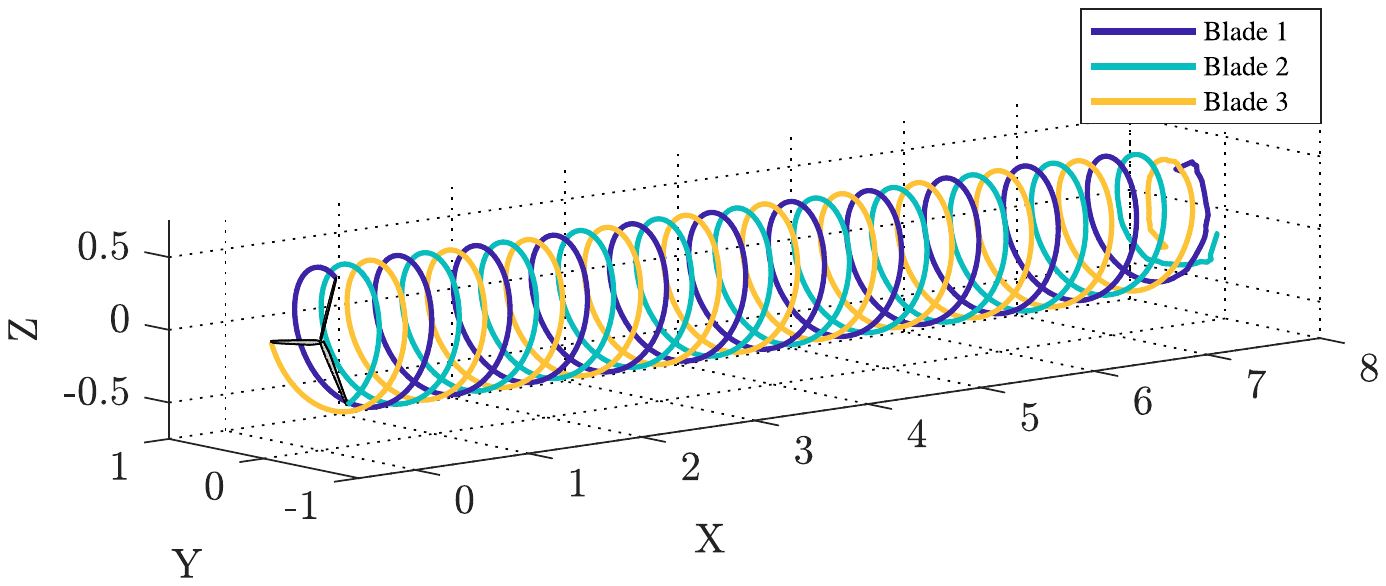}
		\caption[]%
		{ Zero-pitched rigid rotor configuration }    
		\label{snapshot_rsym}
	\end{subfigure}
	\hfill
	\begin{subfigure}[b]{0.495\textwidth}  
		\centering 
		\includegraphics[scale=0.575, trim=4cm 10cm 25 9cm]{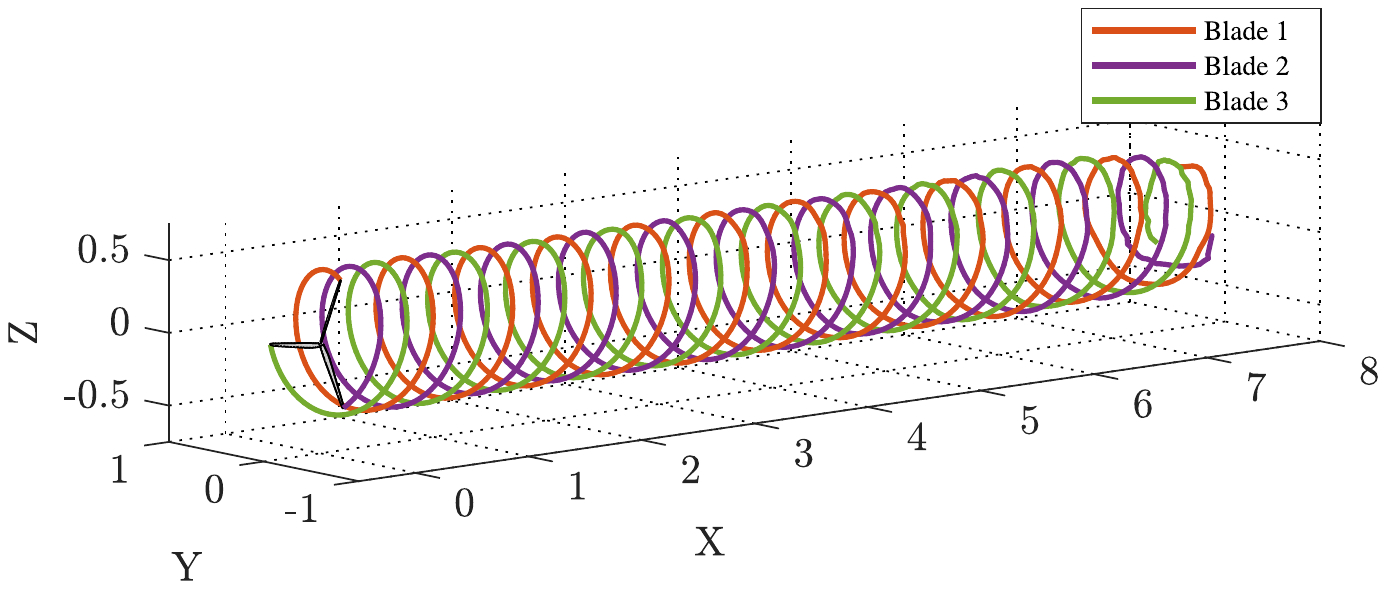}
		\caption[]%
		{Pitched rigid rotor configuration}
		\label{snapshot_rtilt}
	\end{subfigure}
	\hfill
	\begin{subfigure}[b]{0.495\textwidth}  
		\centering 
		\includegraphics[scale=0.575, trim=4cm 10cm 25 8cm]{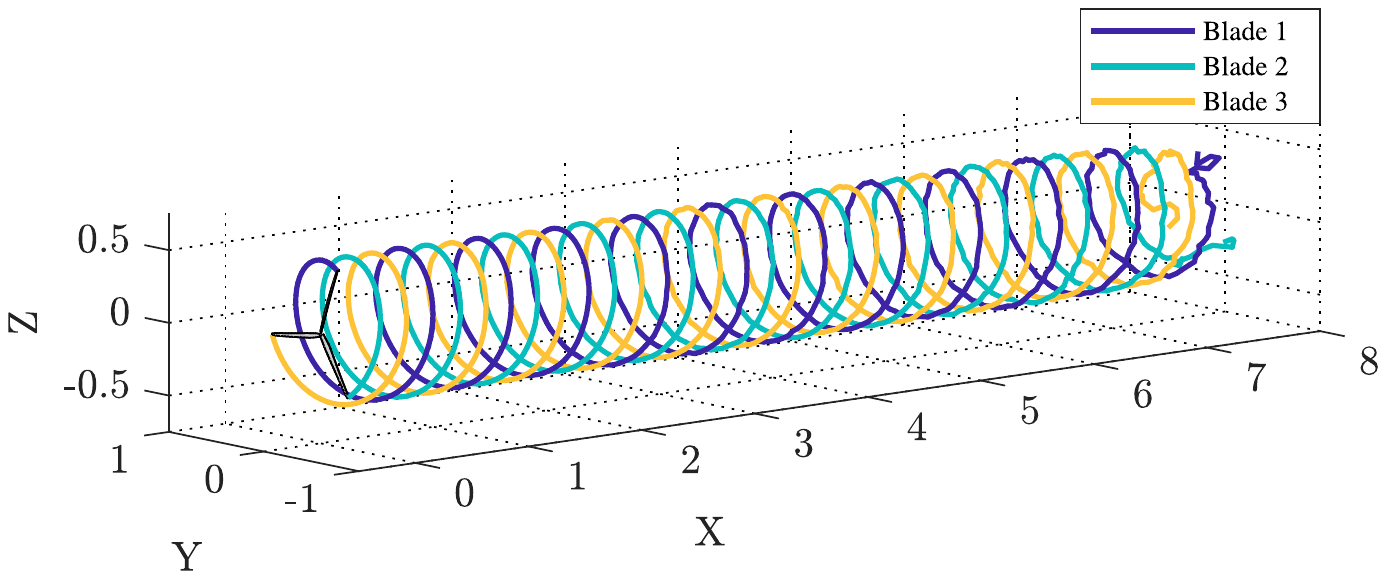}
		\caption[]%
		{Zero-pitched flexible rotor configuration}
		\label{snapshot_rtilt}
	\end{subfigure}
	\hfill
	\begin{subfigure}[b]{0.495\textwidth}  
		\centering 
		\includegraphics[scale=0.575, trim=4cm 10cm 25 8cm]{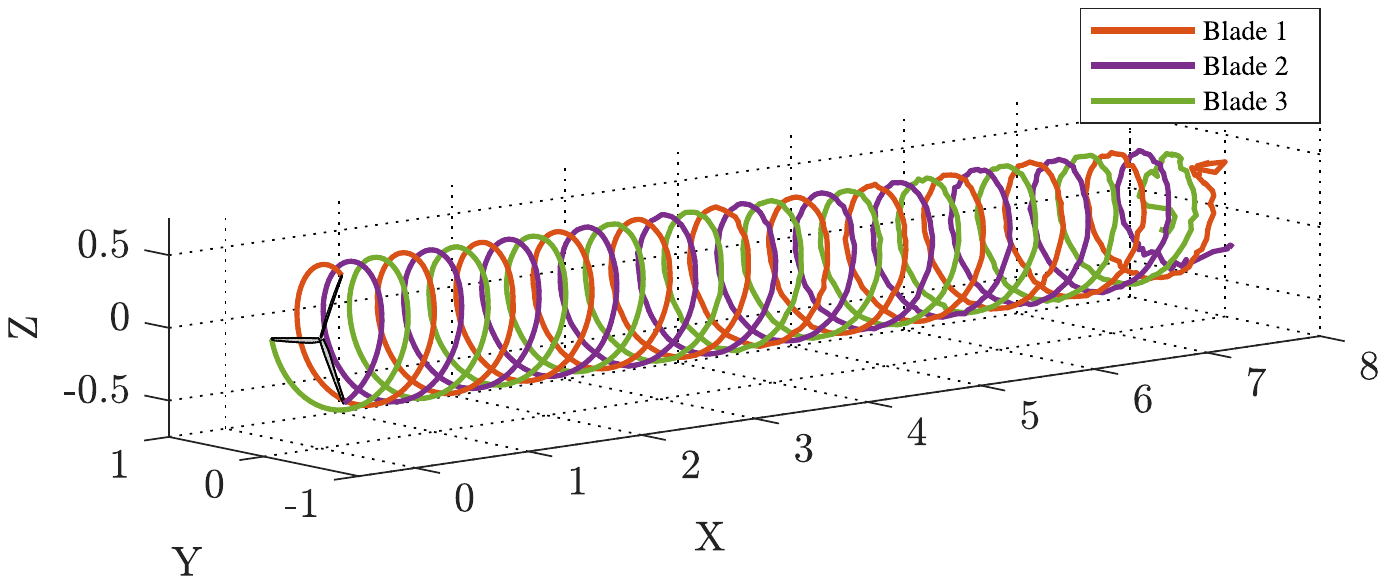}
		\caption[]%
		{Pitched flexible rotor configuration}
		\label{snapshot_rtilt}
	\end{subfigure}
	\caption[]
	{Case 2 tip-vortex snapshot at $t=$ 60 s for rigid and flexible rotor operation} 
	\label{snapshot_arsymtilt}
\end{figure*}

Case 3 of the NREL 5MW wind turbine is now investigated. Snapshots of the ZPR and PR configurations at $t=60$ s are shown in Fig.~\ref{snapshot_arsymtilt}. The generation of tip vortices presented in Fig.~\ref{snapshot_arsymtilt} exhibit much more coherent and smooth helical vortex geometry than any of the other cases for both ZPR and PR  configurations. The coherence of the tip vortex geometry is attributed to the decrease in tip-speed ratio, which increases the distance between adjacent vortices (helical pitch) that mitigates the influence of perturbation propagation. Unlike prior cases 1 and 2, it is difficult to qualitatively highlight in case 3 the influence of rotor-plane tilt on the stability of tip vortices as both ZPR and PR configurations maintain a very coherent tip vortex structure. Therefore, no clear qualitative conjecture can be made about the influence of operational configurations on numerical and qualitative wake break down. Similarly, the influence of blade flexibility on the stability of tip vortices generated in ZPR and PR configurations is not clearly observed qualitatively, as the $15^{\circ}$ blade pitch has reduced aerodynamic loading of the blades and blade deformation. Thus, unlike prior cases, no clear qualitative hypothesis can be made on the influence of blade elasticity on the stability of tip vortices.  

\begin{figure*}[h!]
	\centering
	\begin{subfigure}[b]{0.495\textwidth}
		\centering
		\includegraphics[scale=0.55, trim=7cm 8cm 6cm 9cm]{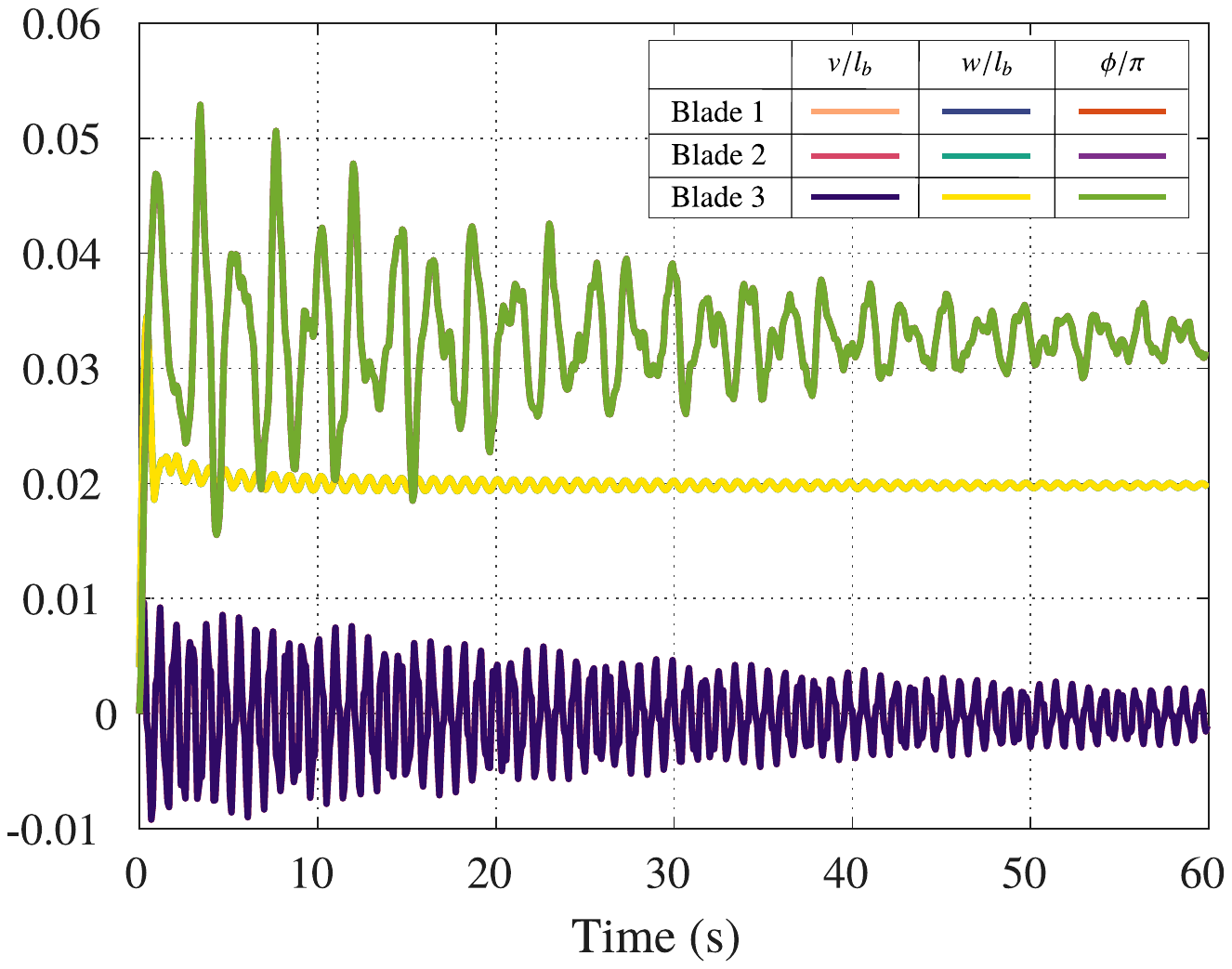}
		\caption[]%
		{ Zero-pitched rotor configuration}    
	\end{subfigure}
	\hfill
	\begin{subfigure}[b]{0.495\textwidth}  
		\centering 
		\includegraphics[scale=0.55, trim=3cm 8cm 25 9cm]{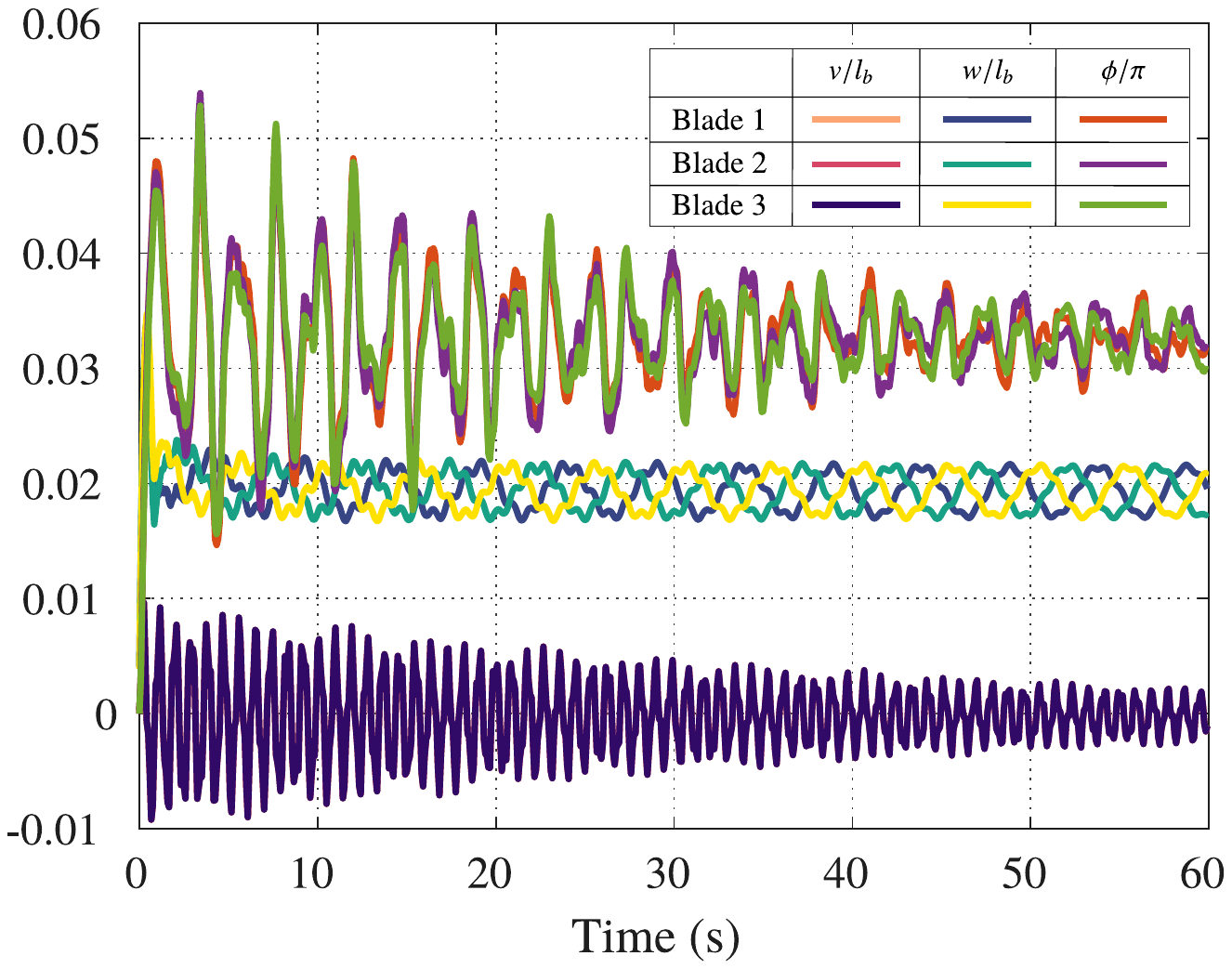}
		\caption[]%
		{Pitched rotor configuration}
	\end{subfigure}
	\caption[]
	{Case 3 tip-vortex snapshot at $t=$ 60 s for rigid and flexible rotor operation } 
	\label{AR_aeroelastic_results}
\end{figure*}

Figure \ref{AR_aeroelastic_results} presents the aeroelastic responses at the tips of the rotor-blades for both ZPR and PR configurations in case 3. It is important to note that case 3 corresponds to an above-rated operational mode, in which the NREL 5MW reference wind turbine introduces blade pitch to reduce aerodynamic loading resulting from high wind-speeds \cite{jonkman:6}. In addition, work presented by \citet{rodriguezJRE} (Fig.~16 in \cite{rodriguezJRE}) show that case 3 exhibits negative angles-of-attack from mid-length to the tips of rotor-blades, which in turn can impact stability of tip vortices. In this case, the dominant aeroelastic response for both configurations remains the flapwise deformation. Here, the flapwise degree-of-freedom exhibits an initial transient deformation of about 3.5\% the blade length and begins converging to a steady-state deformation of 2\% the blade length at about $t=15$ s into the simulation. The ZPR configuration shows identical behavior for all rotor-blades across and degrees-of-freedom, while the PR configuration shows blades exhibiting an out-of-phase periodic behavior in the flapwise degrees-of-freedom as previously seen in cases 1 and 2. The time-histories of the edgewise deformation exhibit negligible responses, and torsional responses reflect about 3.3\% of $\pi$. Torsional responses are again effectively negligible with regard to its aerodynamic impact on the wake dynamics as the torsional frequency response amplitude is much lower relative to that of the flapwise degree-of-freedom, which again is the dominant kinematic response.

\begin{figure}
	\centering
	\begin{subfigure}[b]{0.475\textwidth}
		\centering
		\includegraphics[scale=0.5, trim=6.5cm 8cm 6cm 10cm]{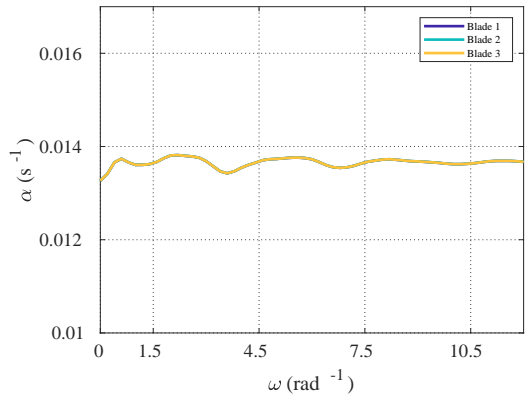}
		\caption[]%
		{Rigid zero-pitched rotor configuration}    
		\label{avt_ar_symtilt_oms}
	\end{subfigure}
	\hfill
	\begin{subfigure}[b]{0.475\textwidth}  
		\centering 
		\includegraphics[scale=0.5, trim=7.5cm 8cm 6cm 10cm]{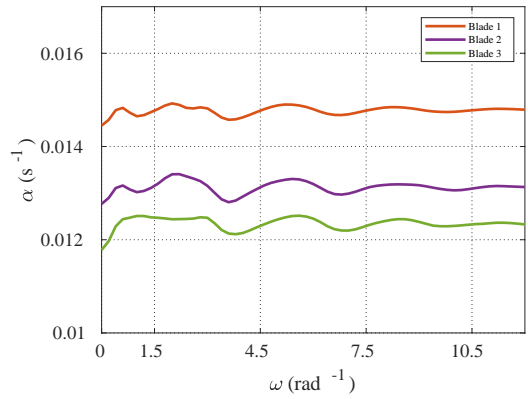}
		\caption[]%
		{Rigid pitched rotor configuration}
		\label{avt_ar_symtilt_blds}
	\end{subfigure}
	\begin{subfigure}[b]{0.475\textwidth}  
	\centering 
	\includegraphics[scale=0.5, trim=8cm 8cm 6cm 8cm]{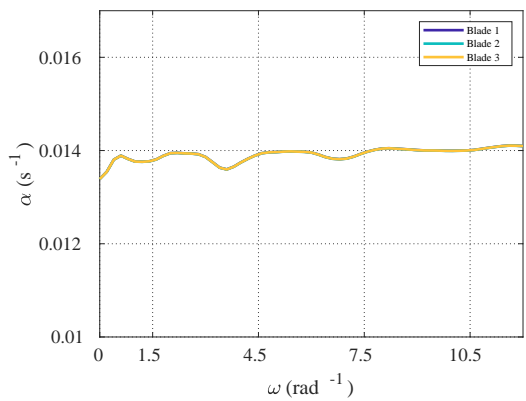}
	\caption[]%
	{Flexible zero-pitched rotor configuration}
	\label{avt_ar_symtilt_blds}
\end{subfigure}
	\begin{subfigure}[b]{0.475\textwidth}  
	\centering 
	\includegraphics[scale=0.5, trim=6cm 8cm 6cm 8cm]{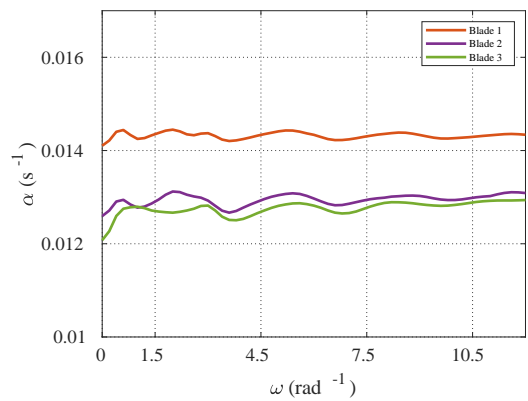}
	\caption[]%
	{Flexible pitched rotor configuration}
	\label{avt_ar_symtilt_blds}
\end{subfigure}
	\caption[]%
	{Stability trend snapshots at $t=40$ s of tip vortices shed from rigid and flexible ZPR rotor (left column) and PR configurations (right column).}
	\label{ar_sym_avo}
\end{figure}

Snapshots of the stability analysis performed at $t=40$ s for tip vortices in a two rotation window ($4\pi$ window) are presented in Fig.~\ref{ar_sym_avo}. Snapshots of the stability analyses qualitatively show that eigenvalues show less variance across a range of perturbation wavenumbers than in other cases. In fact, case 3 appears not to adhere to the classical stability criteria, namely that eigenvalue peaks of stability trends do not correspond to wavenumber perturbations equal to $\omega=N_b\left(k-1/2\right)$. In fact, the stability trend shifts forward some $\epsilon$ amount, i.e. $\omega=N_b\left(k-1/2\right)+\epsilon$. Figure \ref{ar_sym_avo} shows that at $t=40$ s the stability for PR configurations reach higher eigenvalues than ZPR configurations.

\begin{figure*}[h!]
	\centering
	\begin{subfigure}[b]{0.475\textwidth}
		\centering
		\includegraphics[scale=0.6, trim=6cm 8cm 6cm 7cm]{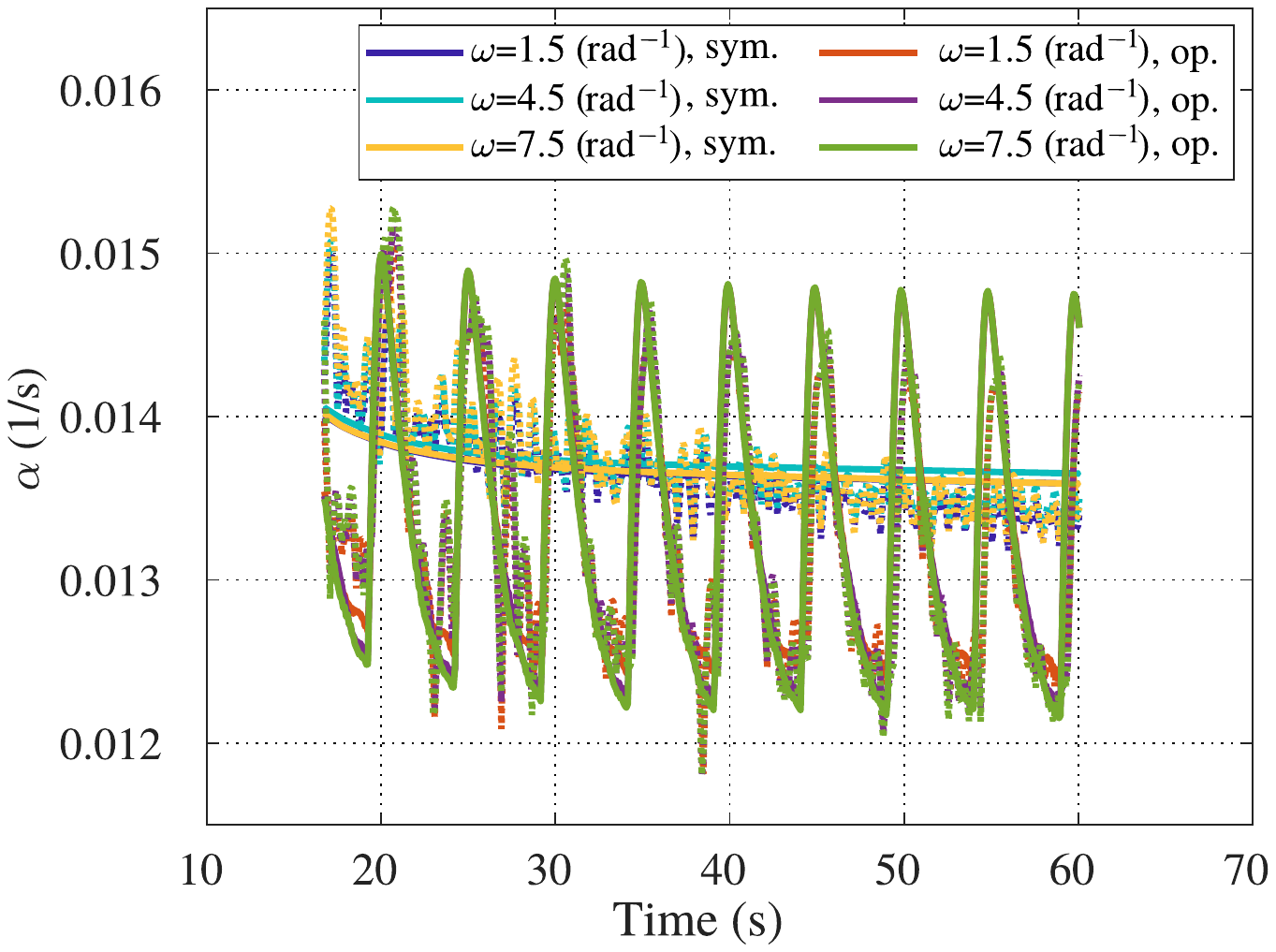}
		\caption[]%
		{Eigenvalue time-histories for blade 1}    
		\label{avt_ar_symtilt_oms}
	\end{subfigure}
	\hfill
	\begin{subfigure}[b]{0.475\textwidth}  
		\centering 
		\includegraphics[scale=0.6, trim=6cm 8cm 6cm 7cm]{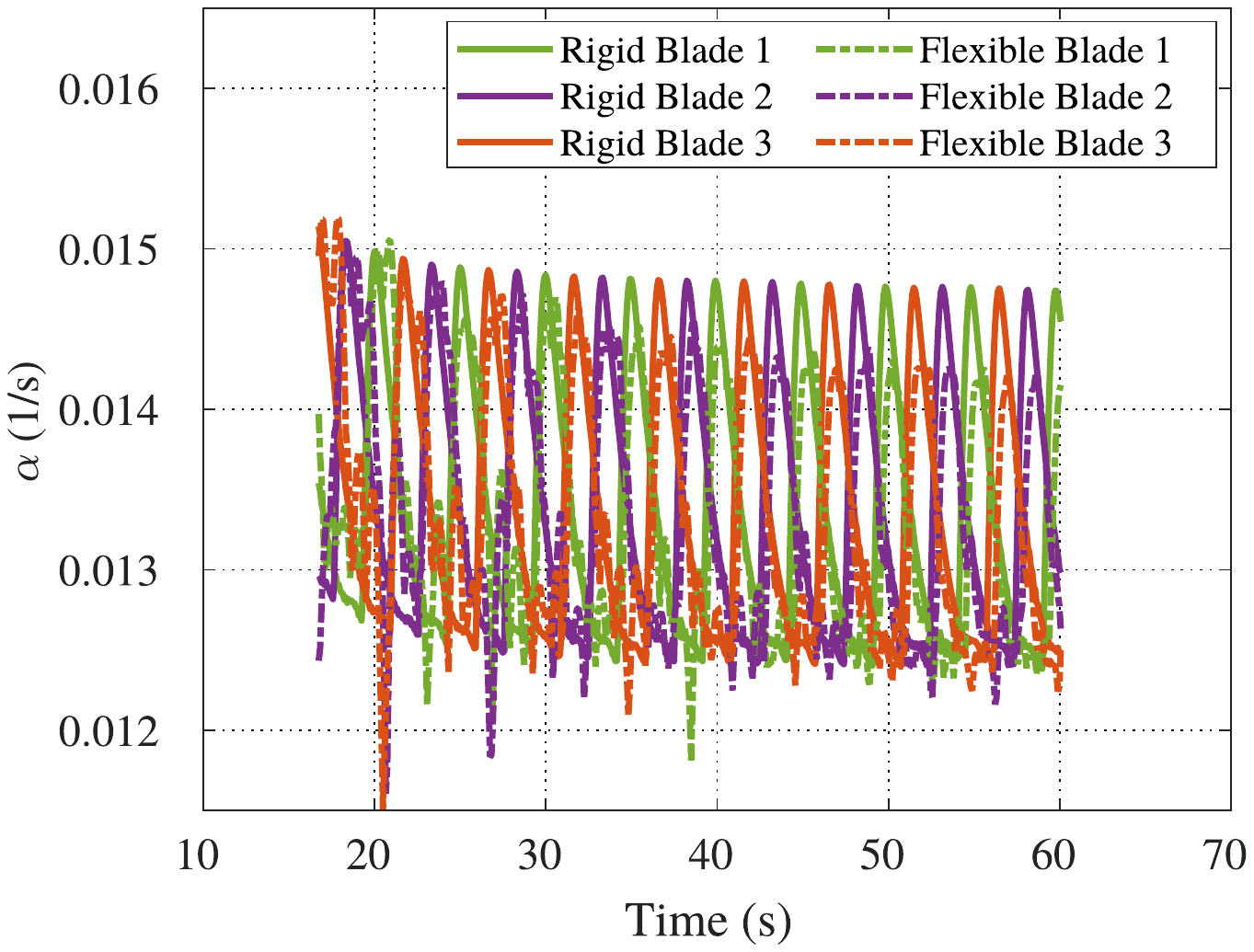}
		\caption[]%
		{Eigenvalue time-history for blade 1, 2, and 3}
		\label{avt_ar_symtilt_blds}
	\end{subfigure}
	\caption[]
	{Case 3 time history operation growth rates tip vortex 1: a) eigenvalues corresponding to perturbation wavenumbers $\omega$=1.5, 4.5, 7.5 rad\textsuperscript{-1} for ZPR and PR configurations (legend provides color scheme of rotor configuration and dashed lines represent time-history of growth-rates from flexible blades); b) operational configuration eigenvalues corresponding to $\omega$=1.5 rad\textsuperscript{-1} for all three blades.} 
	\label{avt_ar_symtilt}
\end{figure*}

The time histories of eigenvalues corresponding to perturbation wavenumbers $\omega=1.5,\:4.5$ and $7.5$ rad$^{-1}$, i.e., $N_b=3$ and $k=1,2,3$ for $\omega=N_b\left(k-1/2\right)$ are now tracked in time for ZPR and PR configurations in Fig.~\ref{avt_ar_symtilt}. Unlike cases 1 and 2 shown in Figs.~\ref{avt_br} and \ref{avt_r}, respectively, the time histories of eigenvalues in case 3 for perturbation wavenumbers $\omega=1.5,\:4.5$ and $7.5$ are nearly identical in both the ZPR and PR configurations. For both configurations, it is seen that as time progresses blade flexibility generates more stable (lower positive eigenvalues) than rigid rotors. Finally, although qualitative differences between tip vortices generated by ZPR and PR configurations in Fig.~\ref{snapshot_arsymtilt} are minor, the underlying physical differences are highlighted by the temporal stability characteristics shown in Fig.~\ref{avt_ar_symtilt}, which demonstrates that the periodic change in angle-of-attack generated by the rotor-plane tilt results in higher eigenvalues of the PR configuration, thereby generating more unstable tip vortices. Figure \ref{avt_ar_symtilt_blds} also shows the out-of-phase behavior of tip vortex stability trends generated by the out-of-phase period change in angle-of-attack for individual rotor blades.

\begin{figure*}[h!]
	\centering
	\begin{subfigure}[b]{0.475\textwidth}
		\centering
		\includegraphics[scale=0.6, trim=6cm 8cm 6cm 7cm]{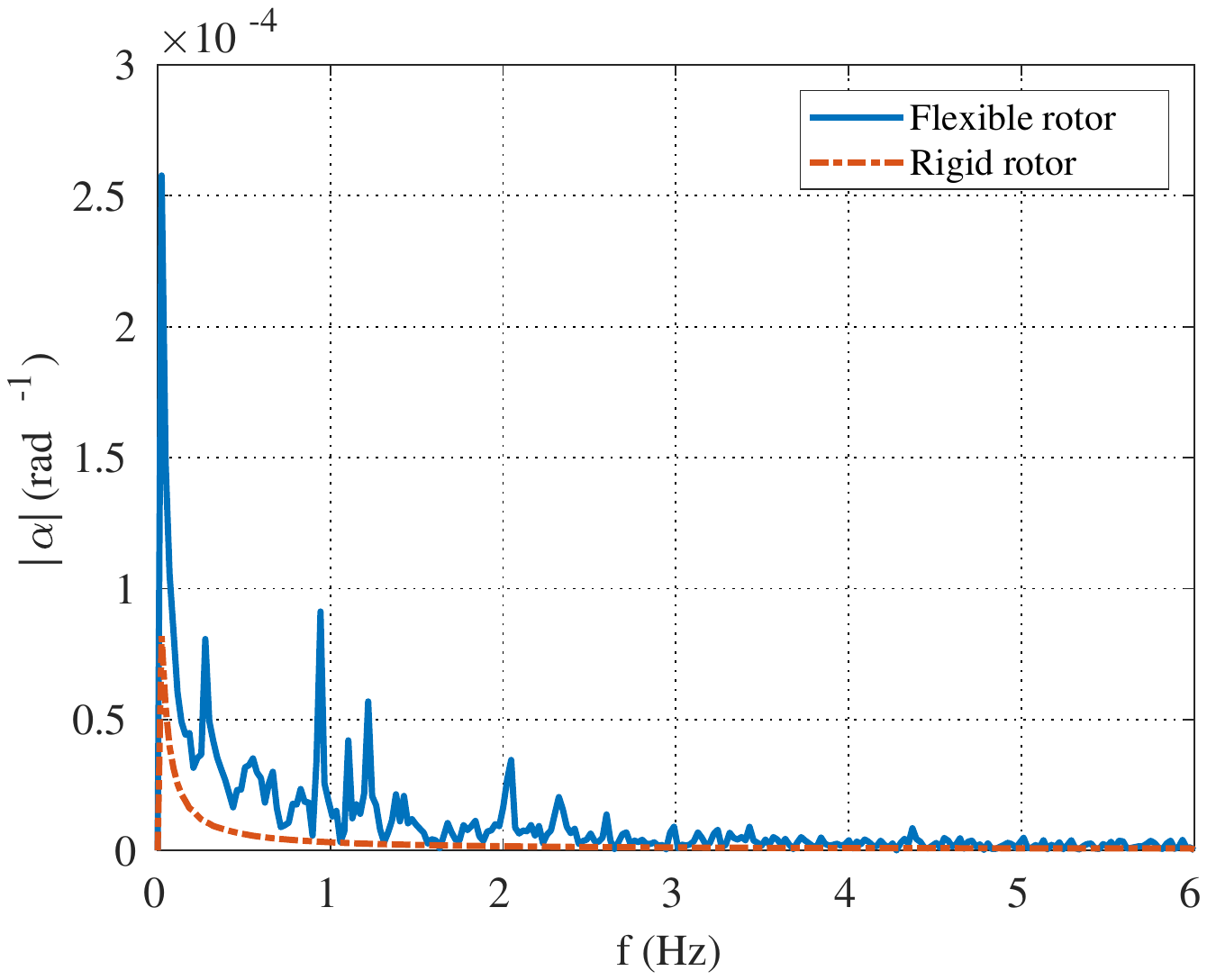}
		\caption[]%
		{ Zero-pitched rotor configuration }    
		\label{ar_sym_FFT}
	\end{subfigure}
	\hfill
	\begin{subfigure}[b]{0.475\textwidth}  
		\centering 
		\includegraphics[scale=0.6, trim=6cm 8cm 6cm 7cm]{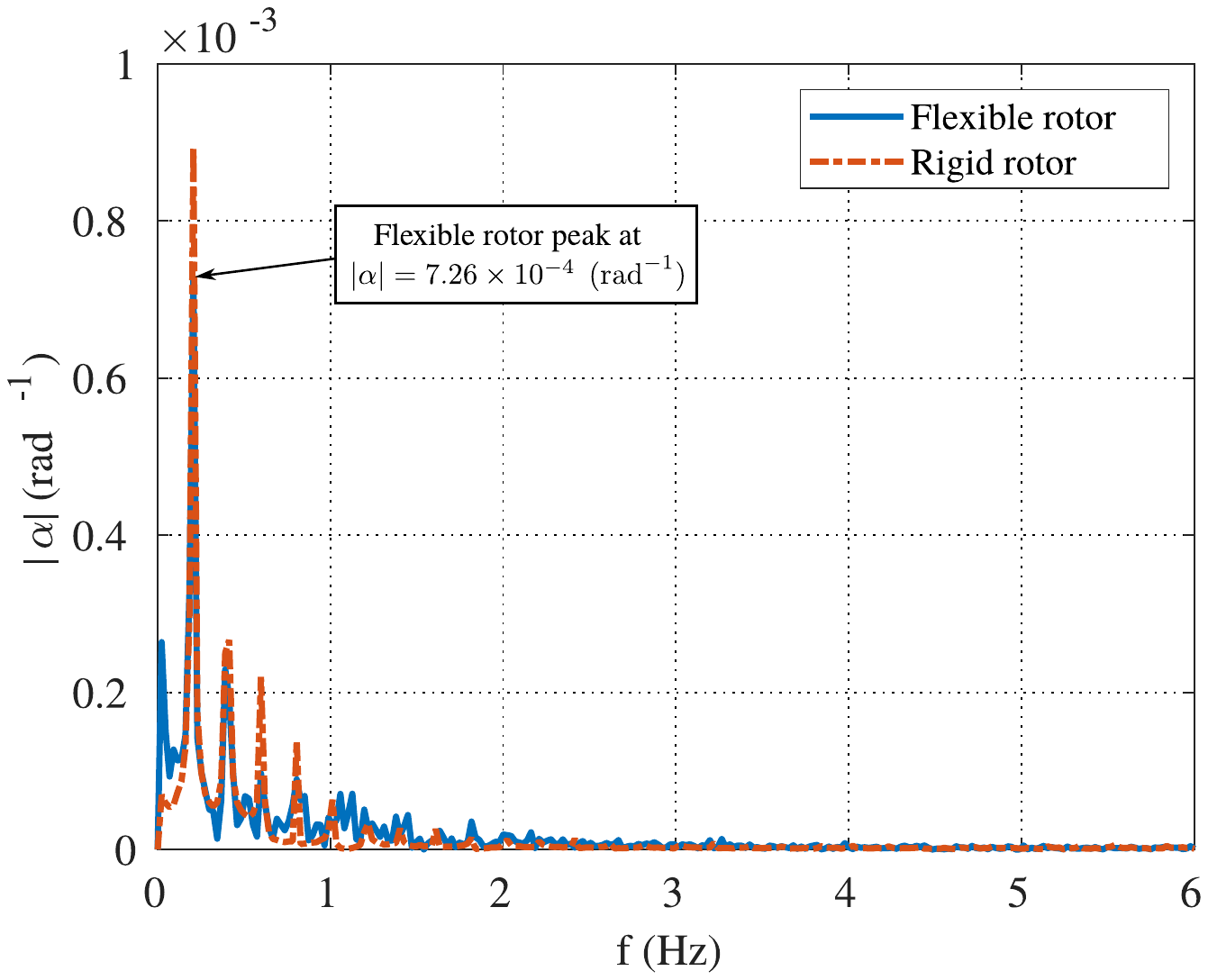}
		\caption[]%
		{Pitched rotor configuration}
		\label{ar_tilt_FFT}
	\end{subfigure}
	\caption[]
	{FFT-derived frequency spectra of eigenvalue time history signal for perturbation wave number $\omega=1.5$ rad\textsuperscript{-1}: a) ZPR configuration, and b) PR configuration} 
	\label{ar_FFT}
\end{figure*}

The frequency spectra of the eigenvalue time-histories for case 3 ZPR and PR rotor configurations is shown in Fig.~\ref{ar_FFT}. For the symmetric rotor configuration, the FFT shows a dominant low frequency contribution that was also seen previously in cases 1 and 2, namely $f\approx0.02$ Hz. It is also interesting to note that for the tip vortices generated by ZPR flexible configurations, this low frequency contribution is much more pronounced than a rigid rotor.  As mentioned previously, identification of this low-frequency contribution is not trivial but seems to be related to the agglomeration of tip vortices downstream, and it may be that blade flexibility amplifies this low-frequency contribution in the amplitude spectrum.  For the flexible rotor ZPR configuration, the FFT is showing a frequency contribution near the first natural frequency of the NREL rotor blade ($f_1\approx 1.2$ Hz), and exactly at the first flapwise natural frequency of the NREL rotor blade. The operational configuration FFT is dominated by the frequency at which the angle-of-attack changes in the tilted rotor plane, which for case 3 is also $f=\Omega/2\pi\approx 0.2$ Hz, since the rotation rate is equal to the rotation rate of case 2 operational conditions. 

\section{Discussion}

The tip vortex analyses presented show that rotor pitch and blade flexibility alters the time-dependent content of tip vortex stability. By introducing the rotor pitch, the time-history of eigenvalues corresponding to perturbation wave numbers $\omega=1.5,\:4.5$ and $7.5$ rad$^{-1}$ fluctuate at a frequency of $f=\Omega/2\pi$. \citet{rodriguezJERT} advanced the claim that stability trends are dependent on the blade passing frequency corresponding to $f=\Omega/2\pi$. However, their investigation incorrectly identified time-varying stability trends for tip vortices shed from different blades despite their base simulation being a zero-pitched rotor configuration. \citet{rodriguezphd} also identified the same frequency content in time-history trends for cases 1-3, but it was also assumed that the stability trend fluctuation was a byproduct of the blade passing frequency. The current investigation shows that the stability trend fluctuation is not a byproduct of the blade passing frequency, but a result of the periodic change in angle-of-attack as the rotor blade travels around the tilted rotor-plane. It was also found that blade flexibility introduces spectral content to the stability trends near or equal to the first natural frequency of the rotor blade, which corresponds to the flapwise mode. In fact, for both ZPR and PR cases it was found in corresponding FFTs that the flapwise mode and its natural frequency were the highest aeroelastic contribution in the stability trend. The presence of the flapwise spectral content in the stability trends suggest that the flapwise mode can be the leading aeroelastic contributor to wake breakdown. However, the impact that both blade flexibility and rotor-pitch have had on the stability trends indicate that tip vortices are most influenced by the largest dynamic or kinematic response in the rotor dynamical system, i.e., stability trends are shaped by the dominant spectral content present in the rotor dynamical system,  such as blade deformation or periodic changes of angle-of-attack due to rotor configuration. This conclusion may be an important aspect to further investigate tip vortex stability of floating offshore wind turbines, where the dominant operational dynamics may interchange from rotor dynamics to environmental dynamics (wave-induced loading and rigid body motion).

The analyses presented herein have also shown the impact that blade flexibility has on tip vortex stability for an aeroelastic rotor based on the NREL 5MW reference wind turbine rotor. \citet{rodriguezJERT} investigated initially the role of flexibility on the stability of tip vortices and concluded that blade deformation destabilizes tip vortices. However, their stability analyses were conducted on the entire tip vortex geometry, which included perturbing tip vortex locations that were already initially perturbed by artificial transients due to the impulse loading of the inflow conditions and numerical instabilities. To avoid performing stability analyses on tip vortices perturbed by numerical artifacts, \citet{rodriguezphd} performed a windowed analysis to identify regions where numerical instabilities would not corrupt stability analyses.  \citet{rodriguezphd} employed the windowed stability analysis and found that by evaluating time-history of eigenvalues corresponding to $\omega=1.5$ rad$^{-1}$ wavenumbers, blade flexibility can generate less unstable (lower positive eigenvalues) tip vortices than rigid rotors. However, no time history analyses were performed on higher wavenumber pertubations. The present analyses align well with results presented in \cite{rodriguezphd} for wavenumbers $\omega=1.5$ rad$^{-1}$. Furthermore, it was also observed that for cases 1 and 2, where no rotor blade pitch is present, blade flexibility can reduce sensitivity to low wavenumber perturbations and increase sensitivity to higher wavenumbers. 

Stability analyses of tip vortices for case 3 showed the lowest eigenvalues across a range of wavenumber perturbations compared to cases 1 and 2. These relatively low eigenvalues reinforce the observation that the coherent and long tip vortex structures in Fig.~\ref{snapshot_arsymtilt} show minimal signs of wake break-down, i.e. case 3 is the least unstable. Case 3 configurations (i.e.~rotor configurations with a $\theta_{\textup{bl}}=15^{\circ}$ blade pitch) also showed no agreement with the classical stability trend for $\alpha$ v.~$\omega$. In fact, it was seen that peak eigenvalues were located at forward shifted pertubation wavenumbers. Furthermore, eigenvalues demonstrated monotonic values across the range of perturbation wavenumbers relative to case 1 and 2 stability trends. The monotonic behavior of tip vortex stability was highlighted in the time history analysis in Fig.~\ref{avt_ar_symtilt_oms}, where tip vortices perturbed at $\omega=1.5,\:4.5$ and $7.5$ rad$^{-1}$ showed almost identical eigenvalues. An overview of the mathematical framework of the stability analysis highlights that by introducing blade pitch the local induced velocity field may introduce stability characteristics that are vastly different than the classical stability trends of the tip vortices. However, further investigations are required to identify the underlying mechanisms at play that cause the change in stability trends due to blade pitch.

\section{ Conclusions}\label{conclusion_sec}

The stability of tip vortices shed from flexible rotors have been investigated numerically. In contrast to prior work, the presented effort, for the first time in the literature, employed a strongly-coupled aeroelastic FVM numerical framework to generate the tip-vortex structure and perform corresponding tip-vortex linear-eigenvalue stability analyses based on two vortex core models: the Vatistas finite core model and the cutoff model. It was found that the Vatistas model used to desingularize the Biot-Savart was susceptible to divergent numerical artifacts that are dependent on vortex shedding frequency and were not present in the cutoff modeling. Further investigation into the impact of numerical time integration with Vatistas and cutoff finite core modeling are necessary to understand the numerical issues at hand. Nevertheless, it was found that by employing a windowed stability analysis, whereby only the early-aged segments of tip vortices are considered, classical stability trends were recovered on a three-bladed canonical rotor.

The validated stability analysis was then employed on the tip vortices generated by the NREL 5MW reference wind turbine for rigid and flexible blades under zero-pitched and pitched rotor (5$^{\circ}$ rotor-plane tilt) configurations for three distinct operational conditions: below-rated (case 1), rated (case 2), and above-rated (case 3), all of which, to the best of the authors' knowledge, is the first time presented in the literature. The tip vortex stability analyses and corresponding time-history analyses presented in the investigation demonstrated three key findings and contributions to the literature of vortex dynamics: 1) tip vortex stability trends are shaped by the dominant spectral content present in the rotor dynamical system; 2) blade flexibility may generate tip vortices that are less sensitive to low wavenumber perturbations but more sensitive to higher wavenumber perturbations; and 3) introducing blade pitch alters the local induced velocity field and alters tip vortex stability such that peak eigenvalue trends do not adhere to classical stability criteria. 

Though the current work has presented new contributions into the field of tip-vortex dynamics and stability from an aeroelastic perspective, there remains a need to further address the limitations of the numerical framework and employ its utilities to draw more general conclusions about aeroelastic mechanisms present in tip-vortex stability. Specifically, future work will entail a comparative study between stability analyses conducted by windowed truncation and an FVM-tailored time-integration approach to highlight each method's limitations and/or benefits. Future work will also include performing parametric investigations of general rotorcraft design and configurations to more conclusively determine the mechanism aeroelasticity plays on tip vortex stability.

\section{Acknowledgments}
Most of this research was performed while the first author was at Lehigh University. The first and second authors wish to acknowledge the support of the Air Force Office of Scientific Research under Award No. FA9550-15-1-0148 monitored by Dr. Gregg Abate. The first author also wishes to acknowledge the support from the Naval Research Laboratory's Karles Fellowship. The third author acknowledges the support from Technical Data Analysis Inc.~through the Small Business Innovation Research Topics N171-027 and N171-0416 under Phase II agreements NRL-2019-051  and NRL-2019-024 and the support from the Office of Naval Research through the Naval Research Laboratory’s core
funding.

\bibliography{mybibfile}

\end{document}